\documentclass[review,3p,times]{elsarticle}

\usepackage[ruled,linesnumbered]{algorithm2e}
\usepackage{algpseudocode}
\usepackage{array}
\usepackage{comment}
\usepackage{multirow}
\usepackage{subcaption}
\usepackage{caption}
\usepackage{booktabs}
\usepackage{float}
\usepackage{mathtools}
\usepackage{bm}
\usepackage{xcolor}
\usepackage{hyperref}
\hypersetup{
    colorlinks=true,
    linkcolor=blue,
    urlcolor=blue,
}

\usepackage{amssymb}
\usepackage[numbers]{natbib}
\journal{Journal of Computational Physics}

\begin{document}

\begin{frontmatter}



\title{Efficient Solution of Bimaterial Riemann Problems for Compressible Multi-Material Flow Simulations}


\author{Wentao Ma}
\author{Xuning Zhao}
\author{Shafquat Islam}
\author{Aditya Narkhede}
\author{Kevin Wang\corref{mycorrespondingauthor}}
\cortext[mycorrespondingauthor]{Corresponding author}
\ead{kevinwgy@vt.edu}

\address{Department of Aerospace and Ocean Engineering, 
	Virginia Polytechnic Institute and State University,
	Blacksburg, Virginia 24061, USA}

\begin{abstract}
When solving compressible multi-material flow problems, an unresolved challenge is the computation of advective fluxes across material interfaces that separate drastically different thermodynamic states and relations. A popular idea in this regard is to locally construct bimaterial Riemann problems, and to apply their exact solutions in flux computation. For general equations of state, however, finding the exact solution of a Riemann problem is expensive as it requires nested loops. Multiplied by the large number of Riemann problems constructed during a simulation, the computational cost often becomes prohibitive. The work presented in this paper aims to accelerate the solution of bimaterial Riemann problems without introducing approximations or offline precomputation tasks. The basic idea is to exploit some special properties of the Riemann problem equations, and to recycle previous solutions as much as possible. Following this idea, four acceleration methods are developed, including (1) a change of integration variable through rarefaction fans, (2) storing and reusing integration trajectory data,  (3) step size adaptation, and (4) constructing an R-tree on the fly to generate initial guesses. The performance of these acceleration methods is assessed using four example problems in underwater explosion, laser-induced cavitation, and hypervelocity impact. These problems exhibit strong shock waves, large interface deformation, contact of multiple ($>2$) interfaces, and interaction between gases and condensed matters. For all the problems, the acceleration methods are able to significantly reduce the computational cost without affecting solver robustness or solution accuracy. In different cases, the solution of bimaterial Riemann problems is accelerated by $37$ to $87$ times. As a result, the total cost of advective flux computation, which includes the exact Riemann problem solution at material interfaces and the numerical flux calculation over the entire computational domain, is accelerated by $18$ to $81$ times.
\end{abstract}

\begin{comment}
    \begin{highlights}
\item Acceleration technique for the iterative exact Riemann solver, increasing the computational speed by {\color{red}XXX times (hopefully)} by saving and reusing intermediate solutions from previous iterations.
\item Multiphase CFD simulations are accelerated by {\color{red}XXX} times without any sacrifice in accuracy.
\item {\color{red}EOS1, EOS2, and EOS3} have been tested in this paper.
\item Comparing to the published acceleration technique using sparse grid methods, the proposed acceleration method is more user-friendly because it does not need users to provide additional information to the solver.
% \item utilizing inherent property of rarefaction waves
\end{highlights}
\end{comment}


\end{frontmatter}

%\newpage
%\tableofcontents
%\newpage

\section{Introduction}
\label{sec:intro}

Fluid flows involving multiple materials or phases arise in many scientific and engineering disciplines. In this paper, we consider a class of problems in which different materials (or phases) are separated by sharp interfaces, and the flow needs to be treated as being compressible. Many fluid dynamics problems involving bubbles, droplets, phase transitions, and chemical reactions fall into this category (e.g.,~\cite{Zhao2023, xiang2017, Houim2013}). Some problems in solid and soft matter mechanics also belong to this category, when the material behaves like a compressible fluid due to some special loading condition (e.g.,~\cite{krimmel2010simulation,Islam2023}). Because of compressibility, the physical model typically involves a thermodynamic equation of state (EOS) for each material. Across a material interface, discontinuity arises not only in some state variables (e.g.~density, internal energy), but also in the EOS. These discontinuities can be quite significant, if the interface separates gaseous and condensed phases.

Different numerical methods have been developed to represent material interfaces. In Lagrangian~\cite{Neumann1950, Vilar2016} and Arbitrary Lagrangian-Eulerian (ALE)~\cite{Loubere2010, MARBOEUF2019} frameworks, the computational mesh is updated to conform to the material interface. In the Eulerian framework, the material interface is tracked using numerical methods such as interface capturing~\cite{Banks2007, kitamura2014}, level set~\cite{levelset2018}, volume of fluid~\cite{vof2002}, and front tracking~\cite{Tryggvason2001frontTracking}. Regardless of the method used to represent interfaces, the spatial discretization of the governing equations (e.g.~Navier-Stokes equations) at an interface presents new challenges not encountered in single-phase flow simulations. Many popular numerical flux functions (e.g.~Roe, HLLC, HLLE) assume a single EOS. Therefore, they cannot be directly applied at material interfaces. It is also well known that spurious oscillations often occur near material interfaces, even if the applied numerical scheme performs well for single-phase flows (e.g.,~\cite{Liu2005_newMGFM, Johnsen2012}). For example, Brummelen {\it et al.} showed that when an interface capturing method is used, spurious pressure oscillation may occur due to the loss of pressure-invariance property in discretization~\cite{Brummelen2003}. Fedkiw {\it et al.} developed the ghost fluid method, in which ghost states on both sides of a material interface are constructed~\cite{Fedkiw1999}. In order to suppress an ``overheating'' phenomenon, one needs to apply the ``isobaric fixing'' technique that requires solving another auxiliary partial differential equation~\cite{Fedkiw1999,Fedkiw1999AnIF}. Some authors have also reported that the original ghost fluid method may produce inaccurate results when an interface is hit by a shock wave, or the density or EOS jumps significantly across it~\cite{Liu2003_MGFM, Liu2005_newMGFM, Sambasivan2009}.

In the past two decades, many authors have proposed the idea of constructing bimaterial Riemann problems across material interfaces, and applying their exact solutions in flux computation~\cite{Liu2005_newMGFM,Wang2006_rGFM,Aslam2003, Sambasivan2009,Terashima2010,Cocchi1997,Igra2002,kitamura2014,Farhat2008ExactRiemann, Farhat2012FIVER,Bo2011, Houim2013, Jafarian2017}. For example, Liu {\it et al.} incorporated exact Riemann solvers in their ghost fluid method, in order to correctly predict the double rarefaction waves in nearly cavitating flow conditions~\cite{Liu2005_newMGFM}. Later, Wang {\it et al.} used the exact Riemann solutions to populate not only ghost cells but also the real fluid cells near material interfaces, thereby eliminating the need of isobaric fix~\cite{Wang2006_rGFM}. Cocchi and Saurel developed a prediction-correction scheme in which the solutions of exact Riemann problems are used to correct flow states near contact discontinuities~\cite{Cocchi1997}. Igra and Takayama modified this method to make it work in conjunction with the level-set method~\cite{Igra2002}. Also, Kitamura {\it et al.} conducted a comparative study of the AUSM (Advection Upstream Splitting Method)  family of schemes, and found that when combined with the exact Riemann problem solver, AUSM schemes are applicable to more challenging problems with larger pressure ratios~\cite{kitamura2014}. Recently, Farhat {\it et al}. developed the FIVER (FInite Volume method with Exact multiphase Riemann solvers) method, in which the exact solutions of bimaterial Riemann problems are used directly to compute advective fluxes, instead of populating ghost or real fluid cells~\cite{Farhat2008ExactRiemann, Farhat2012FIVER, wang_lea_farhat_2015,main_zeng_avery_farhat_2017,ho2021discrete}. FIVER has been applied to study several shock-dominated multiphase flow and fluid-structure interaction problems, including underwater explosion and implosion~\cite{farhat_wang_main_kyriakides_lee_ravi-chandar_belytschko_2013, ma2022computational}, pipeline explosion~\cite{wang_lea_farhat_2015}, cavitation erosion~\cite{wang_2017,cao2019shock, cao2021shock,xiang_ma_liang_yu_liao_sankin_cao_wang_zhong_2021}, laser-induced vaporization~\cite{Zhao2023}, and hypervelocity impact~\cite{Islam2023}. Furthermore, some authors have proposed to extend exact Riemann problem solvers to account for additional interfacial physics, such as surface tension and phase transition~\cite{Bo2011, Houim2013, Jafarian2017}.

A unique advantage of the constructed bimaterial Riemann problems is that they account for not only the discontinuity of state variables (e.g.~density, internal energy) across a material interface, but also the change of thermodynamic EOS. With this in mind, one may expect this idea to be more impactful for problems that require sophisticated, highly nonlinear EOS. In reality, however, most of the efforts reviewed above deal with simple EOS, such as perfect and stiffened gases. For general EOS, computational cost becomes a major issue, as finding the exact solution of a Riemann problem requires nested loops~\cite{toro2013riemann,kamm2015exact}. In particular, when the solution contains a rarefaction fan, each iteration in the outer loop requires numerically integrating two state variables through the fan, with the upper bound of the integral being an unknown variable~\cite{kamm2015exact}. As a result, finding the exact solution of a Riemann problem often requires evaluating the EOS thousands of times. In contrast, approximate Riemann solvers (e.g.~Roe, HLLC) only call the EOS a few times. Therefore, although the exact Riemann problem solver is only applied within a lower-dimensional subset (i.e.,~material interfaces) of the computational domain, its cost can be much higher than the approximate Riemann solver that is applied in the entire domain. This dramatic increase of computational cost often makes the simulation unfeasible.

In this work, our objective is to accelerate the solution of bimaterial Riemann problems without introducing approximations or requiring additional precomputation. Our basic idea is to exploit some special properties of the Riemann problem equations, and to recycle previous solutions (obtained during the same simulation) as much as possible. Following this idea, we present four acceleration methods that can be used either in combination or separately. Three of the four methods are designed to accelerate the numerical integration through rarefaction fans, as it is the most expensive part of the solution procedure. We show that through a change of integration variable, the upper bound of the integral can be determined. We also show that the integral curve (i.e.,~isentrope) is uniquely determined by the inputs of each constructed Riemann problem. Therefore, the numerical integration performed within each iteration of the solution procedure is along the same curve, only with a different end point. In light of this observation, we present a method that eliminates repeated calculations by storing and reusing previous integration trajectories. The conventional solution procedure uses a fixed integration step size. We show that the nonlinearity of the EOS and the high degree of automation needed here allow us to benefit from step size adaptation. Furthermore, the cost of the exact Riemann problem solver also depends on the quality of the initial guesses. The fourth acceleration method is designed to generate better initial guesses, by storing previous Riemann solutions in a five-dimensional R-tree~\cite{Guttman1984Rtrees,rtreeBook} and finding initial guesses through nearest-neighbor search. To assess the performance of these acceleration methods, we solve four example problems in underwater explosion, laser-induced cavitation, and hypervelocity impact. These problems exhibit strong shock waves, large interface deformation, contact of multiple ($>2$) interfaces, and interaction between gases and condensed matters. The robustness, accuracy, and computational efficiency of the acceleration methods are put to the test.

A major difference between the current work and previous acceleration efforts (e.g.,~\cite{Colella1985,Liu2005_newMGFM,Rallu2009,Farhat2012FIVER}) is that here we do not introduce approximations or require any offline precomputation. Previously, Colella and Glaz accelerated single-material Riemann problem solvers by developing a local parametrization for the EOS~\cite{Colella1985}. This method assumes slow variation of the thermodynamic state, which may not be true in challenging multi-material flow problems. Liu {\it et al.} found that their ghost fluid method equipped with a specifically designed approximate two-phase Riemann problem solver cannot provide correct solutions under nearly cavitating flow conditions~\cite{Liu2005_newMGFM}. Exact solutions of Riemann problems were thus employed to address the two strong rarefaction waves in this situation. Farhat {\it et al.} avoided online numerical integration by tabulating and interpolating the Riemann invariants on sparse grids~\cite{Rallu2009,Farhat2012FIVER}. For the problem they considered, which involves the stiffened gas EOS and the Jones-Wilkins-Lee (JWL) EOS, only one two-dimensional sparse grid needs to be constructed. But for arbitrary EOS, two three-dimensional sparse grids need to be constructed for each material. This offline precomputation can be expensive. It also requires a high level of expertise from the user, who is expected to specify, before the simulation, the resolution of the sparse grids as well as the lower and upper bounds of each tabulated variable. These complexities motivated us to pursue a different path, that is, to accelerate the Riemann solver without introducing approximations or requiring offline precomputation.

The remainder of this paper is organized as follows. In Sec.~\ref{sec:equation}, we present the governing compressible Navier-Stokes equations and the bimaterial Riemann problem with general convex and smooth EOS. We also briefly describe the aforementioned FIVER method, which is adopted in this work to perform numerical tests. In Sec.~\ref{sec:acceleration}, we discuss the computational efficiency of the exact Riemann problem solver in detail. The acceleration methods mentioned above are presented. In Sec.~\ref{sec:numericalExperiments}, we present the numerical tests and discuss the performance of the acceleration methods. Finally, we provide a summary in Sec.~\ref{sec:summary}. 

\section{Governing equations and solution framework}
\label{sec:equation}

\subsection{Navier-Stokes equations with general equations of state}
In general, the dynamics of compressible multi-material flows are governed by Navier-Stokes equations, which enforce the conservation of mass, momentum, and energy. In the literature, the exact solutions of bimaterial Riemann problems are used mostly to compute the advective fluxes (e.g.,~\cite{Farhat2012FIVER, Houim2013}). Therefore, we write the Navier-Stokes equations as
\begin{equation}
\frac{\partial \bm{U}}{\partial t} + \nabla \cdot \mathcal{F}(\bm{U}) = \bm{\mathcal{S}},\quad\quad\text{in}~ \Omega\times(0,~t_{\text{max}}],
\label{eq:NS_3d_1}
\end{equation}\
which emphasizes the advective fluxes $\mathcal{F}(\bm{U})$. Here, $\bm{U}$ denotes the conservative state vector. $t$ denotes time, and $\bm{\mathcal{S}}$ represents collectively the additional terms that may need to be included in the conservation laws, depending on the specific flow being analyzed. Examples include viscous fluxes, heat diffusion fluxes, heat sources, and body forces. Because bimaterial Riemann problems are neither constructed based on these terms nor used to evaluate them, we do not formulate $\bm{\mathcal{S}}$ explicitly. $\Omega$ denotes the spatial domain of the flow, and $t_{\text{max}}$ the maximum physical time of interest. $\bm{U}$ and $\mathcal{F}(\bm{U})$ are given by
\begin{equation}
\bm{U} = \begin{bmatrix}
\rho \\
\rho \bm{u} \\
\rho e_t
\end{bmatrix},
\quad
\mathcal{F}(\bm{U}) = \begin{bmatrix}
\rho \bm{u}^T \\
\rho \bm{u} \otimes \bm{u} + p \bm{I} \\
(\rho e_t + p)\bm{u}^T
\end{bmatrix},
\label{eq:NS_fluxes}
\end{equation}
where $\rho$, $e_t$, and $p$ denote the density, total energy per unit mass, and pressure, respectively.
 $\bm{u}=[u, v, w]^T$ is the velocity vector. $\bm{I}$ denotes the $3 \times 3$ identity matrix. In addition,
 \begin{equation}
 e_t = e + \dfrac{1}{2}|\bm{u}|^2,
 \end{equation}\
 where $e$ is the internal energy per unit mass.

\begin{comment}
    \begin{equation}
\bm{U} = \begin{bmatrix}
\rho \\ \rho u \\ \rho v \\ \rho w \\ E
\end{bmatrix},
\quad
\bm{\mathcal{F}}(\bm{U}) = \begin{bmatrix}
\rho u  & \rho v  & \rho w \\ 
\rho u^2 + p  &  \rho uv  &  \rho u w \\
\rho u v  &  \rho v^2 + p  &  \rho v w \\
\rho u w &  \rho v w  &  \rho w^2 + p\\
\rho H u  &  \rho H v  &  \rho H w
\end{bmatrix},
\label{eq:NS_fluxes}
\end{equation}\
where $\rho$ and $p$ denote fluid density and pressure, $\bm{u} = [u, v, w]^T$ denotes the velocity vector, $E$ denotes the total energy per unit volume, i.e.,
\begin{equation}
E = \rho e + \frac{1}{2} \rho \|\bm{u}\|_2^2,
\label{eq:total_energy_per_volume}
\end{equation}\
where $e$ denotes the internal energy per unit mass. $H$ denotes the total enthalpy per unit mass, given by
\begin{equation}
H = \dfrac{1}{\rho}\big(E + p).
\label{eq:total_entalphy_per_mass}
\end{equation}\
\end{comment}

We assume that the fluid domain $\Omega$ consists of multiple non-overlapping open sets (i.e.~subdomains) occupied by different materials, i.e.,
\begin{equation}
\Omega = \bigcup\limits_{m=1}^M \text{cl}(\Omega_{m}),\quad \Omega_{m}\cap\Omega_n = \emptyset,~\forall m,~n~(m\neq n),
\end{equation}\
where $\text{cl}(\Omega_m)$ denotes the closure of $\Omega_m$. Within each subdomain $\Omega_m$, a thermodynamic equation of state (EOS) of the form
\begin{equation}
p = p(\rho, e)
\label{eq:eos_all}
\end{equation}\
is needed to close the governing equations. The speed of sound, $c$, is given by
\begin{equation}
c = \sqrt{\dfrac{\partial p}{\partial \rho}\Big|_e + \frac{p}{\rho^2}\dfrac{\partial p}{\partial e}\Big|_\rho}.
\label{eq:speed_of_sound}
\end{equation}\

Here, we introduce a few specific examples that are used in the numerical tests in Sec.~\ref{sec:numericalExperiments}. Nonetheless, the algorithms presented in this paper are designed to be applicable to general convex and smooth equations of state in the form of~\eqref{eq:eos_all}.

As a generalization of perfect gas, the noble-Abel stiffened gas equation of state~\cite{NobelAbel2016} is given by
\begin{equation}
p=(\gamma-1) \dfrac{\rho(e  - e_c)}{1  - \rho b}  - \gamma p_{c},
\label{eq:stiffened_EOS}
\end{equation}\
where $\gamma$, $e_c$, $b$, and $p_c$ are constant parameters. The sound speed is given by 
\begin{equation}
    c\left(p, \rho\right) = \sqrt{\dfrac{\gamma (p + p_c)}{\rho - \rho^2 b}}.
    \label{eq:soundSpeedSG}
\end{equation}\

The Jones-Wilkins-Lee (JWL) equation of state is widely used to model the thermodynamics of detonation products~\cite{menikoff2015jwl}. It has the expression
\begin{equation}
p=\omega \rho e + f(\rho),
\label{eq:JWL_EOS}
\end{equation}\
where
\begin{equation}
f(\rho) = A_{1}\left(1-\frac{\omega \rho}{R_{1} \rho_{0}}\right)  \exp \Big(-\frac{R_{1} \rho_{0}}{\rho}\Big)+A_{2}\left(1-\frac{\omega \rho}{R_{2} \rho_{0}}\right) \exp \Big(-\frac{R_{2} \rho_{0}}{\rho} \Big),
\end{equation}\
Here, $\rho_0$, $\omega$, $A_1$, $A_2$, $R_1$, and $R_2$ are constant parameters. The sound speed is given by
\begin{equation}
c(p, \rho)=\sqrt{\frac{\gamma p-f(\rho)+\rho f^{\prime}(\rho)}{\rho}}.
\label{eq:eq:soundSpeedJWL}
\end{equation}\

The Mie-Gr\"{u}neisen equation of state is often used to model solids and liquids under high pressure. In this work, we adopt the formulation given in~\cite{Robinson2019mie}, i.e.,
\begin{equation}
p = \dfrac{ \rho_0 c_0^2 \eta}{(1 - s \eta)^2} \Big(1 - \dfrac{1}{2}\Gamma_0\eta\Big) + \rho_0\Gamma_0 e,
\label{eq:MG_EOS}
\end{equation}\
with $\eta = 1 - \rho_0 / \rho$. $\rho_0$ and $c_0$ denote the density and bulk speed of sound in the ambient condition. $s$ is the slope of the Hugoniot curve. $\Gamma_0$ is the Grüneisen parameter in the ambient condition. In this case, the sound speed is given by
\begin{equation}
c(p, \rho)= \sqrt{ \frac{\rho_0^2 c_0^2}{\rho^2} \frac{1+(s-\Gamma_0)\eta}{(1-s\eta)^3} + \frac{p}{\rho^2}\Gamma_0\rho_0}.
\label{eq:eq:soundSpeedMG}
\end{equation}\

For these examples, both $e$ and $c$ can  be evaluated analytically in closed-form. There are more complicated EOS for which one or both of them can only be evaluated through numerical iteration (e.g.~\cite{ANEOS2021}). As a result, the computational cost may increase drastically.

If the term $\bm{\mathcal{S}}$ in \eqref{eq:NS_3d_1} involves temperature, a temperature law is also needed for each material. However, the temperature law is not involved in the construction and solution of exact Riemann problems. Therefore, additional details are omitted for the sake of conciseness.

\subsection{Formulation of bimaterial Riemann problems at material interfaces}
\label{sec:exactRiemann}

\begin{figure}[H]
    \centering
    \includegraphics[width=160mm,trim={0cm 0cm 0cm 0cm},clip]{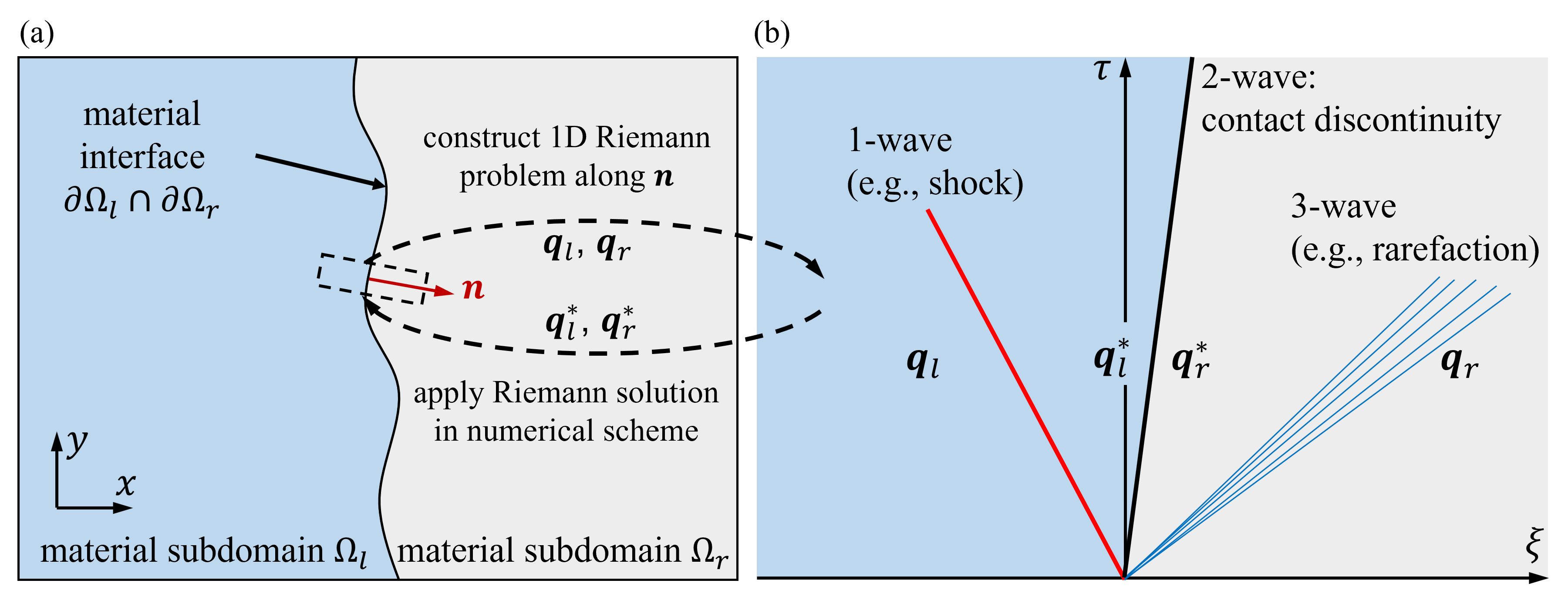}
    \caption{Construction and solution of a bimaterial Riemann problem at  material interface.}
    \label{fig:interfaceRiemann}
\end{figure}

At the interface between two adjacent material subdomains, normal velocity and pressure are assumed to be continuous, i.e.,
\begin{equation}
\begin{matrix}
\left( \bm{u}_l - \bm{u}_r \right) \cdot \bm{n} = 0, \\
p_l - p_r = 0,
\end{matrix}
~~~~~\text{on}~\partial\Omega_l\cap\partial\Omega_r,
\label{eq:interface_ff}
\end{equation}\
where $\bm{n}$ denotes the unit normal to the material interface, and the subscripts $l$ and $r$ indicate the one-sided limits of the state variable ($\bm{u}$ or $p$) obtained within the two subdomains. At each point on the interface, the bimaterial Riemann problem is defined as the 1D Euler equations~\cite{Cocchi1997, Igra2002, Liu2005_newMGFM, Aslam2003, Sambasivan2009, Terashima2010, kitamura2014, Farhat2008ExactRiemann, Farhat2012FIVER},
\begin{equation}
\frac{\partial \bm{q}}{\partial \tau} + \frac{\partial \mathcal{F} \left( \bm{q} \right)}{\partial \xi} = 0, 
\label{eq:FF_Riemann}
\end{equation}\
with a piecewise constant initial condition obtained from the original 2D or 3D flow field (Fig.~\ref{fig:interfaceRiemann}). Here, $\tau$ is a fictitious time coordinate, $\xi$ is the spatial coordinate, which in principle is along the direction of $\bm{n}$. The initial condition, which features a discontinuity at the material interface, is given by
\begin{equation}
    \bm{q} \left(\xi, 0\right) = 
    \begin{dcases}
        \bm{q}_l,&\text{if}~\xi \leq 0,\\           \bm{q}_r,&\text{if}~\xi > 0,
    \end{dcases}
    \label{eq:riemannIC}
\end{equation}\
with
\begin{equation}
\bm{q}_{l} \equiv \begin{bmatrix}
\rho_{l}\\
\rho_{l} u_l\\
\rho_l e_{tl}
\end{bmatrix}
=
\begin{bmatrix}
\rho_{l}\\
\rho_{l} \big(\bm{u}_{l} \cdot \bm{n}\big)\\
\rho_l e_l + \dfrac{1}{2} \rho_l \big(\bm{u}_l\cdot\bm{n}\big)^2
\end{bmatrix},\quad
\bm{q}_{r} \equiv \begin{bmatrix}
\rho_{r}\\
\rho_{r} u_r\\
\rho_r e_{tr}
\end{bmatrix}
= \begin{bmatrix}
\rho_{r}\\
\rho_{r} \big(\bm{u}_{r} \cdot \bm{n}\big)\\
\rho_r e_r + \dfrac{1}{2} \rho_r \big(\bm{u}_r\cdot\bm{n}\big)^2
\end{bmatrix}.
\end{equation}\

Solving the Riemann problem exactly generally means finding a weak entropy solution that consists of three waves, as illustrated in Fig.~\ref{fig:interfaceRiemann}(b). The one in the middle (i.e.,~the 2-wave) is a contact discontinuity that tracks the motion of the material interface. It separates the two material subdomains in which different EOS should be applied. Each of the other two waves can be either a shock wave or a rarefaction fan. These three waves divide the $\xi$-$\tau$ space into four regions, plus the interior of any rarefaction fan. The outer two regions have not been affected by the shock and/or rarefaction yet, so the initial fluid states $\bm{q}_l$ and $\bm{q}_r$ sill hold there (Fig.~\ref{fig:interfaceRiemann}(b)). The state variables in the two middle regions next to the material interface are unknowns that need to be solved for. Following the convention in the literature (``star states''), these  variables are identified by a superscript $*$. It should be noted that the aforementioned three-wave solution structure requires the EOS to be convex and smooth~\cite{Menikoff1989}. Given the convexity and smoothness of the EOS, the Riemann solution is unique if the EOS in the meantime satisfies the “medium condition” in~\cite{Menikoff1989}.

Across the contact discontinuity, pressure and velocity are continuous, i.e.,
\begin{equation}
\begin{matrix}
p^*_l = p^*_r, \\
u^*_l = u^*_r.
\end{matrix}
\label{eq:interface_ERPS}
\end{equation}\
Therefore, the solution of the exact Riemann problem automatically satisfies the interface conditions~\eqref{eq:interface_ff}.

Across a shock wave, the Rankine-Hugoniot jump conditions can be derived from~\eqref{eq:FF_Riemann}. After some algebraic manipulations, we get the following two  equations.
\begin{equation}
e_K(p_K^*,\rho_K^*) - e_K(p_K,\rho_K) + \dfrac{1}{2}(p_K + p_K^*)\Big(\dfrac{1}{\rho_K^*} - \dfrac{1}{\rho_K}\Big)  = 0
\label{eq:RH1_final}
\end{equation}\
and 
\begin{equation}
u_K^* = u_K \mp \sqrt{-\big(p_K^* - p_K\big)\Big(\dfrac{1}{\rho_K^*} - \dfrac{1}{\rho_K}\Big)},
\label{eq:RH2_final}
\end{equation}\
where $K = l, r$. Equation~\eqref{eq:RH1_final} only involves the thermodynamic variables, and is known as the Hugoniot equation. The function $e_K(p,\rho)$ evaluates internal energy using the EOS that holds in material subdomain $\Omega_K$. In~\eqref{eq:RH2_final}, the minus sign holds for $K=l$, and the plus sign holds for $K=r$.

Through a rarefaction fan, two Riemann invariants can be derived from the 1D Euler equations~\eqref{eq:FF_Riemann}. They are
\begin{equation}
    s - s_K = 0
    \label{eq:invariant1}
\end{equation}\
and 
\begin{equation}
    u - u_K \pm \int_{\rho_K}^{\rho} \dfrac{c_K(s_K,\rho)}{\rho} d\rho = 0,
    \label{eq:invariant2}
\end{equation}\
where $s$, $u$, $\rho$ denote the entropy, velocity, and density evaluated at any point within a rarefaction fan, including its boundaries. In~\eqref{eq:invariant2}, the plus sign  applies to the $1$-wave (i.e.,~$K=l$), and the minus sign applies to the $3$-wave (i.e.,~$K=r$). The first Riemann invariant~\eqref{eq:invariant1} indicates that entropy is constant through a rarefaction fan. Therefore, the sound speed can be evaluated from $c^2 = \displaystyle \frac{\partial p}{\partial \rho}\Big|_s$. Integrating this relation through the rarefaction fan, i.e.,~from the ambient state to the star state, gives
\begin{equation}
p_K^* = p_K + \int_{\rho_K}^{\rho_K^*} c_K^2(s_K,\rho) d\rho,
\label{eq:invariant1_final}
\end{equation}\
where the function $c_K(s,\rho)$ evaluates the sound speed using the EOS for material $K$. Similarly, substituting $\rho=\rho^*_K$ and $u = u^*_{K}$ into~\eqref{eq:invariant2} yields
\begin{equation}
    u^*_K  = u_K \mp \int_{\rho_K}^{\rho_K^*} \dfrac{c_K(s_K,\rho)}{\rho} d\rho,
    \label{eq:invariant2_final}
\end{equation}\
where the minus sign applies to the $1$-wave (i.e.,~$K=l$) and the plus sign applies to the $3$-wave (i.e.,~$K=r$). Although these two equations involve entropy in the evaluation of sound speed, the actual computation does not necessarily need it. We can write the sound speed $c$ as a function of $p$ and $\rho$, as given in equations~\eqref{eq:soundSpeedSG},~\eqref{eq:eq:soundSpeedJWL}, and~\eqref{eq:eq:soundSpeedMG}.

To determine whether each of the $1$ and $3$ waves is a shock or a rarefaction, we enforce an entropy condition. Each of the two waves is determined to be a shock wave if and only if $p_K^* > p_K$, where $K = l$ for the $1$-wave, and $K=r$ for the $3$-wave.

In summary, the solution of the exact Riemann problem, $\bm{q}^*_l$ and $\bm{q}^*_r$, is determined by Eqs.~\eqref{eq:interface_ERPS},~\eqref{eq:RH1_final},~\eqref{eq:RH2_final},~\eqref{eq:invariant1_final},~\eqref{eq:invariant2_final}, and  the above entropy condition. The shock speed, $u_s$, and the state variables through any rarefaction fan can be easily derived based on $\bm{q}^*_l$ and $\bm{q}^*_r$. For a general equation of state, the exact Riemann problem cannot be solved analytically. The numerical solution methods are presented in Sec.~\ref{sec:acceleration} in detail.

\subsection{Interface tracking and discretization of governing equations}
\label{sec:numerical_scheme}
We briefly describe the interface tracking and spatial discretization methods adopted in this work, thereby providing a clear context for the construction and solution of bimaterial Riemann problems. Nonetheless, the algorithms for solving the Riemann problems  --- to be introduced in Sec.~\ref{sec:acceleration}  --- are not restricted to these methods.

\begin{figure}[H]
    \centering
       \centering
       \includegraphics[width=100mm,trim={0cm 0cm 0cm 0cm},clip]{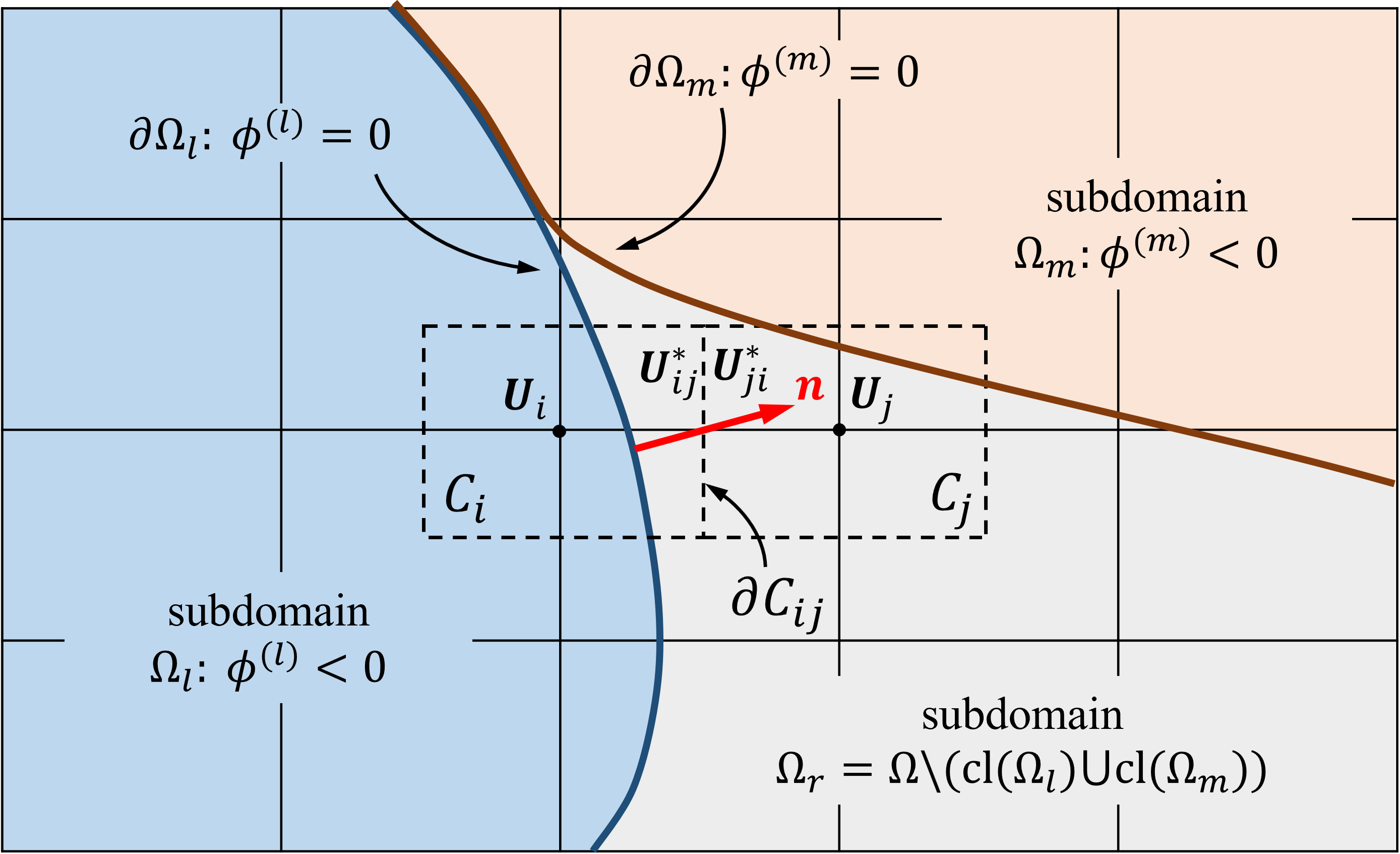}
    \caption{Interface tracking and discretization of governing equations: The level set method and FIVER.}
    \label{fig:FIVER}
\end{figure}

For a problem that involves $M~(\geq 2)$ materials, there can be as many as $C_2^M = M(M-1)/2$ material interfaces. Instead of tracking each one of them explicitly, we solve $M-1$ level set equations, i.e.
\begin{equation}
\frac{\partial \phi^{(m)} \left( \bm{x}, t \right)}{\partial t} + \bm{u} \cdot \nabla \phi^{(m)} = 0,\quad m=1,2,\cdots M-1,
\label{eq:levelSet1}
\end{equation}\
to track the boundaries of the first $M-1$ subdomains (Fig.~\ref{fig:FIVER}). Here, $\phi^{(m)} \left( \bm{x}, t \right)$ denotes a level set function, initialized to be the signed shortest distance from $\bm{x}$ to $\partial \Omega_m$. Any point $\bm{x}\in \Omega$ is within subdomain $\Omega_{m}$ ($m=1,2,\cdots M-1)$ at time $t$ if and only if $\phi^{(m)}(\bm{x},t)<0$. The last subdomain is simply 
\begin{equation}
\Omega_M = \Omega \setminus \Big(\bigcup\limits_{m=1}^{M-1} \text{cl}(\Omega_m)\Big).
\end{equation}\
Notably, all the level set equations share the same velocity field $\bm{u}$, which prevents spurious overlapping and separation of subdomains. The interface between each pair of materials  --- e.g.~$m$ and $n$  ($1\leq m<n \leq M$)  --- is simply given by
\begin{equation}
\partial\Omega_m \cap \partial\Omega_n =
\begin{cases}
\Bigl\{\bm{x}\in\Omega,~~\phi^{(m)}(\bm{x},t)=\phi^{(n)}(\bm{x},t)=0\Bigr\}, & n<M,\\
\Bigl\{\bm{x}\in\Omega,~~\phi^{(k)}(\bm{x},t)=0~~\text{iff}~~ k=m\Bigr\}, & n=M.
\end{cases}
\end{equation}\

We discretize the governing equations \eqref{eq:NS_3d_1} using the FIVER method~\cite{Main2017, Zhao2023}. Applying a standard finite volume method to the left hand side of \eqref{eq:NS_3d_1} yields
\begin{equation}
\frac{\partial \bm{U}_i}{\partial t} + \frac{1}{\left \| C_i \right \|}\sum_{j \in N(i)}\int_{\partial C_{ij}} \mathcal{F}(\bm{U}) \cdot \bm{n_{ij}} dS = \bm{\mathcal{S}}^h,
\label{eq:integrating_NS_equation}
\end{equation}\
where $C_i$ denotes a control volume in $\Omega$, $\left \| C_i \right \|$ being its volume. $\bm{U}_i$ denotes the cell-average of $\bm{U}$. $N(i)$ is the set of nodes that are connected to node $i$ by an edge. $\partial C_{ij} = \partial C_i \cap \partial C_j$ is the cell interface between  $C_i$ and $C_j$. $\bm{\mathcal{S}}^h$ represents a numerical approximation of $\bm{\mathcal{S}}$, which does not need to be based on the same finite volume method. Between cell $C_i$ and each neighboring cell $C_j$, a numerical advective flux is computed to approximate the surface integral in~\eqref{eq:integrating_NS_equation}. If $i$ and $j$ belong to the same material subdomain, a conventional flux function (e.g.~local Lax-Friedrichs (LLF), Roe, HLLC) is directly applied. If they belong to different material subdomains (e.g.~$\Omega_l$ and $\Omega_r$ in Fig.~\ref{fig:FIVER}), a 1D bimaterial Riemann problem is constructed using the method described in Sec.~\ref{sec:exactRiemann}. Its solution, $\bm{q}^*_l$ and $\bm{q}^*_r$, are mapped back to 2D or 3D, generating interface states $U_{ij}^*$ and $U_{ji}^*$. Then, the advective fluxes on the two sides of $\partial C_{ij}$ are computed as
\begin{equation}
\begin{matrix}
        F_{ij} = \left \| \partial C_{ij} \right \| \Phi\left(\bm{U}_{ij}, \bm{U}^*_{ij}, n_{ij}, \text{EOS}{(\Omega_l)}\right) \\
        F_{ji} = \left \| \partial C_{ij} \right \| \Phi\left(\bm{U}_{ji}, \bm{U}^*_{ji}, n_{ji}, \text{EOS}{(\Omega_r)}\right), \\
\end{matrix}
\label{eq:flux_riemann}
\end{equation}\
where $\left \| \partial C_{ij} \right \|$ is the area of $\partial C_{ij}$, and $\Phi$ denotes a conventional numerical flux function. $\bm{U}_{ij}$ and $\bm{U}_{ji}$ are interface states reconstructed using the MUSCL (monotonic upwinding scheme for conservation laws) method and a slope limiter. Additional details about FIVER can be found in~\cite{Main2017, Zhao2023}.

\section{Efficient solution of bimaterial Riemann problems}
\label{sec:acceleration}

\subsection{Conventional solution procedure and its efficiency issues}
\label{sec:efficiency}

We start with summarizing the conventional method for solving Riemann problems (e.g.,~\cite{toro2013riemann,kamm2015exact}), while at the same time, emphasizing its high computational cost. This cost may not be a practical issue if the user only needs to solve one or a few Riemann problems. Nonetheless, it can dominate the entire simulation cost, when bimaterial Riemann problems are constructed and solved along each edge in the mesh that penetrates a material interface, although the union of all the interfaces is still a lower-dimensional subset of the computational domain.

The equations that govern the three-wave entropy solution of the Riemann problem are Eqs.~\eqref{eq:interface_ERPS} to~\eqref{eq:invariant2_final}, together with the entropy condition described in Sec.~\ref{sec:exactRiemann}. The unknowns are $p^*~(=p_l^*=p_r^*)$, $u^*~(=u_l^*=u_r^*)$, $\rho_l^*$, and $\rho_r^*$. We define function $f(p)$ as
\begin{equation}
f(p) = u_l^*(p;~\rho_l, u_l, p_l, \text{EOS}_l)  - u_r^*(p;~\rho_r, u_r, p_r, \text{EOS}_r),
\label{eq:1dRiemann_overall_function}
\end{equation}\
where the subscripts $l$ and $r$ indicate the material subdomains separated by the interface. For an arbitrary input $p$, the first term of $f(p)$ evaluates the interface velocity $u_l^*$ that complies with an interface pressure $p^*=p$, and the left ambient state $\rho_l$, $u_l$, and $p_l$. Specifically, the entropy condition is first examined to determine whether the 1-wave is a shock or a rarefaction. If it is a shock, Eqs.~\eqref{eq:RH1_final} and \eqref{eq:RH2_final} are solved for $u_l^*$, with $K=l$ and $p^*_K=p$. If it is a rarefaction, Eqs.~\eqref{eq:invariant1_final} and \eqref{eq:invariant2_final} are solved for $u_l^*$, also with $K=l$ and $p^*_K=p$. Similarly, the second term of $f(p)$ evaluates the interface velocity $u_r^*$ that complies with an interface pressure $p^*=p$, and the right ambient state $\rho_r$, $u_r$, and $p_r$. Again, it starts with enforcing the entropy condition, followed by the solution of either \eqref{eq:RH1_final} and \eqref{eq:RH2_final}, or \eqref{eq:invariant1_final} and \eqref{eq:invariant2_final}, depending on whether the 3-wave is a shock or a rarefaction. Because of \eqref{eq:interface_ERPS}, the true solution of interface pressure, $p^*$, is a root of $f(p)$, i.e.
\begin{equation}
f(p^*) = 0.
\label{eq:1dRiemann_overall}
\end{equation}\

\begin{figure}[H]
\centering
    \includegraphics[width=100mm,trim={0cm 0cm 0cm 0cm},clip]{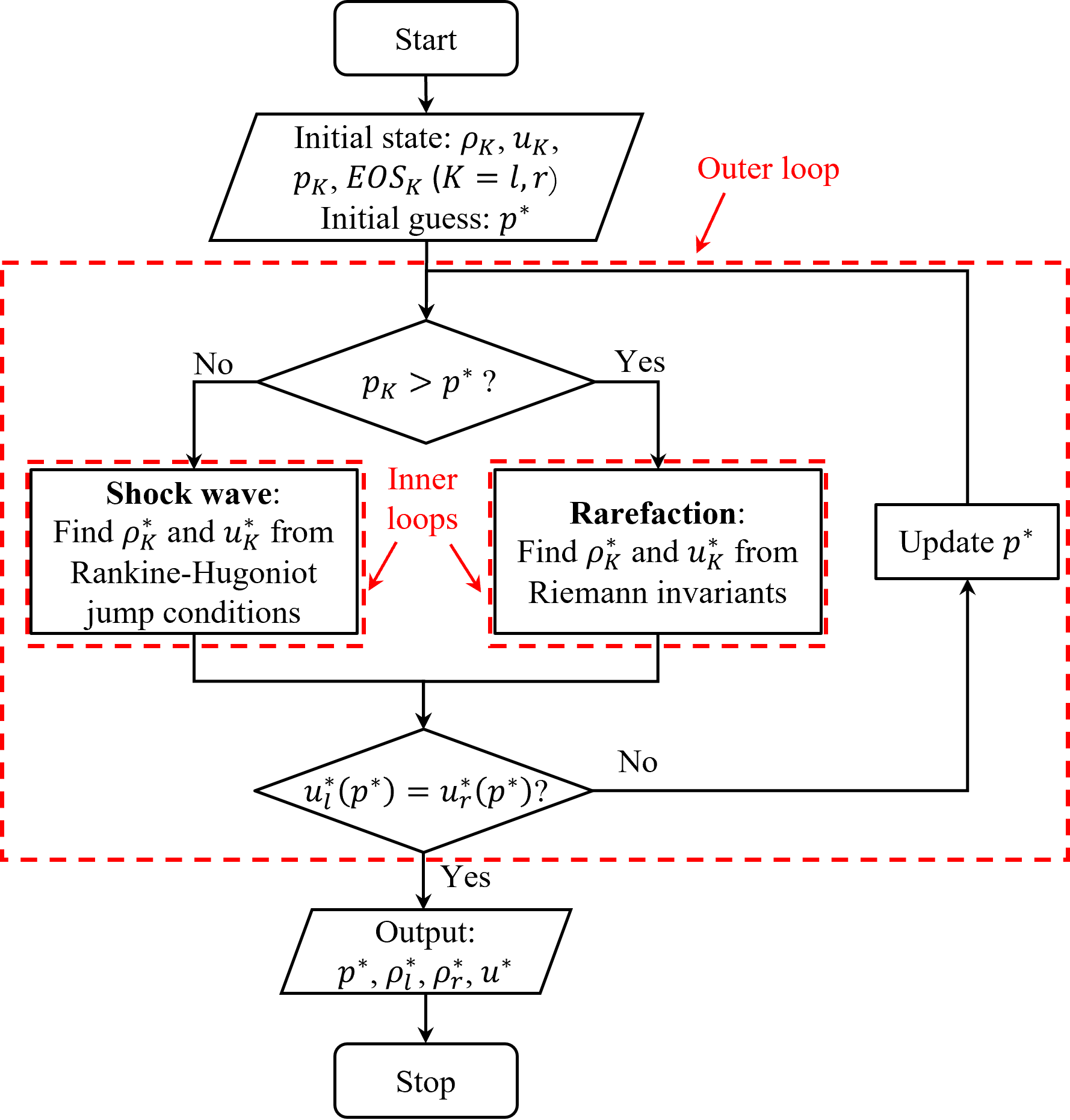}
    \caption{The conventional procedure for solving bimaterial Riemann problems.}
    \label{fig:RiemannSolver}
\end{figure}
Equation~\eqref{eq:1dRiemann_overall} is nonlinear, and for general equations of state it cannot be solved analytically. The standard method is to solve it numerically through an iterative procedure, such as the one illustrated in Fig.~\ref{fig:RiemannSolver}. Starting with an initial guess of $p^*$, in each iteration, we first apply the entropy condition to determine whether each of the $1$ and $3$ waves is a shock or a rarefaction. If it is a shock wave, $\rho^*_K$ can be first obtained by solving Eq.~\eqref{eq:RH1_final} using an iterative method (e.g.~the secant method). Then, $u^*_K$ can be obtained using Eq.~\eqref{eq:RH2_final}. 

If it is a rarefaction fan, $u^*_K$ and $\rho^*_K$ are computed from Eqs.~\eqref{eq:invariant1_final} and~\eqref{eq:invariant2_final}. For simple equations of state such as perfect gas, the integrals therein can be evaluated analytically in closed-form (e.g.~see~\cite{Farhat2012FIVER}). However, for an arbitrary equation of state, the two integrals need to be evaluated using a numerical integration method. For example, using the fourth-order Runge-Kutta method, in each step, the pressure and velocity are evaluated as
\begin{equation}
p^{(i+1)} = p^{(i)}+\frac{1}{6}\left[ \hat{c}_{1}^2 + 2\left(\hat{c}_{2}^2+\hat{c}_{3}^2\right)+\hat{c}_{4}^2\right] \Delta \rho,
\end{equation}\
\begin{equation}
u^{(i+1)}=u^{(i)} \mp \frac{1}{6} \left[ \frac{\hat{c}_1}{\hat{\rho}_1} + 2\left(\frac{\hat{c}_2}{\hat{\rho}_2} + \frac{\hat{c}_3}{\hat{\rho}_3} \right) + \frac{\hat{c}_4}{\hat{\rho}_4}\right] \Delta \rho,
\end{equation}\
where $p^{(i)}$ and $u^{(i)}$ are the pressure and density evaluated from the previous step. In the second equation, the minus sign applies to the $1$-wave, and the plus sign applies to the $3$-wave. $\hat{c}_i$ and $\hat{\rho}_i$ $(i = 1,2,3,4)$ are sound speed and density evaluated at intermediate stages within the step, specifically,
\begin{equation*}
\begin{aligned}
\hat{c}_{1}=&~c_K\left(p^{(i)},~\hat{\rho}_1\right),~\text{with}~\hat{\rho}_1 =~\rho^{(i)},\\
\hat{c}_2 =& ~c_K\left(p^{(i)} + 0.5 \Delta \rho \hat{c}_1^2,~\hat{\rho}_2\right),~\text{with}~\hat{\rho}_2 =~\rho^{(i)} + 0.5 \Delta \rho, \\
\hat{c}_3 =& ~c_K\left(p^{(i)} + 0.5 \Delta \rho \hat{c}_2^2,~\hat{\rho}_3\right),~\text{with}~\hat{\rho}_3 =~\rho^{(i)} + 0.5 \Delta \rho, \\
\hat{c}_4 =& ~c_K\left(p^{(i)} + \Delta \rho \hat{c}_3^2,~\hat{\rho}_4\right),~\text{with}~\hat{\rho}_4 =~\rho^{(i)} + \Delta \rho.
\end{aligned}
\end{equation*}\

It should be noted that in Eq.~\eqref{eq:invariant1_final}, the unknown variable is the integral's upper bound, $\rho^*_K$, while the integration result ($p^*$) on the left hand side is given. In order to find $\rho^*_K$ that yields the given $p^*$, a trial-and-error procedure is needed. For example, Kamm~\cite{kamm2015exact} proposed an iterative procedure to find a constant $\Delta \rho$ that would allow the numerical integration to end at $p^*_K$. In each iteration, the entire course of integration is repeated with an updated $\Delta \rho$. A more efficient method is to update $\Delta \rho$ in each step of the numerical integration process, based on the difference between the current pressure $p^{(i)}$ and the given $p^*$\footnote{This method is adopted in the baseline simulations presented in Sec.~\ref{sec:numericalExperiments}.}.  Either way, the computational cost is higher than standard numerical integration, where the lower and upper bounds of the integral are both given.

If the computed interface velocities, $u_l^*$ and $u_r^*$, satisfy
\begin{equation}
    \left| u^*_l(p^*)- u^*_r(p^*) \right| < \epsilon_u,
    \label{eq:riemannConverge}
\end{equation}\
an approximate solution of $p^*$ has been obtained. Here, $\epsilon_u$ denotes the error tolerance. $\rho_l^*$ and $\rho_r^*$ have also been obtained through the solution of the Hugoniot equation~\eqref{eq:RH1_final} and/or the Riemann invariant~\eqref{eq:invariant1_final}. $u^*$ can be set by
\begin{equation}
    u^* = \dfrac{1}{2}\left( u^*_l+ u^*_r \right).
\end{equation}\

The propagation speed of any shock wave can be calculated easily using the Rankine-Hugoniot jump conditions. If needed, the solution of $\rho$, $u$, and $p$ within any rarefaction fan can be obtained from the numerical integration procedure as a byproduct. 

If \eqref{eq:riemannConverge} is not satisfied, the nonlinear equation solver (i.e.,~the outer loop in Fig.~\ref{fig:RiemannSolver}) enters the next iteration with an updated guess of $p^*$, which can be obtained using the secant method.
\vspace{5mm}

Compared to approximate Riemann problem solvers, such as LLF~\cite{rusanov1970difference}, Roe and HLLC~\cite{hu2009hllc}, the cost of the exact Riemann problem solver presented above is significantly higher, as it involves nested loops (Fig.~\ref{fig:RiemannSolver}). This difference can be examined by counting and comparing the number of times the EOS are evaluated. For the exact Riemann problem solver, let $N_{\text{out}}$ denote the number of iterations in the outer loop. Let $N_{\text{sh}}$ be the average number of iterations in the inner loop for solving the shock Hugoniot equation. Let $N_{\text{rf}}$ be the number of integration steps required to go through any rarefaction fan. The number of EOS evaluations  --- including the evaluation of $c$, $e$, and $p$  --- is $2$ within each step of the shock Hugoniot equation solver, and $4$ within each integration step. Therefore, executing the algorithm shown in Fig.~\ref{fig:RiemannSolver} requires at least
\begin{equation}
N_{\text{exact}} = N_{\text{out}}(2N_{\text{sh}} + 4N_{\text{rf}})
\end{equation}\
calls to the EOS, if we assume the solution structure consists of one shock wave and one rarefaction fan. In comparison, the LLF, Roe, and HLLC solvers require $4$, $3$, and $5$ calls to the EOS, respectively. Table~\ref{tab:EOS_times}  shows a typical scenario with $N_{\text{out}}=N_{\text{sh}}=5$ and $N_{\text{rf}}=200$, in which the cost of the exact Riemann problem solver is roughly $1000$ times that of approximate Riemann problem solvers, in terms of the number of EOS evaluations.

\begin{table}[H]
\centering
\caption{Number of EOS evaluations required by exact and approximate Riemann problem solvers.}
\begin{tabular}{m{3.5cm}>{\centering\arraybackslash}m{3.3cm}>{\centering\arraybackslash}m{1cm}>{\centering\arraybackslash}m{1cm}>{\centering\arraybackslash}m{1cm}}
\toprule
 & \multirow{2}{*}{Exact Riemann solver} & \multicolumn{3}{c}{Approx. Riemann solver} \\ \cline{3-5} 
         && LLF & Roe & HLLC \\ \midrule
Shock ($=2N_{\text{out}}N_{\text{sh}}$)      & 50   & -   & -   & -    \\
Rarefaction ($=4N_{\text{out}}N_{\text{rf}}$)& 4000 & -   & -   & -    \\
Total$^*$      & 4050 & 4   & 3   & 5    \\ \bottomrule
\end{tabular}
    \begin{tabular}{ll}
    {\footnotesize $*$} & {\footnotesize Assuming one shock wave and one rarefaction fan in the solution structure (Fig.~\ref{fig:interfaceRiemann}(b))} 
    \end{tabular}
\label{tab:EOS_times}
\end{table}

The exact Riemann problem solver is only applied across or near material interfaces, which is a lower-dimensional subset of the computational domain. In comparison, approximate Riemann solvers are generally applied within the entire domain. Although the number of times the exact Riemann problem solver is used is smaller, its cost can still dominate the entire simulation cost. As an example, consider a square domain in 2D, discretized uniformly into $n$ elements in both directions. The total number of edges along which numerical fluxes need to be computed is roughly $2n^2$. Let us assume that $2n$ of them penetrate material interfaces. In this case, the ratio between the number of approximate Riemann problems and that of exact Riemann problems is approximately $n$. Based on the statistics shown in Table~\ref{tab:EOS_times}, for $n=200$, exact Riemann problem solutions account for more than $80 \%$ of the computational cost, measured by the number of EOS evaluations.

In numerical tests, we have found that during flux calculations, over $95\%$ of the computation time can be consumed by the exact Riemann problem solver. This number is higher than the estimate obtained above, for at least two reasons. First, in  some cases, the numerical integration procedure requires more than $200$ steps (i.e.,~$N_{rf}>200$) to reach a sufficient degree of accuracy (e.g.~see Fig.~\ref{fig:HVI_curve} in Sec.~\ref{sec:HVI}). Second, in parallel computations, the mesh is usually partitioned without knowledge of the location of material interfaces, which results in load imbalance. Some processors may need to solve a larger number of exact Riemann problems, if the subdomain assigned to it covers a larger fraction of material interface.  For the most burdened processor, the ratio of approximate to exact Riemann problems is often lower than $200$, which was assumed in the above estimation.   

Next, we propose several techniques that can significantly reduce the computational cost of the exact Riemann problem solver, without sacrificing accuracy or introducing precomputation efforts (e.g.~sparse-grid tabulation~\cite{Farhat2012FIVER}).

\subsection{Change of integration variable} 
\label{sec:change_variable}
As mentioned in Sec.~\ref{sec:efficiency}, when evaluating the integrals in~\eqref{eq:invariant1_final} and~\eqref{eq:invariant2_final}, the upper bound $\rho^*_K$ is unknown. A trial-and-error procedure is required to find the correct $\rho^*_K$ that makes Eq.~\eqref{eq:invariant1_final} hold for the given $p^*$. We show that this trial-and-error procedure can be eliminated by rewriting~\eqref{eq:invariant1_final} and~\eqref{eq:invariant2_final} in an alternative form, in which the integration variable is $p$ instead of $\rho$. Then, the upper bound of the integral becomes $p^*$, which is specified within the outer loop.

The two Riemann invariants, \eqref{eq:invariant1_final} and~\eqref{eq:invariant2_final}, can be expressed in a differential form, 
\begin{equation}
\left\{\begin{array}{ccc}
\dfrac{ dp}{d\rho} & = & c_K^2(s_K, \rho),\vspace{2mm} \\
\dfrac{du}{d\rho} & = & \mp \dfrac{c_K(s_K, \rho)}{\rho},\\
p(\rho_K) & = & p_K,\\
u(\rho_K) & = & u_K,
\label{eq:ODE1}
\end{array}\right.
\end{equation}\
for all $\rho \in \left[\rho^*, \rho_K \right]$ and $p \in \left[p^*, p_K \right]$. In the second equation, the minus sign applies to the $1$-wave ($K=l$), and the plus sign applies to the $3$-wave ($K=r$). Through the rarefaction fan, entropy does not change, so pressure $p(s_K, \rho)$ is essentially a single variable function of density, i.e.,~$p(\rho)$. Also, the first equation in~\eqref{eq:ODE1} indicates that $p(\rho)$ increases monotonically with respect to $\rho$ since its derivative is $c^2(s_K, \rho)$. Therefore, the function $p(\rho)$ has an inverse function $\rho(p)$. Its derivative, $\dfrac{d\rho}{dp}$, can be obtained easily from the first equation of~\eqref{eq:ODE1}, i.e.,
\begin{equation}
\dfrac{d\rho}{dp} = \dfrac{1}{dp/d\rho} = \dfrac{1}{c_K^2(s_K,\rho(p))} = \dfrac{1}{c_K^2(s_K, p)}.
\label{eq:drhodp}
\end{equation}\
Then, $\dfrac{du}{dp}$ can be derived from the second equation in~\eqref{eq:ODE1} and~\eqref{eq:drhodp} using the chain rule, i.e.,
\begin{equation}
\dfrac{du}{dp} = \dfrac{du}{d\rho} \dfrac{d\rho}{dp}= \mp \dfrac{c_K(s_K,p)}{\rho(p)} \dfrac{1}{c_K^2(s_K,p)} = \mp \dfrac{1}{c_K(s_K,p)\rho(p)}.
\label{eq:dudp}
\end{equation}\
Using~\eqref{eq:drhodp} and \eqref{eq:dudp}, \eqref{eq:ODE1} can be rewritten as
\begin{equation}
\left\{\begin{array}{ccc}
\dfrac{d\rho}{dp} & = & \dfrac{1}{c_K^2(s_K, p)},\\
\dfrac{du}{dp} & = & \mp \dfrac{1}{c_K(s_K,p)\rho(p)},\\
\rho(p_K) & = & \rho_K,\\
u(p_K) & = & u_K,
\label{eq:ODE2}
\end{array}\right.
\end{equation}\
for all $\rho \in \left[\rho^*, \rho_K \right]$ and $p \in \left[p^*, p_K \right]$. Integrating the ODEs in \eqref{eq:ODE2} from $p_K$ to $p^*$ yields
\begin{equation}
\rho^*_K = \rho_K + \int_{p_K}^{p^*} \dfrac{1}{c_K^2(s_K,p)} dp,
\label{eq:invariant1_new}
\end{equation}\
\begin{equation}
    u^*_K  = u_K \mp \int_{p_K}^{p^*} \dfrac{1}{c_K(s_K,p)\rho(p)} dp.
    \label{eq:invariant2_new}
\end{equation}\
Again, we can express sound speed $c$ as a function of $p$ and $\rho$. Hence, a closed-form formulation of entropy is not needed. We use \eqref{eq:invariant1_new} and \eqref{eq:invariant2_new} to replace~\eqref{eq:invariant1_final} and~\eqref{eq:invariant2_final}. In this way, the integration interval is fixed, and the computational overhead associated with the aforementioned trial-and-error procedure is eliminated.

\subsection{Storing and reusing integration trajectory data}
\label{sec:trajectory}
We show that the integrations needed to resolve rarefaction fans, i.e.~\eqref{eq:invariant1_new} and~\eqref{eq:invariant2_new}, can be accelerated by storing and reusing previous results. Our idea is based on the following two observations.
\begin{enumerate}[(1)]
  \item Although the iterative solution of exact Riemann problems (i.e.,~the outer loop in Fig.~\ref{fig:RiemannSolver}) requires repeated evaluation of~\eqref{eq:invariant1_new} and~\eqref{eq:invariant2_new}, the initial state of the integration is always the same, namely, $\left\{p_K, \rho_K, u_K\right\}$. Only $p^*$ varies from one iteration to the next.
\item Once $p_K$, $\rho_K$, and $u_K$ are fixed, the integrands in ~\eqref{eq:invariant1_new} and~\eqref{eq:invariant2_new} are both single variable functions of $p$.
%    \item Through different outer iterations, the two functions represented by the integrands do not change.
\end{enumerate}
($1$) is true because the initial state, $\left\{p_K, \rho_K, u_K\right\}$, is determined by the construction of the Riemann problem, and the iterative solution procedure does not change it. ($2$) is true because entropy is constant through a rarefaction fan, and its value is determined uniquely by the initial state, $\left\{p_K, \rho_K, u_K\right\}$.

Indeed, throughout the solution procedure (i.e.,~the outer loop), we always integrate along the same isentrope, defined by the initial state, and characterized by a single variable, $p$. Only the final state varies in different iterations. Therefore, from one iteration to the next, the ``overlapped'' part of the integration trajectory is first evaluated, then discarded, and then re-evaluated again.

This redundancy can be shown clearly using an example problem. We consider a Riemann problem (i.e.,~\eqref{eq:FF_Riemann}) with initial condition
\begin{equation}
(\rho, u, p, \mathrm{EOS})=\left\{\begin{array}{ll}
\left(1.2\times10^{-6}~\text{g/mm$^3$}, 0.0~\text{mm/s}, 1.0\times10^{5}~\text{Pa}, \text{stiffened~gas}\right) & \text { if } \quad \xi<0~\text{mm} \\
\left(2.204\times10^{-3}~\text{g/mm$^3$},  4.0\times10^5~\text{mm/s}, 1.0\times10^{5}~\text{Pa},  \text{Mie-Gr\"{u}neisen}\right) & \text { if } \quad \xi>0~\text{mm}.
\end{array}\right.
\label{eq:RiemannExample}
\end{equation}\
The parameters in the stiffened gas EOS are given by $\gamma=1.4$, $e_c=0$, $b=0$, and $p_c=0$. Therefore, it degenerates to a perfect gas. The parameters in the Mie-Gr\"{u}neisen EOS are given by $p_0 = 2.204\times10^{-3}~\text{g/mm$^3$}$, $c_0 = 2.22\times10^{6}~\text{mm/s}$, $s = 1.61$, and $\Gamma_0 = 0.65$. The solution has a double-rarefaction wave structure. $200$ steps are specified for the  numerical integration. 

\begin{figure}[H]
    \centering
    \begin{subfigure}{0.5\textwidth}
       \centering
              \caption{}
       \includegraphics[width=80 mm,trim={0cm 0cm 0cm 0cm},clip]{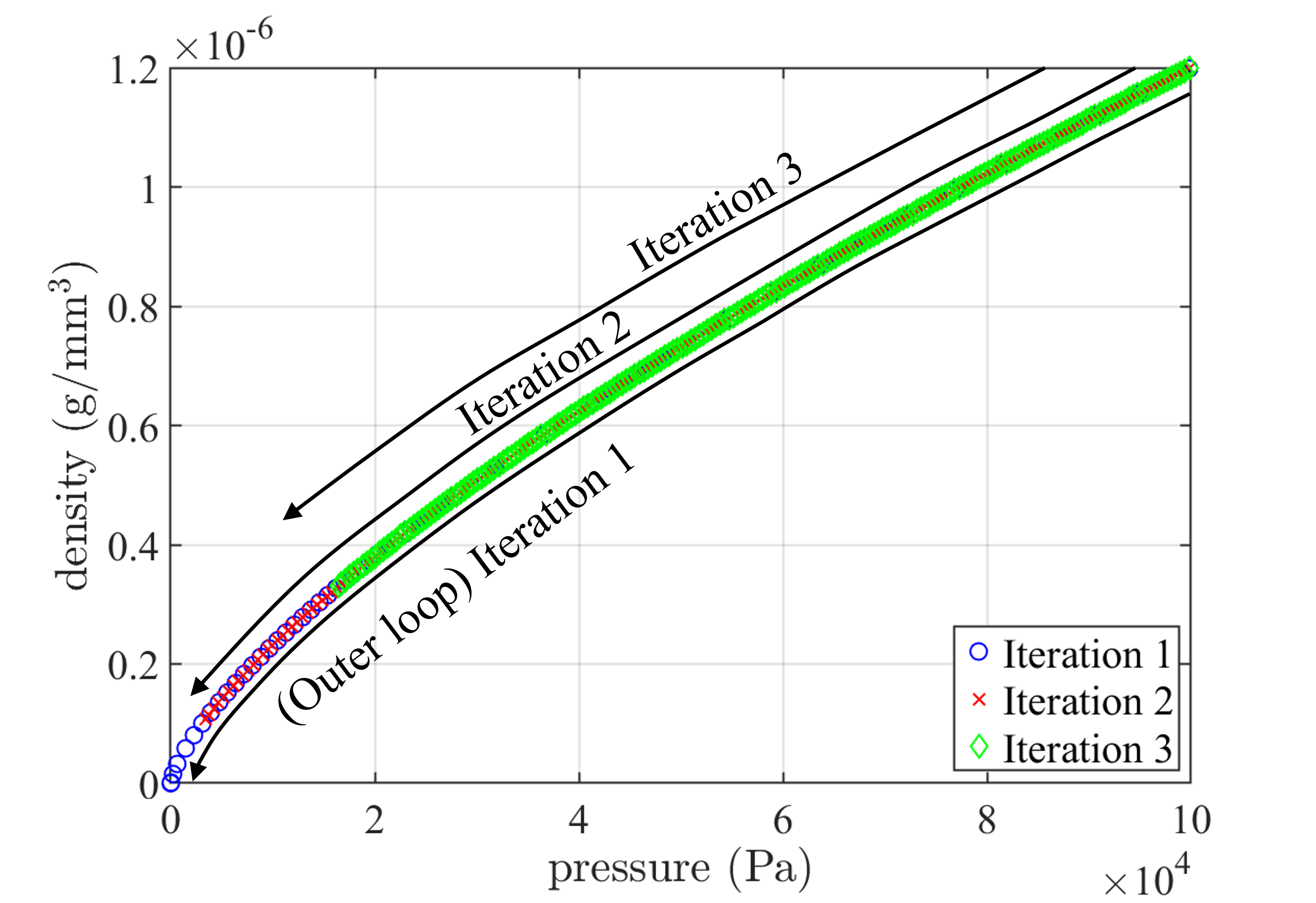}
    \end{subfigure}%
    \begin{subfigure}{0.5\textwidth}
       \centering
              \caption{}
       \includegraphics[width=80 mm,trim={0cm 0cm 0cm 0cm},clip]{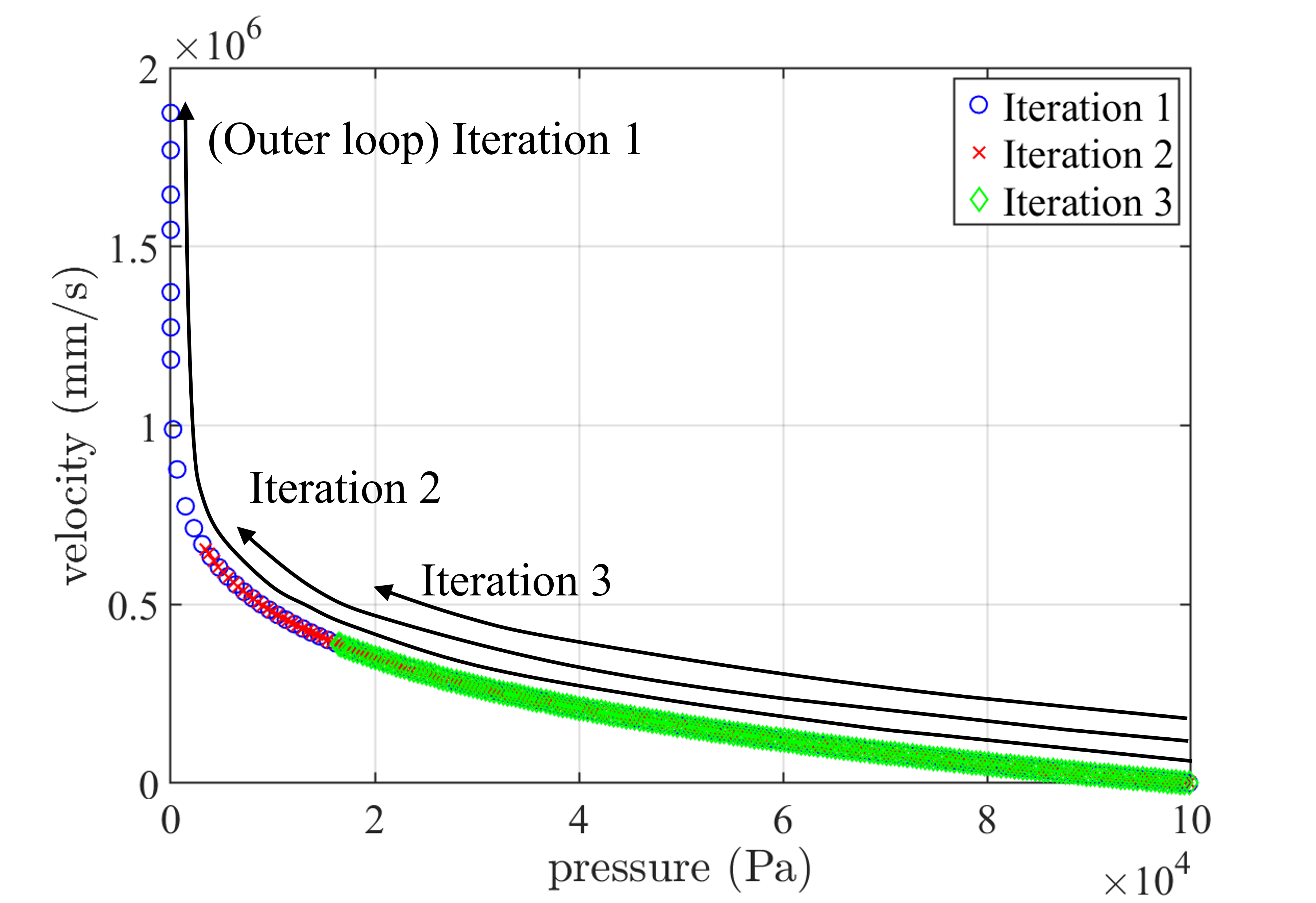}
    \end{subfigure}%
    \caption{Integration trajectories through the $1$-wave of the example Riemann problem in three successive outer loop iterations.}
    \label{fig:integrationTrajectory}
\end{figure}

Figure~\ref{fig:integrationTrajectory} shows the integration trajectories through the $1$-wave, obtained in $3$ successive iterations in the outer loop. Clearly, these curves all start from the same point at $p=1.0\times10^5~\text{Pa}$ (the right end). They follow the same trajectory, only end at different points. It can be observed that the trajectory traversed in the first iteration is the same as that in the next two iterations. In iterations $2$ and $3$, only the last point (i.e.,~the end point) really needs to be evaluated. Nonetheless, the algorithm recalculates the entire trajectory from the right end. Therefore, the vast majority of the computations performed in iterations $2$ and $3$ are redundant and unnecessary. 

To eliminate this redundancy, we propose to store and reuse the numerically obtained integration trajectory. Let $\bm{\chi}$ denote this trajectory, which is a discrete sequence of state variables $\{p,\rho,u\}$ on the isentrope. At the beginning of the outer loop (see Fig.~\ref{fig:RiemannSolver}), $\bm{\chi}$ is initialized to contain only the start point, $\{p_K,\rho_K,u_K\}$. In each iteration of the outer loop, the standard numerical integration procedure is replaced by Algorithm~\ref{alg:trajectory}.

\vspace{0.2cm}
\begin{algorithm}[H]
  \caption{Numerical integration of \eqref{eq:invariant1_new} and \eqref{eq:invariant2_new} with a stored trajectory.}
  \label{alg:trajectory}
        \KwIn{Start point $\left\{p_K, \rho_K, u_K \right\}$, end point $p^*<p_K$, and step size $\Delta p$. \\
        ~~~~~~~~~~~~Stored trajectory $\bm{\chi}$, with $N$ denoting its size.}
        \textbf{Step 1:} Reset the start point based on the stored trajectory $\bm{\chi}$:\\
            \For{$s = N, ~N-1, ~\cdots,~1$}{
                \If{$p^* \leq p_s$}{
                        $\left\{p^{(0)}, \rho^{(0)}, u^{(0)} \right\} \gets \left\{p_s, \rho_s, u_s \right\}$ \Comment{reuse the stored trajectory}\\
                        $\Delta p \gets \min \left(\Delta p, p_s-p^* \right)$ \\     
                        \textbf{break} \\
                }
            }
        \textbf{Step 2:} Integrate, and update the stored trajectory  \\
        $i \gets 0$ \\
        \While{$p^{(i)}>p^*$}{
            $\Delta p \gets \min(\Delta p,~p^{(i)}-p^*)$\\
            $\left\{p^{(i+1)}, \rho^{(i+1)}, u^{(i+1)} \right\} \gets \mathcal{\bm{I}} \left( \left\{p^{(i)}, \rho^{(i)}, u^{(i)} \right\}, \Delta p \right)$ \Comment{integrate one step by 4th-order Runge-Kutta} \\
            \If{$p^{(i+1)} < p_{N}$}{
                $N \gets N + 1$  \\
                $\left\{ p_{N}, \rho_{N}, u_{N} \right\} \gets \left\{p^{(i+1)}, \rho^{(i+1)}, u^{(i+1)} \right\}$  \Comment{add new point to $\bm{\chi}$}\\
            }
            $i \gets i+1$ \\
        }
        \KwOut{$\rho^*_K=\rho^{(i)}$, $u^*_K=u^{(i)}$, and $\bm{\chi}$}
\end{algorithm}
\vspace{0.2cm}

Algorithm~\ref{alg:trajectory} has two main steps. In Step 1, we find the point in $\bm{\chi}$ that is closest to the given $p^*$, but bigger. We start the numerical integration from this point, thereby avoiding redundant calculations as much as possible. In Step 2, we perform the numerical integration, and expand $\bm{\chi}$ with new points obtained on the isentrope.

It is possible that both the $1$-wave and the $3$-wave are rarefactions. Therefore, two separate trajectories, namely $\bm{\chi}_l$ and $\bm{\chi}_r$, should be constructed.

\begin{figure}[H]
    \centering
    \begin{subfigure}{0.5\textwidth}
       \centering
              \caption{}
       \includegraphics[width=80 mm,trim={0cm 0cm 0cm 0cm},clip]{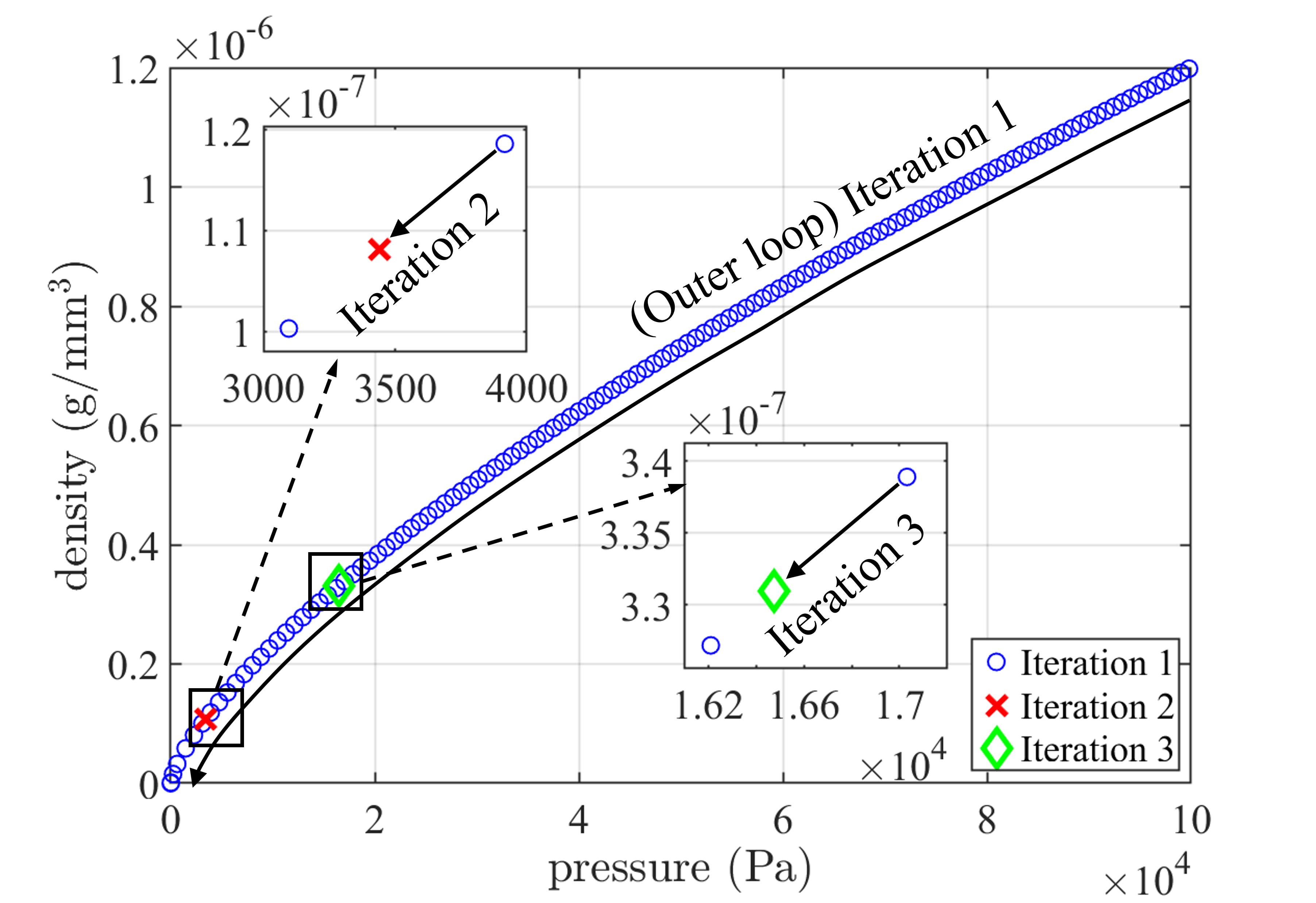}
    \end{subfigure}%
    \begin{subfigure}{0.5\textwidth}
       \centering
              \caption{}
       \includegraphics[width=80 mm,trim={0cm 0cm 0cm 0cm},clip]{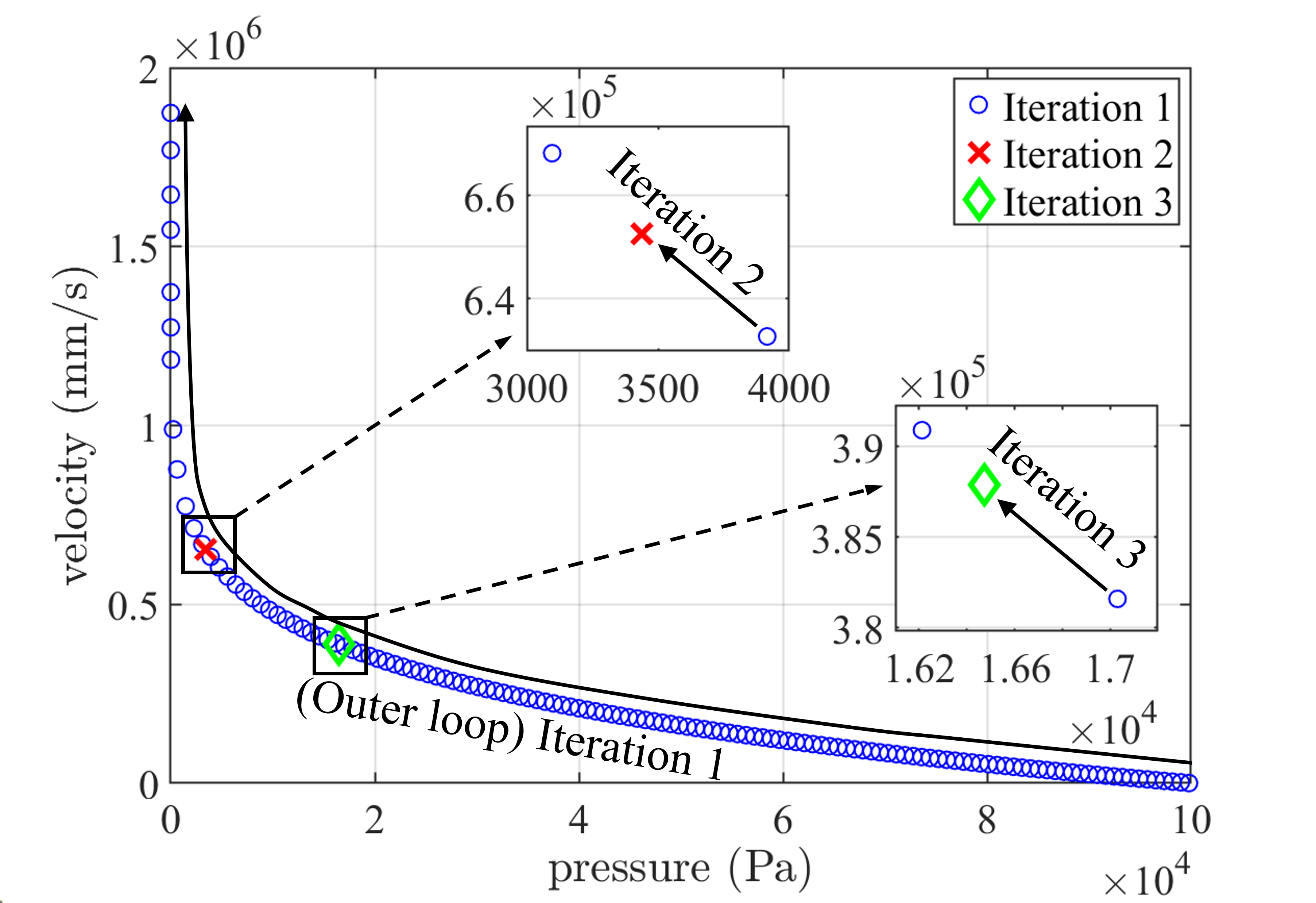}
    \end{subfigure}%
    \caption{Integration trajectories in three successive outer loop iterations, with Algorithm~\ref{alg:trajectory}.}
    \label{fig:integrationTrajectoryM2}
\end{figure}

Consider the same example problem shown in Fig.~\ref{fig:integrationTrajectory}. If we apply Algorithm~\ref{alg:trajectory}, iterations $2$ and $3$ can be completed in one integration step (Fig.~\ref{fig:integrationTrajectoryM2}), as opposed to hundreds. This leads to a significant reduction of computational cost.

\subsection{Adaptive step size}
\label{sec:adaptiveStep}

The accuracy and cost of the numerical integration performed in Step 2 of Algorithm~\ref{alg:trajectory} depend heavily on the step size, $\Delta p$. When a user solves a small number of Riemann problems one by one, it may be possible to specify an appropriate step size for each problem based on the user's experience. Nonetheless, when exact Riemann problems are constructed and solved in 2D and 3D multi-material simulations, manually specifying an integration step size for each Riemann problem is impossible. A common practice is to specify a fixed number of integration steps for all the exact Riemann problems constructed throughout the simulation (e.g.,~\cite{Farhat2012FIVER, Zhao2023, Islam2023}). Clearly, this is not optimal from both accuracy and efficiency standpoints. For different Riemann problems, the number of integration steps required to achieve a certain degree of accuracy varies with the EOS and the integration interval. Moreover, for each Riemann problem, applying a constant $\Delta p$ through the entire rarefaction fan is also problematic, as $\rho(p)$ and $u(p)$ are often highly nonlinear. Their slopes may change rapidly through a rarefaction. For example, in the case shown in Fig.~\ref{fig:integrationTrajectory}(b), the slope of $u(p)$ varies by $6$ orders of magnitude  --- from $-2~\text{mm}/\text{(s Pa)}$ to $-4 \times 10^6~\text{mm}/\text{(s Pa)}$  --- along the trajectory.

\begin{figure}[H]
    \centering
    \includegraphics[width=120 mm,trim={0cm 0cm 0cm 0cm},clip]{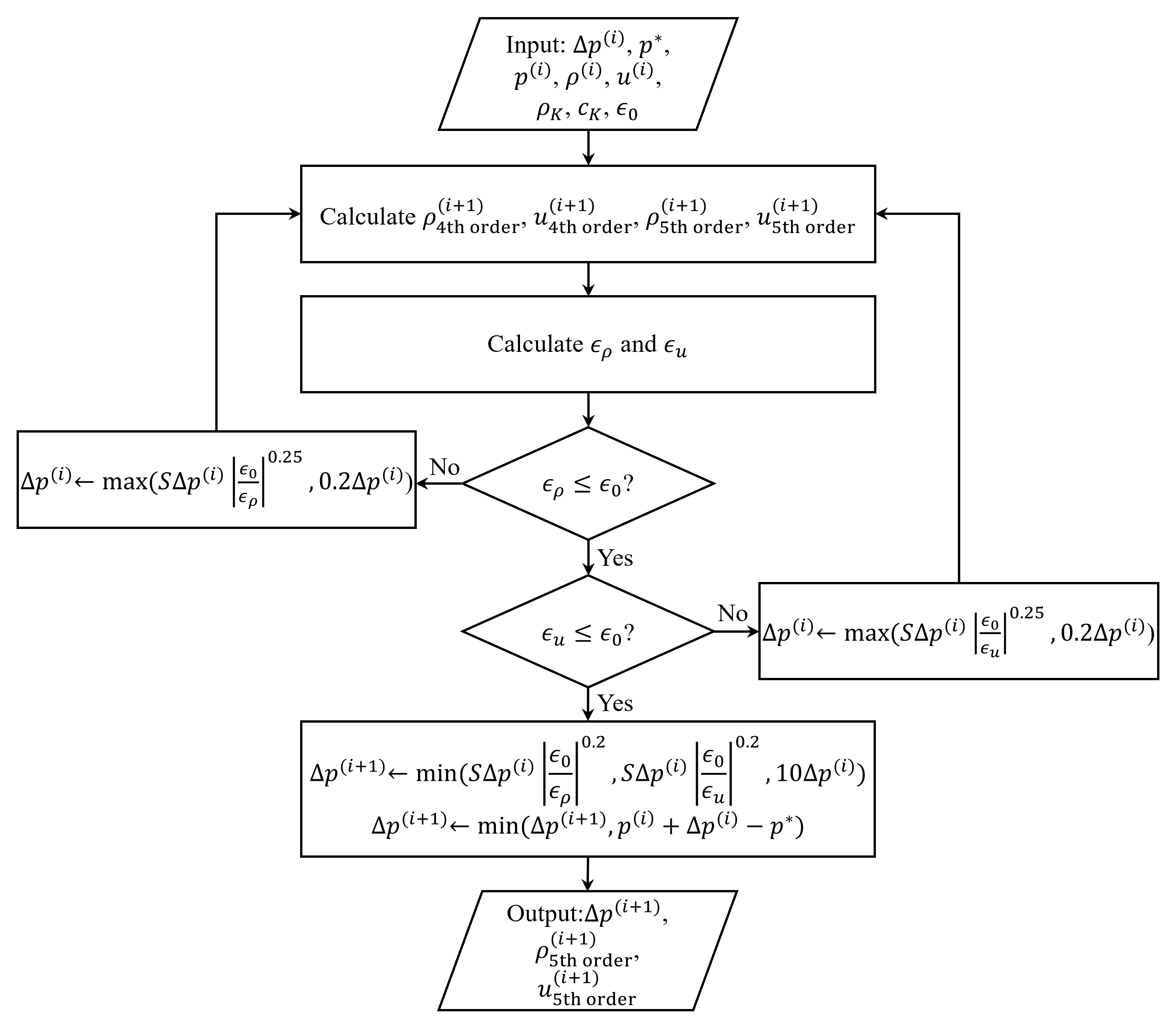}
    \caption{Algorithm 2: One step of numerical integration through a rarefaction fan with adaptive step size.}
    \label{fig:RKF45}
\end{figure}

To resolve this issue, we propose to apply the adaptive Runge-Kutta methods, which automatically adjust the step size based on the local, single-step error of the integration. This method (Algorithm 2) is described in Fig.~\ref{fig:RKF45} in the form of a flowchart. Here, we apply the Runge-Kutta-Fehlberg method~\cite{Fehlberg1969} as an example, which embeds a fourth-order formulation inside a fifth-order one. The difference between the fourth- and fifth-order results serves as an approximation of the local error. Applying this method to~\eqref{eq:invariant1_new} and~\eqref{eq:invariant2_new}, the fourth-order integration formulas for density and velocity are given by
\begin{equation}
\rho^{(i+1)}_{\text{4th order}}=\rho^{(i)} +\left(\frac{2825}{27,648} \frac{1}{\hat{c}_{1}^2} + \frac{18,575}{48,384} \frac{1}{\hat{c}_{3}^2}+\frac{13,525}{55,296} \frac{1}{\hat{c}_{4}^2}+\frac{277}{14,336} \frac{1}{\hat{c}_{5}^2}+\frac{1}{4} \frac{1}{\hat{c}_{6}^2}\right) \Delta p^{(i)},
\end{equation}\
\begin{equation}
u^{(i+1)}_{\text{4th order}}=u^{(i)} \mp \left(\frac{2825}{27,648} \frac{1}{\hat{\rho}_1\hat{c}_1} + \frac{18,575}{48,384} \frac{1}{\hat{\rho}_3\hat{c}_3}+\frac{13,525}{55,296} \frac{1}{\hat{\rho}_4\hat{c}_4}+\frac{277}{14,336} \frac{1}{\hat{\rho}_5\hat{c}_5}+\frac{1}{4} \frac{1}{\hat{\rho}_6\hat{c}_6}\right) \Delta p^{(i)}.
\label{eq:u_4th}
\end{equation}\

Similarly, the fifth-order integration formulas are given by
\begin{equation}
\rho^{(i+1)}_{\text{5th order}}=\rho^{(i)}+\left(\frac{37}{378} \frac{1}{\hat{c}_{1}^2} + \frac{250}{621} \frac{1}{\hat{c}_{3}^2}+\frac{125}{594} \frac{1}{\hat{c}_{4}^2}+\frac{512}{1771} \frac{1}{\hat{c}_{6}^2}\right) \Delta p^{(i)},
\end{equation}\
\begin{equation}
u^{(i+1)}_{\text{5th order}}=u^{(i)} \mp \left(\frac{37}{378} \frac{1}{\hat{\rho}_1\hat{c}_1} + \frac{250}{621} \frac{1}{\hat{\rho}_3\hat{c}_3}+\frac{125}{594} \frac{1}{\hat{\rho}_4\hat{c}_4}+\frac{512}{1771} \frac{1}{\hat{\rho}_6\hat{c}_6}\right) \Delta p^{(i)}.
\label{eq:u_5th}
\end{equation}\

In~\eqref{eq:u_4th} and~\eqref{eq:u_5th}, the minus sign applies to the $1$-wave, and the plus sign applies to the $3$-wave. $\hat{c}_i$ and $\hat{\rho}_i$ $(i = 1,2,\cdots,6)$ are sound speed and density evaluated at intermediate stages, specifically,
\begin{equation}
\begin{aligned}
\hat{c}_{1}=&~c_K\left(p^{(i)},~\hat{\rho}_1\right),~\text{with}~\hat{\rho}_1 =~\rho^{(i)}, \\
\hat{c}_{2}=&~c_K\left(p^{(i)}+\frac{1}{5} \Delta p^{(i)},~\hat{\rho}_2\right),~\text{with}~\hat{\rho}_2 =~\rho^{(i)}+\frac{1}{5} \frac{1}{\hat{c}_{1}^2} \Delta p^{(i)}, \\
\hat{c}_{3}=&~c_K\left(p^{(i)}+\frac{3}{10} \Delta p^{(i)},~\hat{\rho}_3\right),~\text{with}~\hat{\rho}_3 =~\rho^{(i)}+\frac{3}{40} \frac{1}{\hat{c}_{1}^2} \Delta p^{(i)}+\frac{9}{40} \frac{1}{\hat{c}_{2}^2} \Delta p^{(i)},~\\
\hat{c}_{4}=&~c_K\left(p^{(i)}+\frac{3}{5} \Delta p^{(i)},~\hat{\rho}_4\right),~\text{with}~\hat{\rho}_4 =~\rho^{(i)}+\frac{3}{10} \frac{1}{\hat{c}_{1}^2} \Delta p^{(i)} - \frac{9}{10} \frac{1}{\hat{c}_{2}^2} \Delta p^{(i)} + \frac{6}{5} \frac{1}{\hat{c}_{3}^2} \Delta p^{(i)},~ \\
\hat{c}_{5}=&~c_K\left(p^{(i)}+\Delta p^{(i)},~\hat{\rho}_5\right),~\text{with}~\hat{\rho}_5 =~\rho^{(i)}-\frac{11}{54} \frac{1}{\hat{c}_{1}^2} \Delta p^{(i)} + \frac{5}{2} \frac{1}{\hat{c}_{2}^2} \Delta p^{(i)}-\frac{70}{27} \frac{1}{\hat{c}_{3}^2} \Delta p^{(i)} + \frac{35}{27} \frac{1}{\hat{c}_{4}^2} \Delta p^{(i)},~\\
\hat{c}_{6}=&~c_K\left(p^{(i)}+\frac{7}{8} \Delta p^{(i)},~\hat{\rho}_6 \right),~\text{with}~\hat{\rho}_6=~\rho^{(i)}+\frac{1631}{55,296} \frac{1}{\hat{c}_{1}^2} \Delta p^{(i)}+\frac{175}{512} \frac{1}{\hat{c}_{2}^2} \Delta p^{(i)} +\frac{575}{13,824} \frac{1}{\hat{c}_{3}^2} \Delta p^{(i)} +\frac{44,275}{110,592} \frac{1}{\hat{c}_{4}^2} \Delta p^{(i)} +\frac{253}{4096} \frac{1}{\hat{c}_{5}^2} \Delta p^{(i)}.
\end{aligned}
\end{equation}\
The coefficients utilized in the above formulas were developed by Cash and Karp~\cite{Cash1990}. They allow the intermediate stages involved in the fourth-order formula to be re-used in the fifth-order one. 

We normalize the errors in density and velocity using the initial state, i.e.,
\begin{equation}
    \label{eq:rhoError}
    \epsilon_{\rho} = \left| \frac{\rho^{(i+1)}_{\text{4th order}} - \rho^{(i+1)}_{\text{5th order}}}{\rho_K} \right| + \Tilde{\epsilon},
\end{equation}\
\begin{equation}
    \label{eq:uError}
    \epsilon_{u} = \left| \frac{u^{(i+1)}_{\text{4th order}} - u^{(i+1)}_{\text{5th order}}}{c_K} \right| + \Tilde{\epsilon}, 
\end{equation}\
where $\rho_K$ and $c_K$ are density and sound speed in the initial state. Because $\epsilon_{\rho}$ and $\epsilon_{\rho}$ will appear in the denominator, a small positive number, $\Tilde{\epsilon}$, is added to avoid division by zero. In all the numerical tests presented in this paper, it is set to $1\times 10^{-30}$. Then, the integration step size $\Delta p$ is adjusted according to the two errors and an error tolerance $\epsilon_0$. For example, a new step size, $\Delta p^*$, can be calculated based on the density error, $\epsilon_\rho$, by
\begin{equation}
    \label{eq:stepAdapt}
    \Delta p^{*} = S\cdot\Delta p^{(i)}\cdot\left( \frac{\epsilon_0}{\epsilon_{\rho}} \right) ^ {\alpha}.
\end{equation}\
Here, $S\in (0,1)$ is introduced as a safety factor, because $\epsilon_{\rho}$ is only an estimate of the true error. In this work, we set $S=0.9$, following~\cite{Chapra2015numerical}. $\alpha$ is set to $0.2$ if $\epsilon_\rho  \leq \epsilon_0$, and $0.25$ otherwise~\cite{Chapra2015numerical}.

If $\epsilon_{\rho}\leq\epsilon_0$, we set
\begin{equation}
\Delta p^{(i+1)} = \Delta p^{*},
\end{equation}\
and move on to examine the error in velocity (Fig.~\ref{fig:RKF45}). If $\epsilon_{\rho}>\epsilon_0$, we repeat step $i$ with a smaller step size, i.e.
\begin{equation}
\Delta p^{(i)} = \Delta p^{*}.
\end{equation}\
Following~\cite{Press2007}, we cap the variation of $\Delta p$ within one step at a factor of $10$ for increase, and a factor of $5$ for decrease. Next, the same procedure is applied to update $\Delta p$ using the velocity error $\epsilon_u$, as shown in Fig.~\ref{fig:RKF45}.

To apply the adaptive step size within Algorithm~\ref{alg:trajectory}, we note that the input $\Delta p$ in Algorithm~\ref{alg:trajectory} becomes $\Delta p^{(0)}$. It does not need to be determined precisely, as the actual step size will be calculated automatically in each integration step. In practice, we set $\Delta p^{(0)}$ to be a small number to avoid the repetition of the first integration step (i.e.,~$i=0$). Because of step size adaptation, every point (i.e.,~$\left\{p_s~|~s \in [1,N]\right\}$) stored on the integration trajectory $\bm{\chi}$ is quality checked in terms of accuracy. The stored step sizes (e.g.,~$p_{s}-p_{s+1}$) automatically satisfy the requirement on local accuracy. This fact can be utilized to determine the new initial step size $\Delta p^{(0)}$ when we reset the integration start point. If the new start point $p_s$ is not the last stored point on $\bm{\chi}$, the inequality $p_s > p^* > p_{s+1}$ must hold. This means the new initial step size $\Delta p^{(0)}$ can be set by 
\begin{equation}
    \Delta p^{(0)} \gets p_s - p^*.
    \label{eq:newInitialStep1}
\end{equation}\
In this way, the new integration is completed in only one step while satisfying accuracy requirements. If the new start point $p_s$ is the last stored point on $\bm{\chi}$, we set the new initial step size to
\begin{equation}
    \Delta p^{(0)} \gets \min \left(p_{s-1}-p_s, p_s-p^* \right).
    \label{eq:newInitialStep2}
\end{equation}\
Therefore, \eqref{eq:newInitialStep1} and~\eqref{eq:newInitialStep2} are used to replace line $5$ in Algorithm~\ref{alg:trajectory} when adaptive step size is applied. 

\begin{figure}[H]
    \centering
    \begin{subfigure}{0.5\textwidth}
       \centering
              \caption{}
       \includegraphics[width=80 mm,trim={0cm 0cm 0cm 0cm},clip]{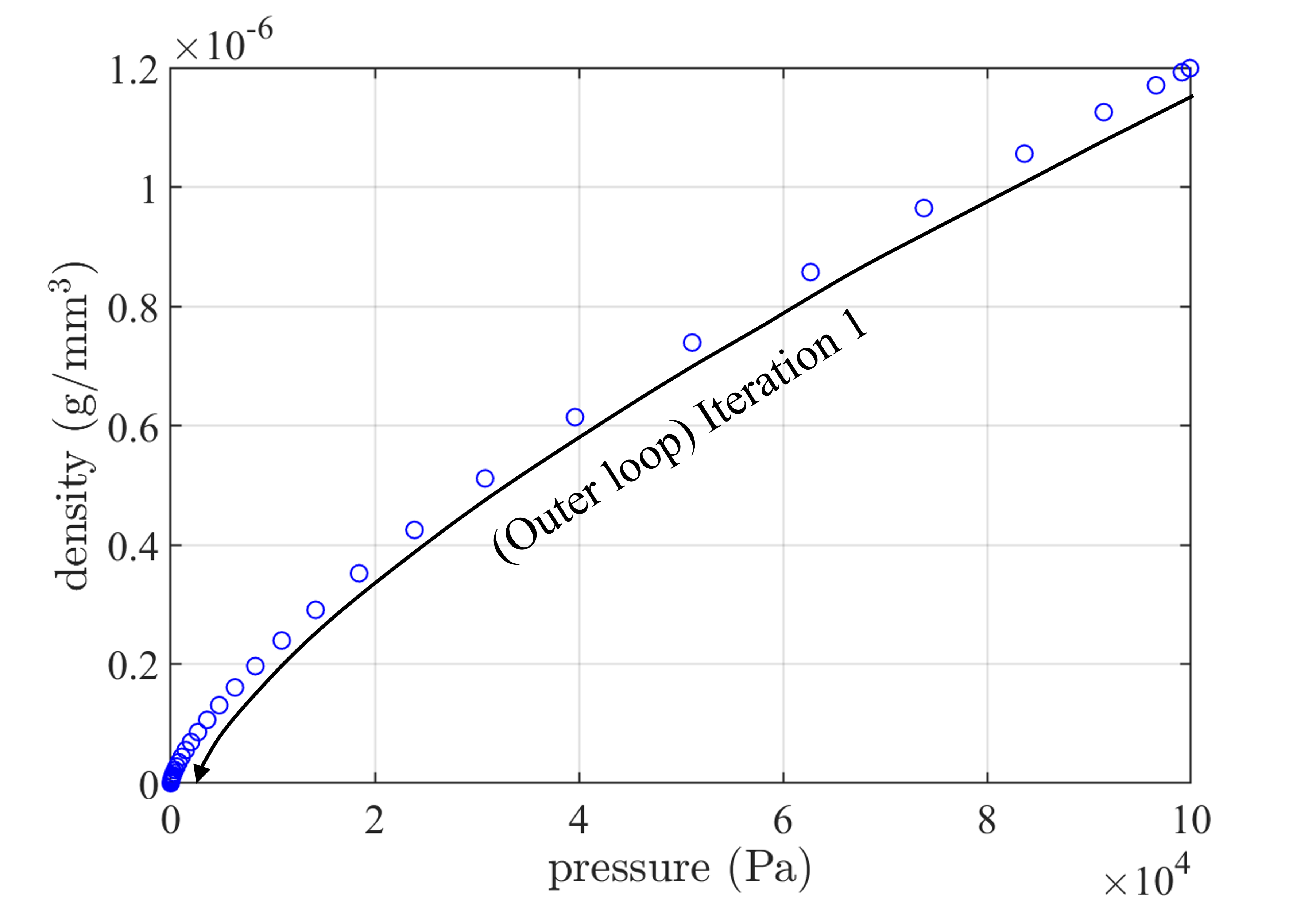}
    \end{subfigure}%
    \begin{subfigure}{0.5\textwidth}
       \centering
              \caption{}
       \includegraphics[width=80 mm,trim={0cm 0cm 0cm 0cm},clip]{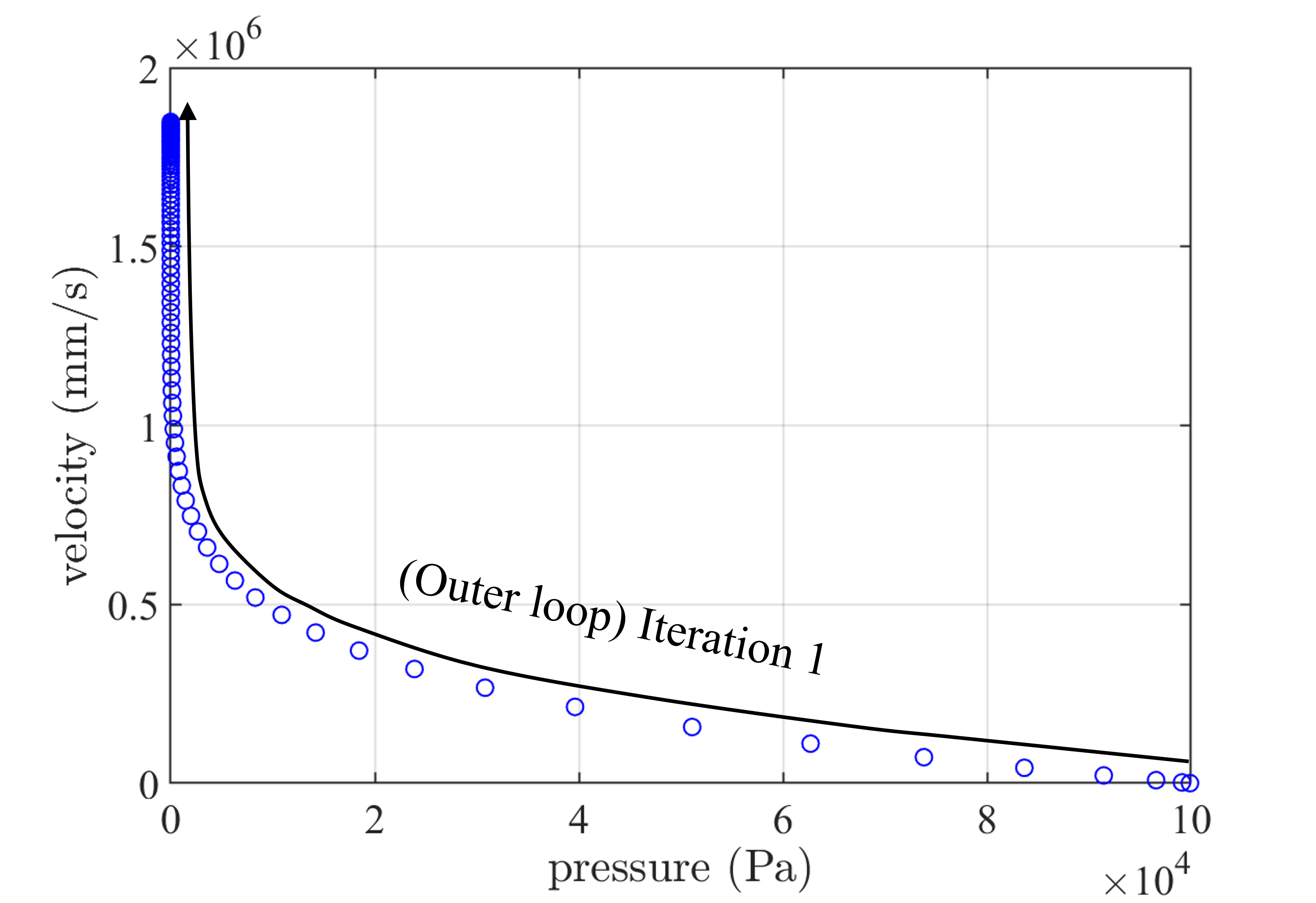}
    \end{subfigure}%
    \caption{Repetition of iteration $1$ in Fig.~\ref{fig:integrationTrajectory} with adaptive step size.}
    \label{fig:integrationTrajectoryNew}
\end{figure}

Repeating the solution of the Riemann problem given in~\eqref{eq:RiemannExample} with adaptive integration step size gives Fig.~\ref{fig:integrationTrajectoryNew}. Comparing with the result shown in Fig.~\ref{fig:integrationTrajectory}, the advantage brought by the adaptive step size is clear. Smaller step sizes are applied near the left end, where the slopes of $\rho(p)$ and $u(p)$ vary rapidly. At the right end, although the starting step size, $\Delta p^{(0)}$ is small, the algorithm is able to quickly increase its value to adapt to the slow variation of slopes. 

\subsection{Obtaining initial guesses using R-tree data structure}
\label{sec:rtree}

The computational cost of exact Riemann problem solution depends on the number of iterations performed in the outer loop (Fig.~\ref{fig:RiemannSolver}). When the error tolerance is fixed, the number of iterations is largely determined by the quality of the initial guesses. A popular method to specify initial guesses is to apply the linear acoustic theory~\cite{kamm2015exact}. Assuming that two initial guesses are needed, such as in the Secant method, the first one, $p^{*(0)}$, is given by
\begin{equation}
p^{*(0)}=\left[C_r p_l+C_l p_r+C_l C_r\left(u_l-u_r\right)\right] /\left(C_l+C_r\right),
\label{eq:initial1}
\end{equation}\
where $l$ and $r$ refer to the two materials, $c$ denotes the sound speed, and $C=\rho c$.  The second initial guess, $p^{*(1)}$, is obtained by applying a different sound speed. Specifically,
\begin{equation}
p^{*(1)}=\left[\overline{C}_r p_l+\overline{C}_l p_r+\overline{C}_l \overline{C}_r\left(u_l-u_r\right)\right] /\left(\overline{C}_l+\overline{C}_r\right),
\label{eq:initial2}
\end{equation}\
where $\overline{C}_K$ ($K=l,r$) is given by
\begin{equation}
\overline{C}_{K} = \begin{cases}\left|p^{*(0)}-p_{K}\right| /\left|u_{K}^{*(0)}-u_{K}\right| & \text { if } \quad u_{K}^{*(0)} \neq u_{K}, \\ \rho_{K} c_{K} & \text { if } \quad u_{K}^{*(0)}=u_{K}.\end{cases}
\end{equation}\
Here, $u_K^{*(0)}$ denotes the velocity obtained with pressure $p^{*(0)}$.

\begin{figure}[H]
    \centering
    \includegraphics[width=110mm,trim={0cm 0cm 0cm 0cm},clip]{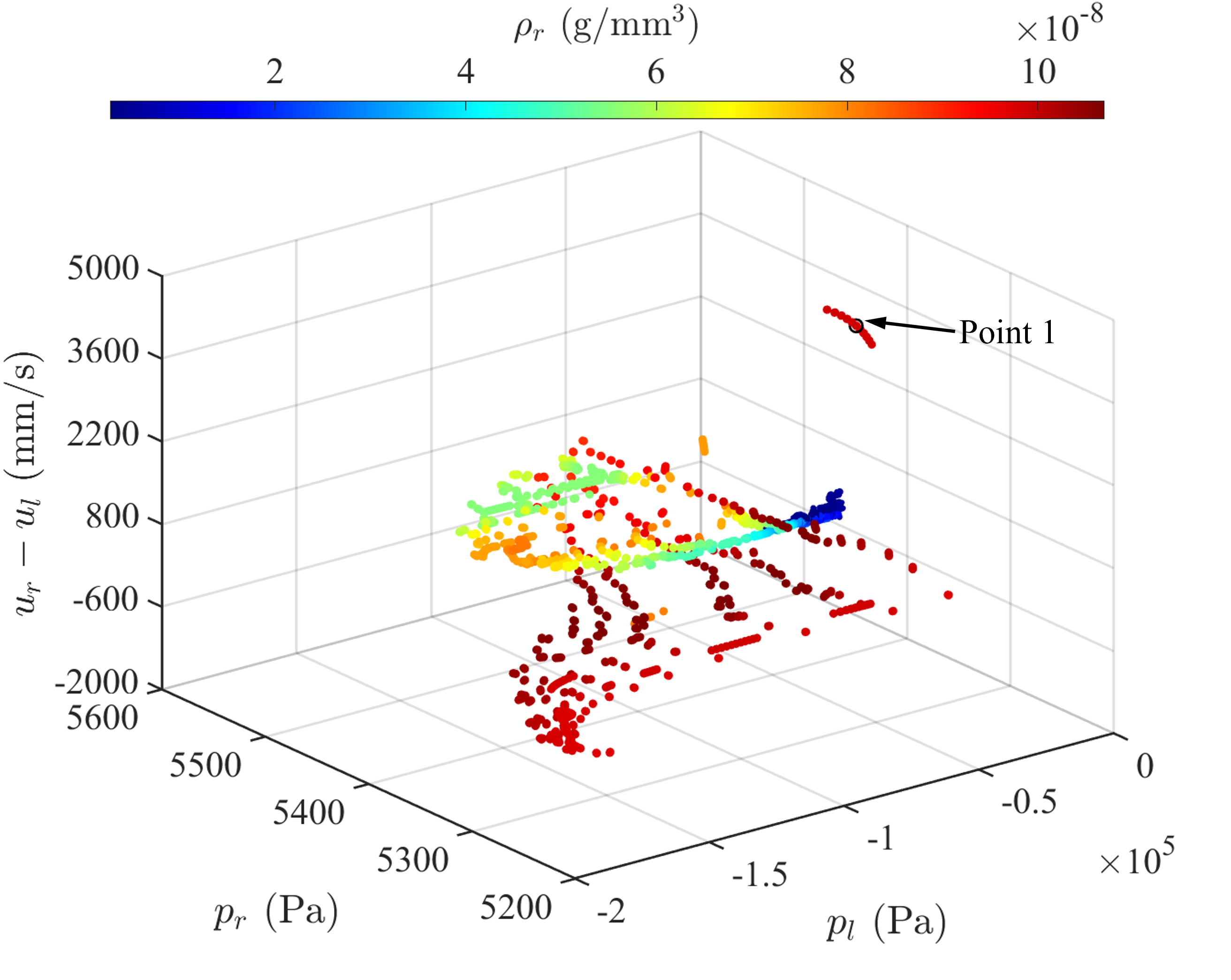}
    \caption{Visualization of the bimaterial Riemann problems solved in the simulation presented in Sec.~\ref{sec:laser} during $10$ time steps. This 5D data set is displayed with $p_l$, $p_r$, and $u_r-u_l$ as the three spatial coordinates. The color of each data point indicates the value of $\rho_r$, and the size of the point indicates the value of $\rho_l$. ``Point 1'' is used to demonstrate the initial guess improvement from the nearest neighbor search.}
    \label{fig:5d_points}
\end{figure}

When the Riemann problem is highly nonlinear, the initial guesses obtained using the linear acoustic theory may not be close to the true solution. To mitigate this issue, we introduce another method based on storing and reusing the solutions of previously solved Riemann problems. The motivation of this data-based method is that in the course of a multi-material simulation, a large number of bimaterial Riemann problems are constructed and solved. Some of them may have similar inputs, hence similar solutions. As an example, Fig.~\ref{fig:5d_points} visualizes the Riemann problems solved in the simulation presented in Sec.~\ref{sec:laser}, within $10$ time steps. Some clusters can be easily identified in this data set. Our basic idea is to store the solutions of solved exact Riemann problems using an efficient data structure, and to generate the initial guesses for a new problem through nearest neighbor search.

Notably, the bimaterial Riemann problem is 5D, with $(\rho_l, p_l, \rho_r, p_r, u_r-u_l)$ as the independent variables. Also, this data set needs to be frequently updated and accessed. For these reasons, we adopt the R-tree data structure, which is specifically designed for efficient indexing and searching of multidimensional data~\cite{Guttman1984Rtrees, rtreeBook}. 

Unlike the idea of constructing a sparse grid before each simulation~\cite{Farhat2012FIVER}, the R-tree data set is designed to be constructed during the simulation (i.e.,~``on the fly''). Specifically, this method involves three components.

\begin{itemize}
    \item {\bf Construction of R-tree.} In the course of a simulation, after solving each bimaterial Riemann problem, we pair the solution $p^*$ with the corresponding 5D inputs, and insert the pair into the R-tree.
    \item {\bf Normalization of the five (5) coordinates.}
    The $5$ coordinates of the database represent density, pressure, and velocity. Their numerical values often differ by several orders of magnitudes, depending on their units (e.g.,~see Fig.~\ref{fig:5d_points}). Directly performing nearest-neighbor search on these dimensional coordinates is problematic, as the evaluation of distance would be biased towards coordinates with larger numerical values. Therefore, we normalize the $5$ coordinates using a lower bound and an upper bound, so that each normalized, non-dimensional coordinate varies between $0$ and $1$. 
    \item {\bf Nearest-neighbor search.} Once a new Riemann problem is constructed, we apply a bounding-box based nearest-neighbor search algorithm~\cite{nearestNeighbor, enhancedNeighbor} to find the data point in the R-tree whose inputs are closest to the new Riemann problem. Its solution,  $p^*$, is taken as the first initial guess for solving the new Riemann problem. The second initial guess, if needed, can be obtained either by Eq.~\eqref{eq:initial2} or by applying a perturbation to the first initial guess.
\end{itemize}

Consider the same example problem shown in Fig.~\ref{fig:5d_points}. In this figure, Point $1$ marks a new Riemann problem that is constructed during the simulation. Table~\ref{tab:rtreeExample} shows the state variables  --- including both the $5$ inputs and the solution $p^*$  --- at this point and its nearest neighbor in the existing R-tree. Clearly, the two points are close to each other. Using the solution at the nearest neighbor as the first initial guess, it takes only $1$ iteration to solve the new Riemann problem at Point $1$. In comparison, it takes $8$ iterations to converge to the same error tolerance, if the first initial guess is generated using the linear acoustic theory, i.e.,~by~\eqref{eq:initial1}.

\begin{table}[H]
\caption{Comparison of the state variables at Point $1$ in Fig.~\ref{fig:5d_points} and its nearest neighbor in the R-tree.}
\label{tab:rtreeExample}
\hspace{-3.5mm}
\begin{tabular}{lcccccc}
\toprule
 &
  $\rho_l~\text{(g/mm}^3$) &
  $p_l~\text{(Pa)}$&
  $\rho_r~\text{(g/mm}^3$) &
  $p_r~\text{(Pa)}$ &
  $u_r-u_l~\text{(mm/s)}$&
  $p^{*}~\text{(Pa)}$ \\ 
\midrule
Point 1 &
  $9.9628\times10^{-4}$ &
  $-4.4378\times10^4$ &
  $9.8425\times10^{-8}$ &
  $5.3338\times10^{3}$ &
  $4.3761\times10^3$ &
  $5.21797\times10^3$ \\
Nearest neighbor &
  $9.9628\times10^{-4}$ &
  $-4.4266\times10^4$ &
  $9.8425\times10^{-8}$ &
  $5.3338\times10^{3}$ &
  $4.3757\times10^3$ &
  $5.21798\times10^3$  \\
\bottomrule
\end{tabular}
\end{table}

This work utilizes the R-tree data structure implemented in the widely used Boost Generic Geometry Library~\cite{boost}. In particular, the insertion of data points, query of the nearest neighbor, and searching of the lower and upper bounds for normalization are performed by built-in functions in the Boost library. More details about the creating, modifying and using R-tree data structure in Boost library can be found in~\cite{boost}. For the numerical experiments in Sec.~\ref{sec:numericalExperiments}, we store the Riemann problems solved in the previous time step, and perform the normalization at the beginning of each time step.

\begin{figure}[H]
    \centering
    \includegraphics[width=160mm,trim={0cm 0cm 0cm 0cm},clip]{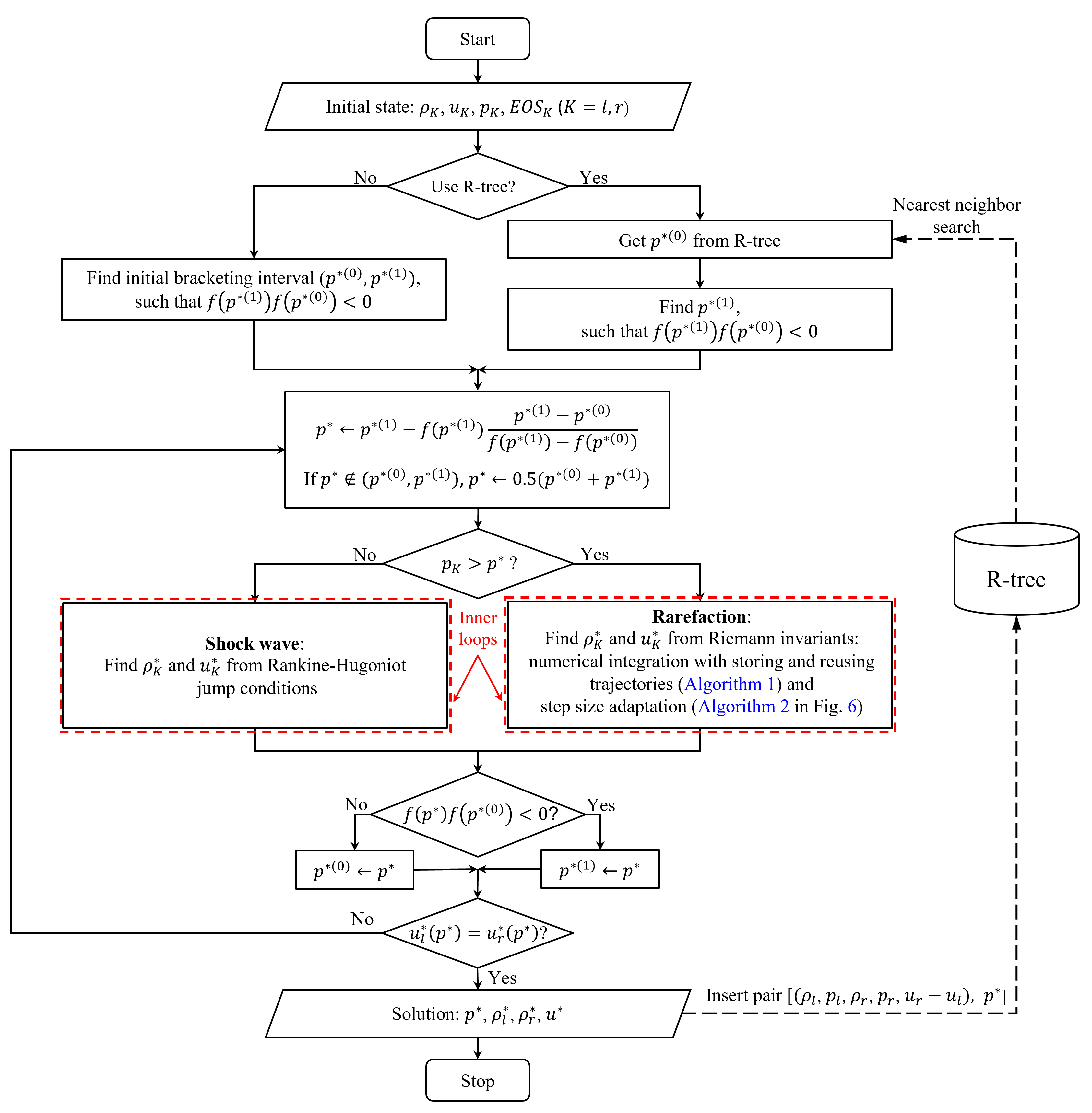}
    \caption{Improved solution procedure for bimaterial Riemann problems in multi-material flow simulaitons.}
    \label{fig:bigFlowChart}
\end{figure}

\vspace{1cm}
In summary, Fig.~\ref{fig:bigFlowChart} presents the new algorithm for solving an exact bimaterial Riemann problem, which incorporates the acceleration methods presented in Sec.~\ref{sec:change_variable},~\ref{sec:trajectory},~\ref{sec:adaptiveStep}, and~\ref{sec:rtree}. 

\section{Numerical experiments}
\label{sec:numericalExperiments}
We assess the performance of the acceleration methods presented in Sec.~\ref{sec:change_variable}  - \ref{sec:rtree} using four example problems that involve multiple (2 or 3) materials and significant discontinuities across their interfaces. For conciseness, the four acceleration methods are referred to as Methods $1$ to $4$ in this section.

\begin{itemize}
    \item Method 1: Change of integration variable (Sec.~\ref{sec:change_variable}).
    \item Method 2: Storing and reusing integration trajectory data (Sec.~\ref{sec:trajectory}).
    \item Method 3: Adaptive step size (Sec.~\ref{sec:adaptiveStep}).
    \item Method 4: Obtaining initial guesses using R-tree data structure (Sec.~\ref{sec:rtree}).
\end{itemize}

The simulations presented in this section are performed using the M2C (Multiphysics Modeling and Computation) solver~\cite{m2c} and the Tinkercliffs computer cluster at Virginia Tech. A standalone version of the exact Riemann problem solver is available at~\cite{riemann}.

\subsection{One-dimensional benchmark problem}
\label{sec:shocktube}
The 1D bimaterial Riemann problem defined by~\eqref{eq:RiemannExample} is solved numerically using the FIVER method described in Sec.~\ref{sec:numerical_scheme}. This problem models a condensed phase (soda lime glass) moving away from a gas (air) at high speed ($400~\text{m}/\text{s}$). The exact solution is used to verify the numerical solver.  At any time $t>0$, density jumps by $4$ orders of magnitude across the material interface, from $0.3~\text{kg}/\text{m}^3$ to $2203.98~\text{kg}/\text{m}^3$. Therefore, this example problem also challenges the robustness of the solver.

To separate the effects of different acceleration methods, we perform four simulations in which the exact Riemann problems constructed during the simulation are solved using different methods.
\begin{itemize}
\item Simulation (\romannumeral 1): using the baseline method presented in Sec.~\ref{sec:efficiency};
\item Simulation (\romannumeral 2): with acceleration Methods (1) and (2);
\item Simulation (\romannumeral 3): with acceleration Methods (1), (2), and (3); and
\item Simulation (\romannumeral 4): with acceleration Methods (1), (2), (3), and (4).
\end{itemize}

The computational domain extends from $x=0$ to $1.0~\text{mm}$. The material interface is initially at $x=0.2~\text{mm}$. $200$ rectangular elements are used to discretize the domain. In Simulations (\romannumeral 3) and (\romannumeral 4), the error tolerance for controlling the adaptive step size (Eqs.~\eqref{eq:rhoError} and~\eqref{eq:uError}) is set to $1.0\times10^{-9}$. In Simulations (\romannumeral 1) and (\romannumeral 2), we need to specify a fixed number of integration steps. To have a fair comparison in terms of solution accuracy, we set the number of integration steps to be the maximum integration steps required in (\romannumeral 3) to reach the error tolerance, which is $5,307$.

\begin{figure}[H]
    \centering
    \begin{subfigure}[b]{75mm}
       \caption{}
       \includegraphics[width=75mm,trim={0cm 0cm 0cm 0cm},clip]{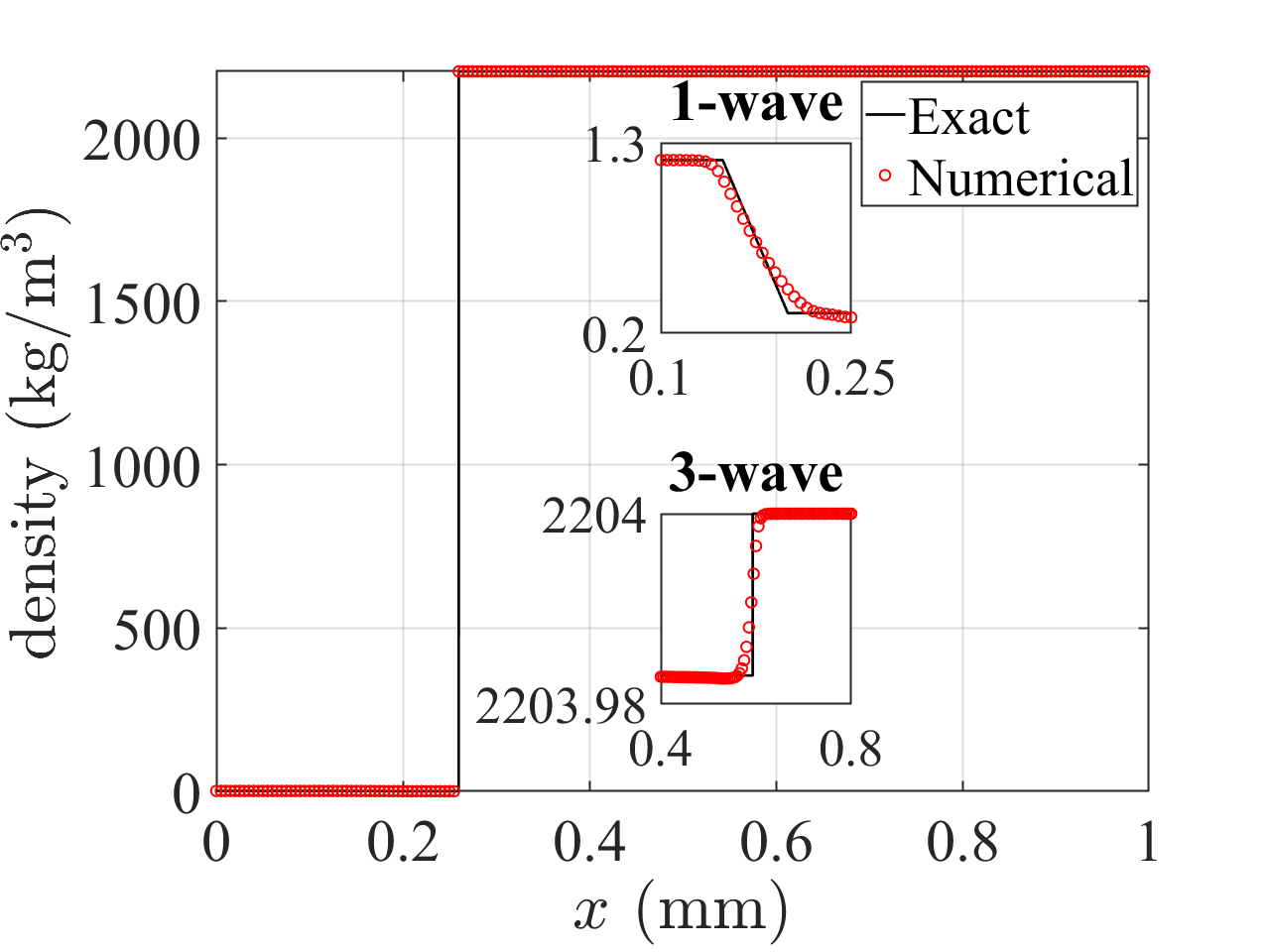}
    \end{subfigure}
    \begin{subfigure}[b]{75mm}
       \caption{}
           \includegraphics[width=75mm,trim={0cm 0cm 0cm 0cm},clip]{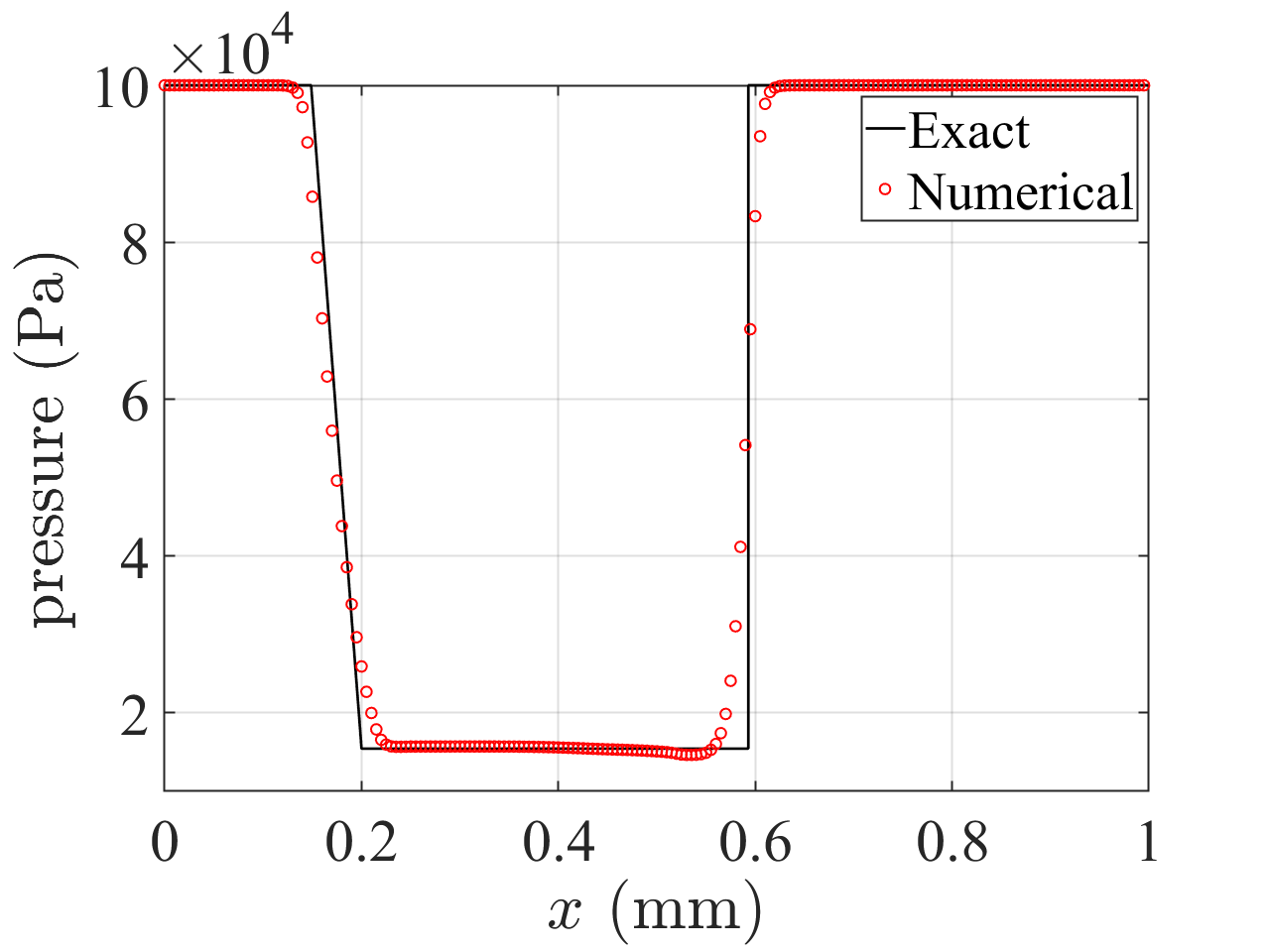}
    \end{subfigure}
    \begin{subfigure}[b]{75mm}
       \caption{}
       \includegraphics[width=75mm,trim={0cm 0cm 0cm 0cm},clip]{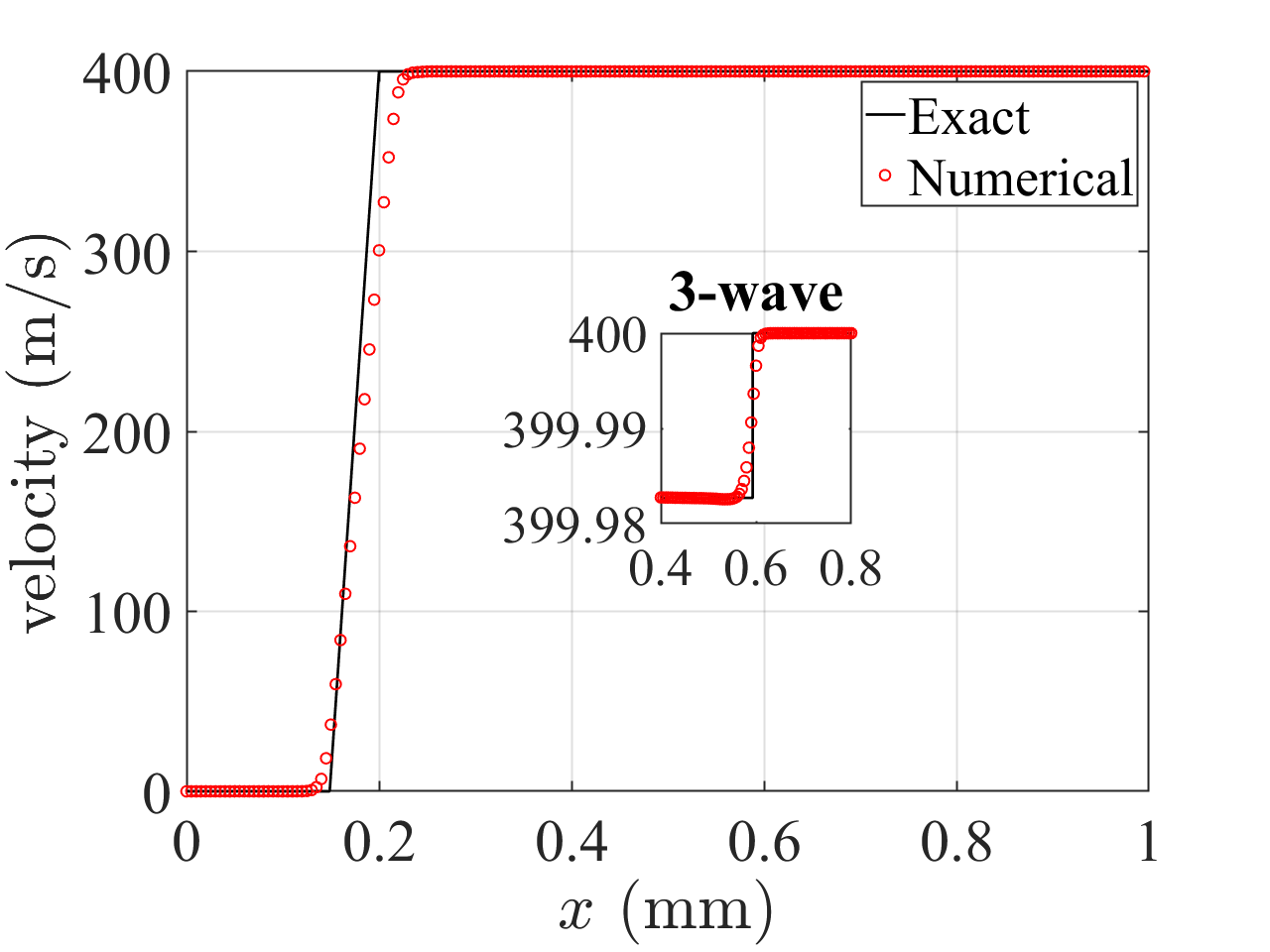}
     \end{subfigure}
    \caption{A 1D bimaterial Riemann problem: Density, pressure, and velocity distributions at $t=0.15~\mu$s.}
    \label{fig:1dCFD}
\end{figure}

The density, pressure, and velocity results obtained from Simulation (\romannumeral 4) at $t = 0.15~\mu\text{s}$ are shown in Fig.~\ref{fig:1dCFD}, in comparison with the exact solution. For all the three variables, the simulation result matches closely the exact solution. The results obtained from Simulations (\romannumeral 1), (\romannumeral 2), and (\romannumeral 3) are essentially the same as that from Simulation (\romannumeral 4), as the acceleration methods do not introduce new approximations. Therefore, their results are not shown separately.

\begin{figure}[H]
    \centering
    \includegraphics[width=80mm,trim={0cm 0cm 0cm 0cm},clip]{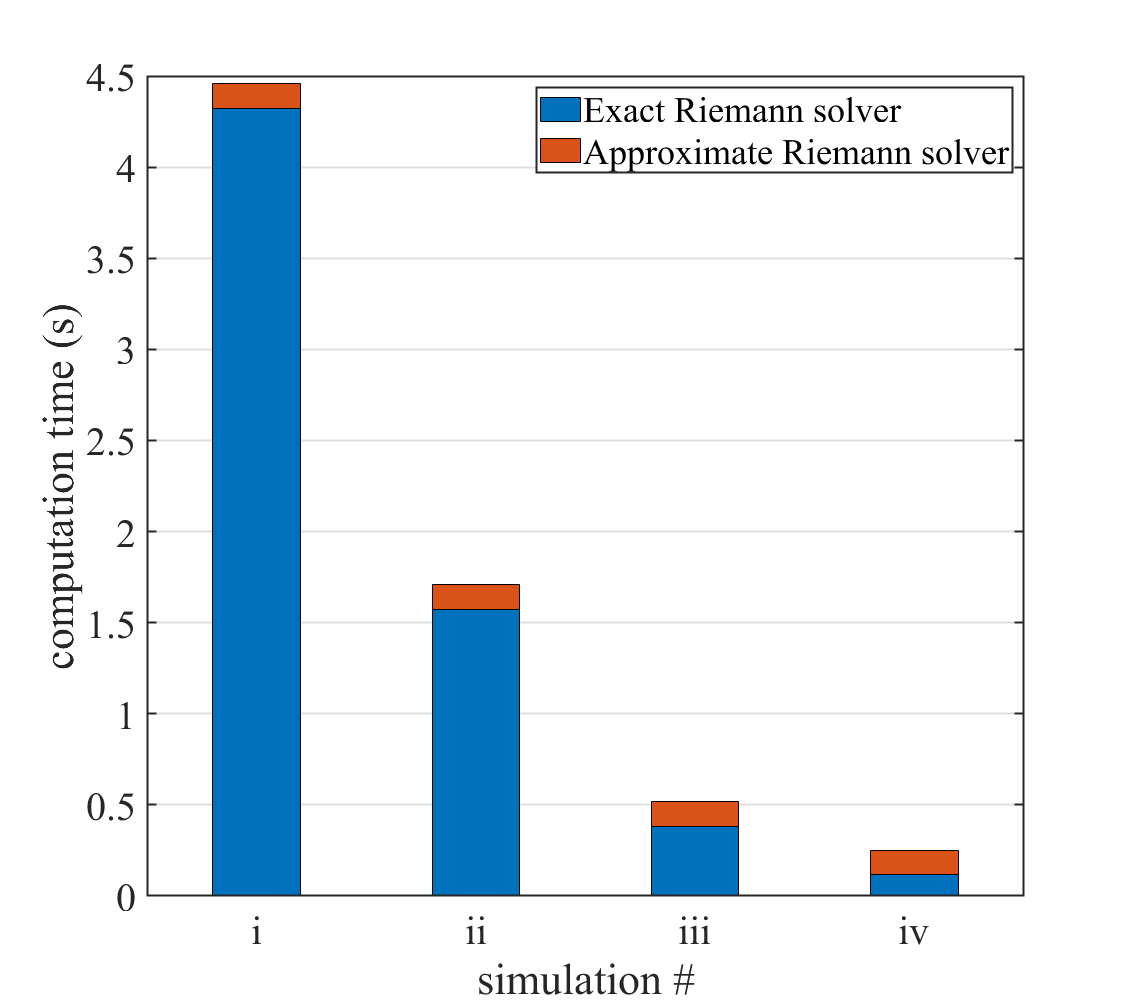}
    \caption{A 1D bimaterial Riemann problem: Comparison of the flux computation time of different simulations.}
    \label{fig:1dRiemann_accel}
\end{figure}

Figure~\ref{fig:1dRiemann_accel} compares the computational cost of the four simulations up to $t = 0.15~\mu\text{s}$. Here, we measure the computation time for evaluating the advective fluxes, which includes two parts: the time spent on solving the exact Riemann problems at the material interface, and the time spent on evaluating the numerical flux functions (LLF in this case) over the entire domain. The effect of the acceleration methods is found to be significant. When all the four methods are adopted (i.e.,~in Simulation (\romannumeral 4)), the time consumed by the exact Riemann solver is reduced by a factor of  $37.6$ compared to the baseline simulation (from $4.32$ s to $0.123$ s) . The total computation time for evaluating the advective fluxes is also significantly reduced, by a factor of $17.8$ (from $4.46$ s to $0.26$ s). 

\begin{figure}[H]
    \centering
       \includegraphics[width=80mm,trim={0cm 0cm 0cm 0cm},clip]{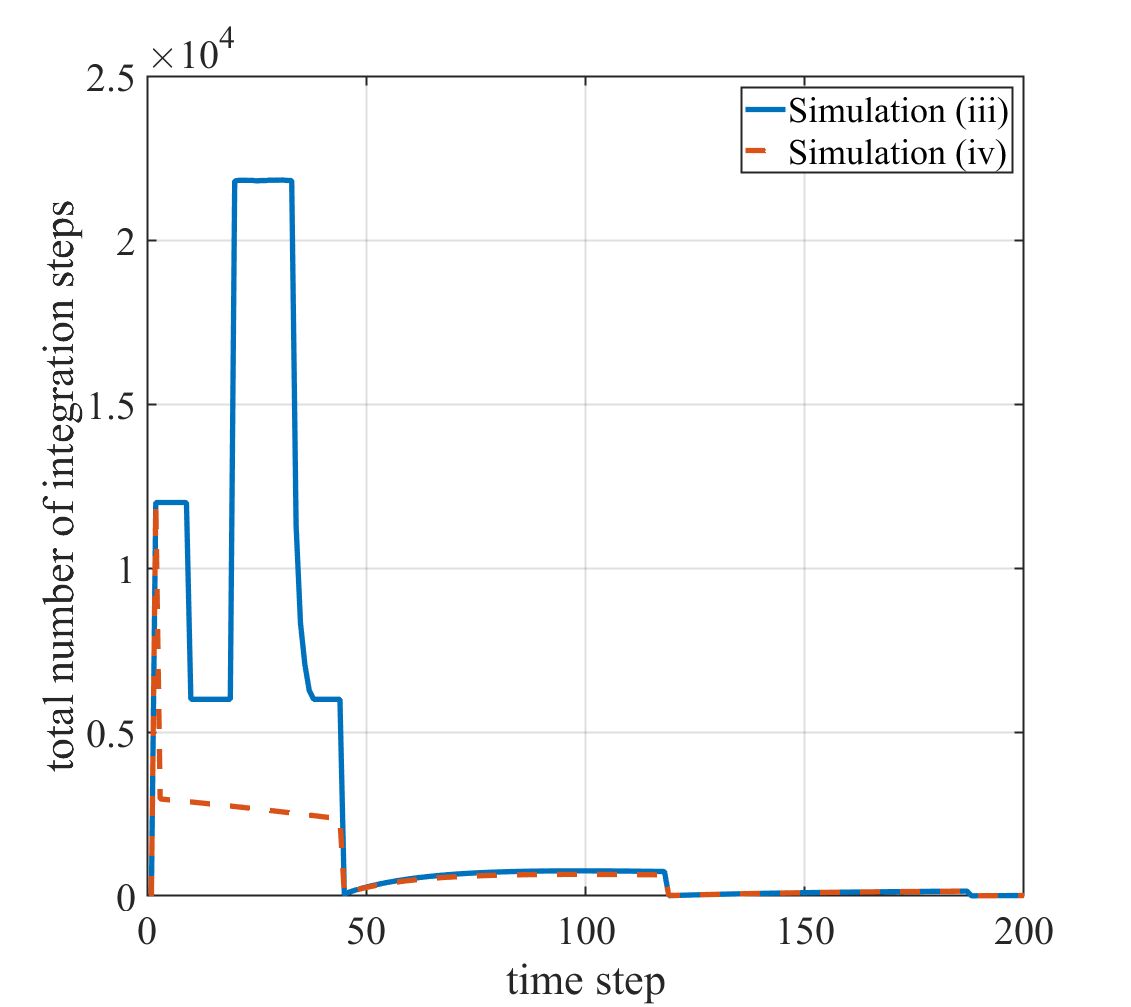}
    \caption{A 1D bimaterial Riemann problem: Effect of R-tree on the total number of integration steps.}
    \label{fig:1dRtree}
\end{figure}

Comparing the computation times of Simulations  (\romannumeral 3) and (\romannumeral 4),  it can be observed that searching for the initial guesses from  the R-tree reduces the cost of exact Riemann problem solutions by about $70\%$. This reduction can be explained by Fig.~\ref{fig:1dRtree}, which shows the number of integration steps in the first $200$ time steps. At each time step, because the initial guesses provided by the R-tree are closer to the true solution, the integration interval is shorter in Simulation (\romannumeral 4), which leads to a smaller number of integration steps. The computational overhead associated with updating, normalizing, and accessing the R-tree is found to be less than $10\%$ of the time spent on solving the exact Riemann problems. This overhead is acceptable, as expected.

\subsection{Underwater explosion}
\label{sec:UNDEX}

We simulate the expansion of a high-pressure gas bubble in a liquid, which can be viewed as a model problem for underwater explosion. Following \cite{Shyue1998}, we specify a bubble with radius $r=0.2$ at $t = 0$. The  (non-dimensional) initial conditions inside and outside the bubble are given by
\begin{equation}
\left(\rho, u, p, EOS\right)= \begin{cases}(0.991,0,3.059 \times 10^{-4} , \text{stiffend gas}) & \text {liquid phase, outside the bubble,} \\ (1.241,0,2.753, \text{stiffened gas}) & \text {gas phase, inside the bubble.}\end{cases}
\end{equation}\
The parameters in the stiffened gas EOS, i.e.,~Eq.~\eqref{eq:stiffened_EOS}, are $\gamma = 5.5$, $e_c=0$, $b=0$ and $p_c = 1.505$ for the liquid phase, and $\gamma=1.4$, $e_c=0$, $b=0$, and $p_c=0$ for the gas phase. The Euler equations are solved, which means in Eq.~\eqref{eq:NS_3d_1}, $\bm{\mathcal{S}}=\bm{0}$.  The computational domain is 2D, extending from $0.0$ to $1.0$ in both $x$ and $y$ directions. It is discretized by a uniform, $500 \times 500$ grid. The Roe's flux function is adopted to compute the advective fluxes. Again, the four simulations designed in Sec.~\ref{sec:shocktube} are performed for this problem. All the simulations are parallelized to run on $256$ CPU cores. Same as in the previous case, the error tolerance for adaptive step size control is set to $1.0\times10^{-9}$ in Simulations (\romannumeral 3) and (\romannumeral 4). Reaching this tolerance requires up to $1,092$ integration steps. This number is then specified as the fixed number of integration steps in Simulations (\romannumeral 1) and (\romannumeral 2), in order to ensure the same level of accuracy.

\begin{figure}[H]
    \centering
    \begin{subfigure}[b]{70mm}
       \caption{}
       \hspace{8mm}
       \includegraphics[width=60.5mm,trim={0cm 0cm 0cm 0cm},clip]{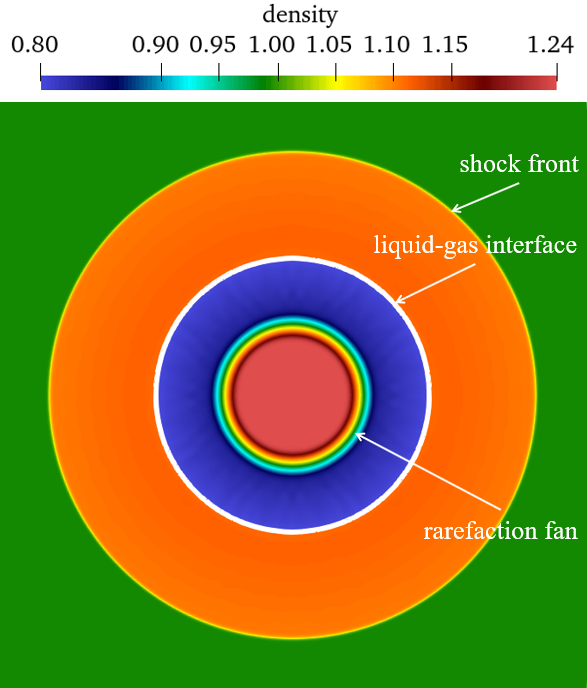}
    \end{subfigure}
    \hspace{10mm}
    \begin{subfigure}[b]{70mm}
       \caption{}
       \hspace{7.5mm}
       \includegraphics[width=60.5mm,trim={0cm 0cm 0cm 0cm},clip]{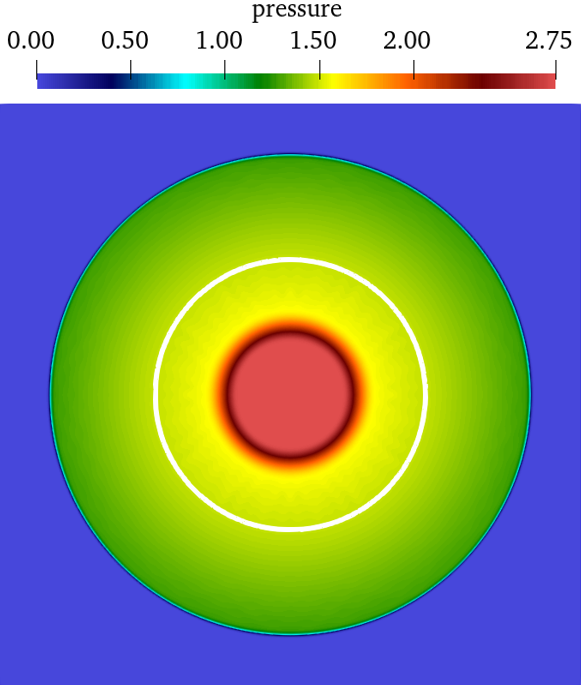}
    \end{subfigure}
    \begin{subfigure}[b]{70mm}
       \caption{}
       \includegraphics[width=70mm,trim={0cm 0cm 0cm 0cm},clip]{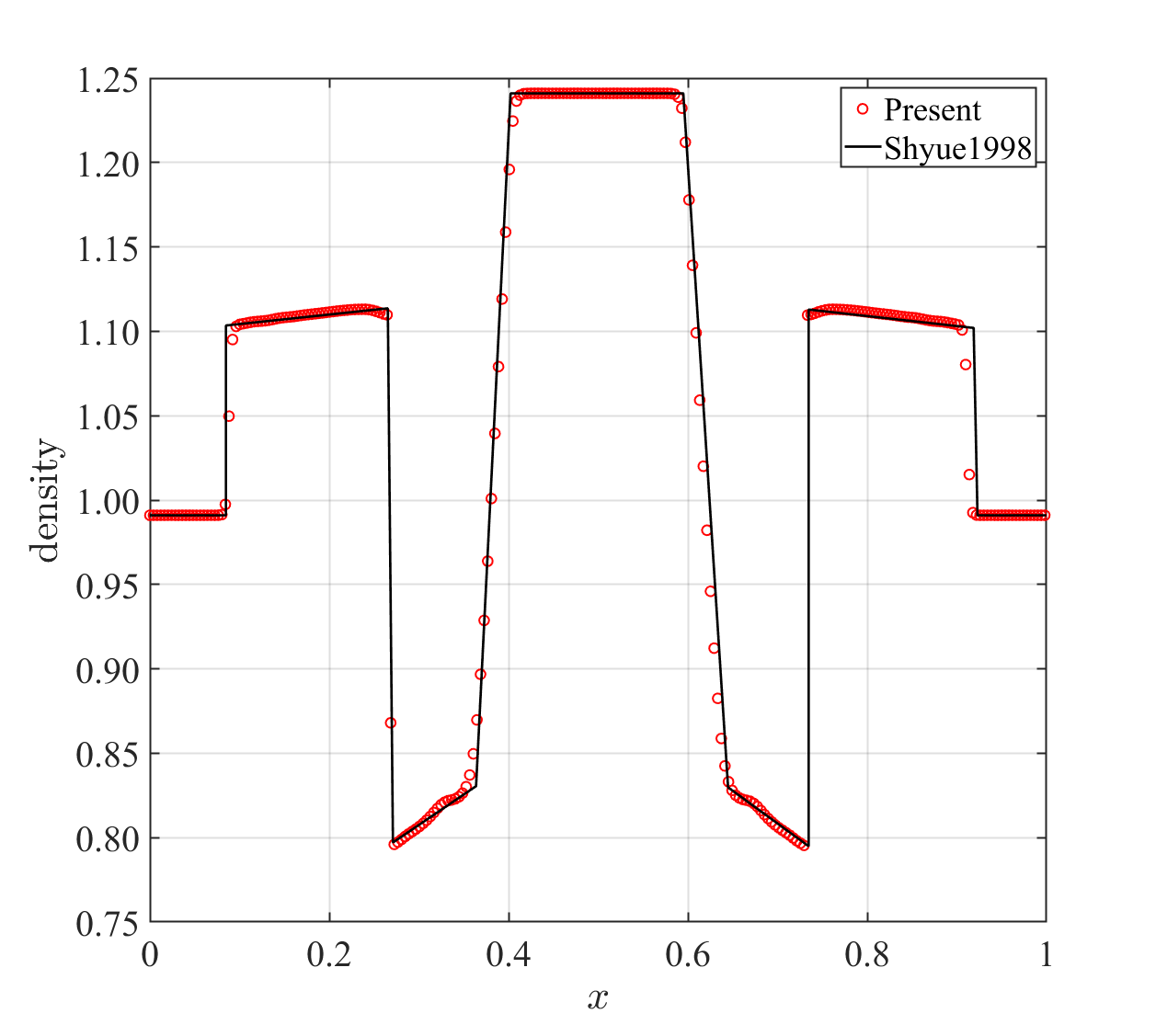}
    \end{subfigure}
    \hspace{10mm}
    \begin{subfigure}[b]{70mm}
       \caption{}
       \includegraphics[width=70mm,trim={0cm 0cm 0cm 0cm},clip]{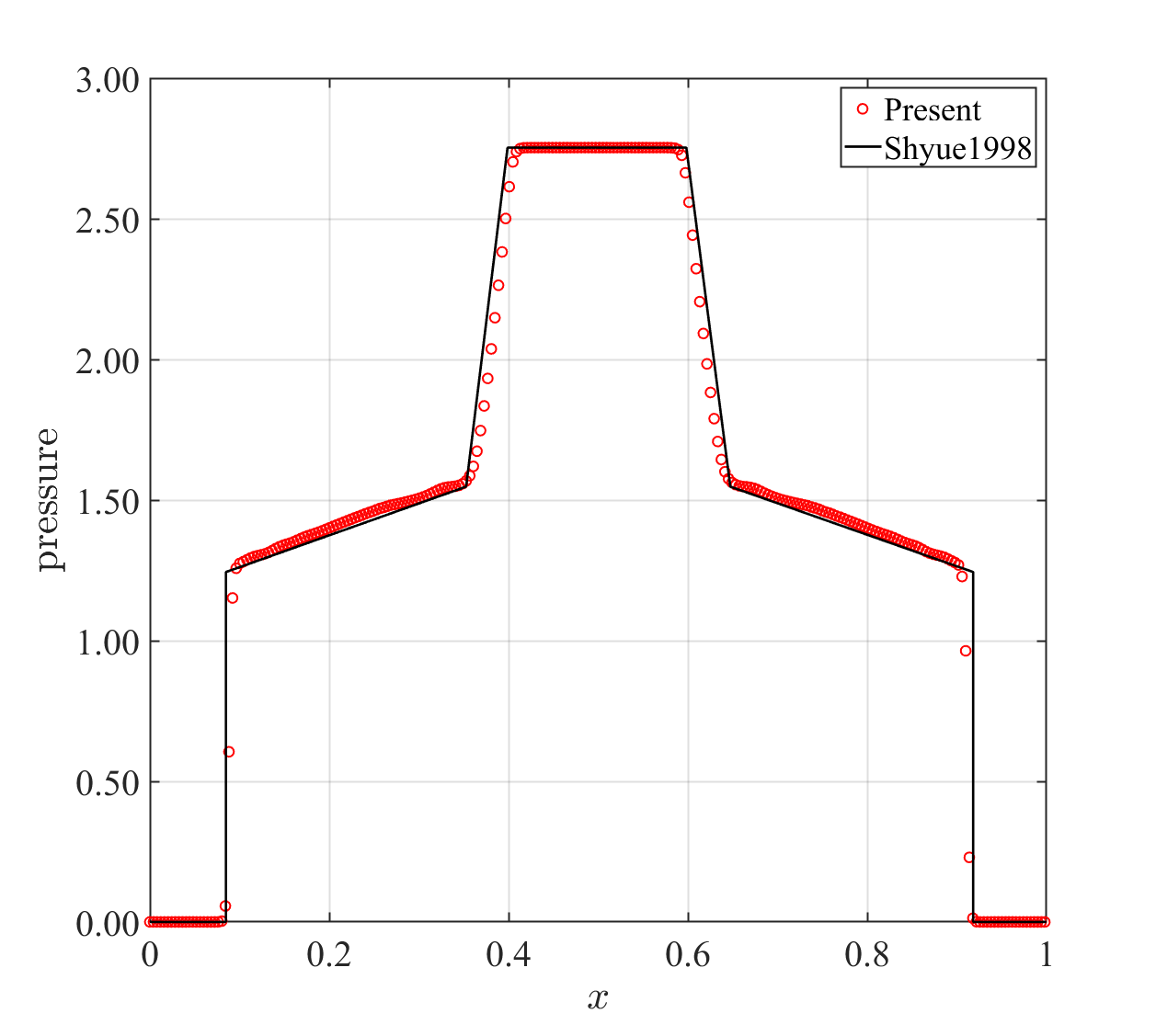}
    \end{subfigure}
    \caption{Underwater explosion: simulation results at $t=0.058$. (a) density filed, (b) pressure filed, (c) density along the horizontal symmetric axis, (d) pressure along the horizontal symmetric axis.}
    \label{fig:2D_SG}
\end{figure}

The results obtained from the four simulations are essentially the same, as the acceleration methods do not introduce new approximations. As an example, Fig.~\ref{fig:2D_SG} shows the result of Simulation (\romannumeral 4) at $t = 0.058$. In this figure, Subfigures (a) and (b) visualize the density and pressure fields.  Subfigures (c) and (d) plot the same solution fields along the horizontal centerline of the computational domain, in comparison with the results given in~\cite{Shyue1998}. For both density and pressure, the present result is in close agreement with the reference. The liquid-gas interface, the outgoing shock wave, and the incoming rarefaction fan are all captured accurately. 

\begin{figure}[H]
    \centering
    \includegraphics[width=80mm,trim={0cm 0cm 0cm 0cm},clip]{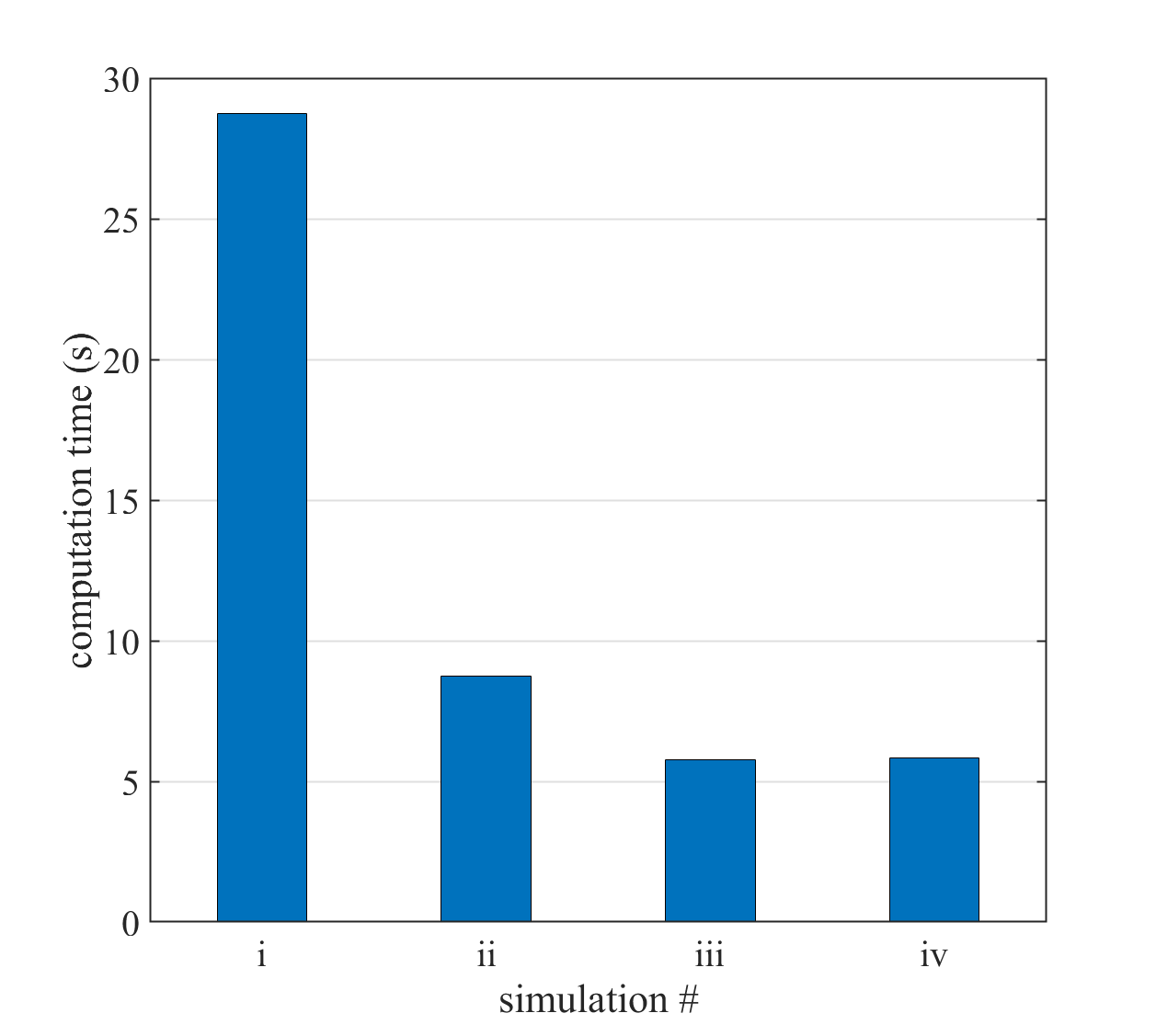}
    \caption{Underwater explosion: Computational cost of exact Riemann problem solutions in the first $200$ time steps.}
    \label{fig:undex_first200}
\end{figure}

Figure~\ref{fig:undex_first200} shows the computational cost of exact Riemann problem solutions within the first $200$ time steps, when the exact Riemann problems are highly nonlinear. After storing and reusing the integration trajectories (i.e.,~Simulation (\romannumeral 2)), the cost of numerical integration becomes trivial after the first iteration in the outer loop. As a result,  the computation time is reduced by a factor of $3$ ($9$ s vs. $29$ s), compared to the baseline Simulation (\romannumeral 1). After adopting step size adaptation (i.e.,~Simulation (\romannumeral 3)), the computation time is further reduced by $34\%$, because fewer integration steps are needed. Comparing the costs of Simulations (\romannumeral 3) and (\romannumeral 1), it is found that the first three acceleration methods, when combined together, provides a $5$-fold cost reduction. By obtaining the initial guesses from R-tree (i.e.,~Simulation (\romannumeral 4)), however, the computational cost increases slightly by $0.86\%$ compared to Simulation (\romannumeral 3). This is because the reduction in Riemann problem solution time is offset by the computational overhead associated with updating and maintaining the R-tree. Specifically, the Riemann problem solution time is reduced by $10.35\%$ compared to Simulation (\romannumeral 3), but the computational overhead amounts to $11.21\%$. The benefit brought by the R-tree is less significant than in the previous case. This can be attributed to the fact that the density jump across the material interface is less than a factor of $2$ in this case, while it was $4$ orders of magnitude in the previous case. Overall, we have found that unless the density and EOS across the material interface are extremely different, the initial guesses provided by the R-tree may not be significantly better than those obtained by the linear acoustic theory. For this reason, Simulation (\romannumeral 4) is excluded from the subsequent test cases, except one in Sec.~\ref{sec:HVI}, where the R-tree's benefits overtake its cost due to the significant discontinuity across the material interface.

\begin{figure}[H]
    \centering
    \includegraphics[width=80mm,trim={0cm 0cm 0cm 0cm},clip]{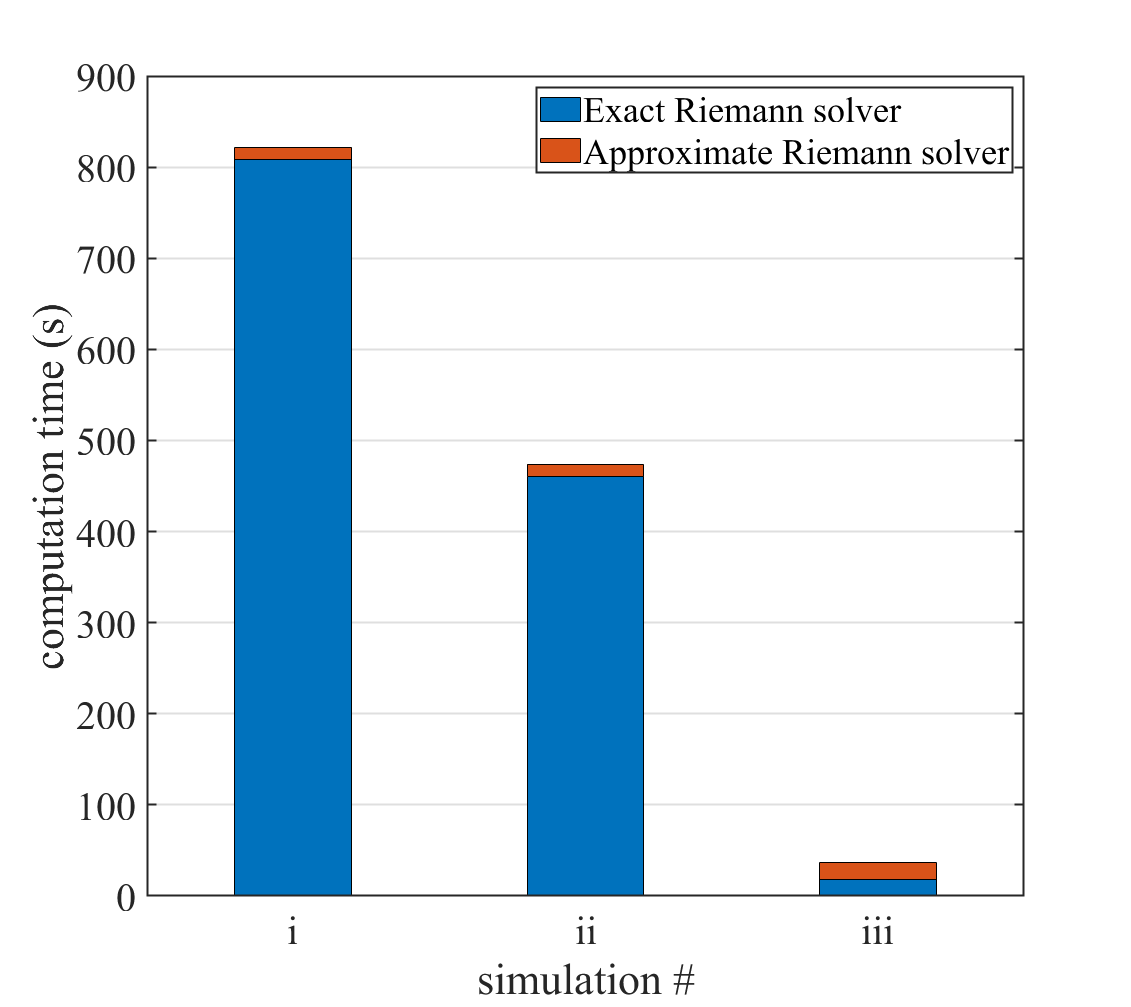}
    \caption{Underwater explosion: Comparison of the flux computation time of different simulations.}
    \label{fig:ShyueSG_accel}
\end{figure}

Furthermore, Fig.~\ref{fig:ShyueSG_accel} compares the computation time of Simulations (\romannumeral 1)(\romannumeral 2)(\romannumeral 3) up to $t = 0.058$, or $13,029$ time steps. It can be observed that when acceleration Methods (1), (2), and (3) are applied, the solution of exact Riemann problems is accelerated by $45$ times (from $808$ s to $18$ s). The total time on advective flux computation is reduced by a factor of $22$ (from $822$ s to $37$ s).

\vspace{5mm}

\begin{figure}[H]
    \centering
    \begin{subfigure}[b]{70mm}
       \caption{}
       \includegraphics[width=70mm,trim={0cm 0cm 0cm 0cm},clip]{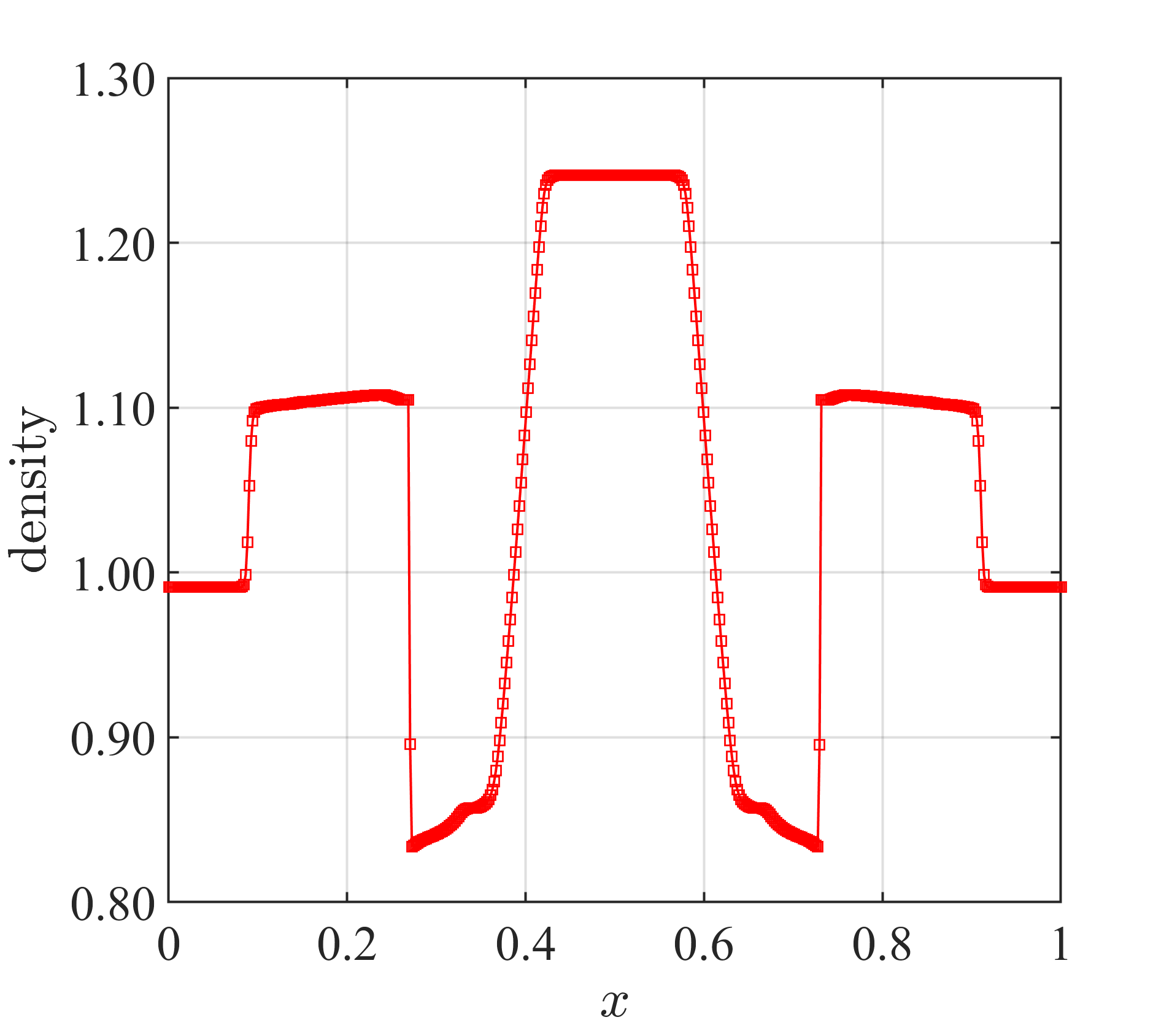}
    \end{subfigure}
    \hspace{10mm}
    \begin{subfigure}[b]{70mm}
       \caption{}
       \includegraphics[width=70mm,trim={0cm 0cm 0cm 0cm},clip]{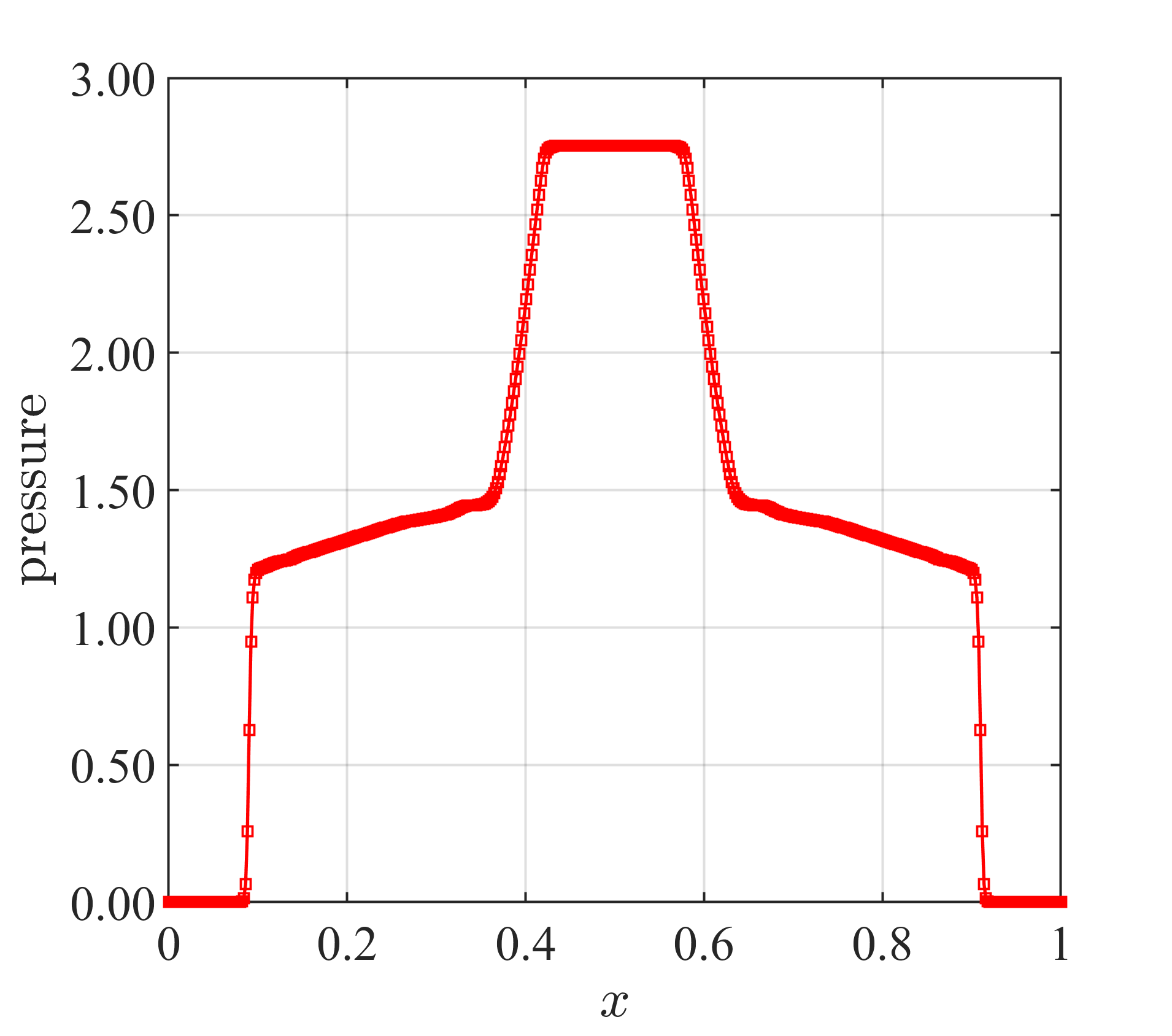}
    \end{subfigure}
    \caption{Underwater explosion with the JWL EOS: Simulation result at $t = 0.058$, along the horizontal centerline of the computational domain.}
    \label{fig:2D_JWL}
\end{figure}

In the aforementioned simulations, the gas inside the bubble is modeled using the stiffened gas EOS, following the example problem described in \cite{Shyue1998}. In general, the thermodynamics of gaseous explosion products can be better described using the JWL EOS~\eqref{eq:JWL_EOS}. Compared to stiffened gas, the JWL EOS is highly nonlinear as it involves exponential functions. Also, $\rho$ can no longer be solved analytically from $p$ and $e$. Instead, an iterative numerical method is needed. To further assess the performance of the acceleration methods, we perform another set of simulations in which the gas inside the bubble is modeled using the JWL EOS. Its parameters are set by $\omega = 0.28$, $A_1 = 37.12$, $A_2 = 0.323$, $R_1 = 4.15$, $R_2 = 0.95$, and $\rho_0 = 1.241$, so that the results are comparable with those obtained using the stiffened gas EOS. Figure~\ref{fig:2D_JWL} shows the density and pressure distributions at $t = 0.058$ along the horizontal centerline of the computational domain.

\begin{figure}[H]
    \centering
    \includegraphics[width=80mm,trim={0cm 0cm 0cm 0cm},clip]{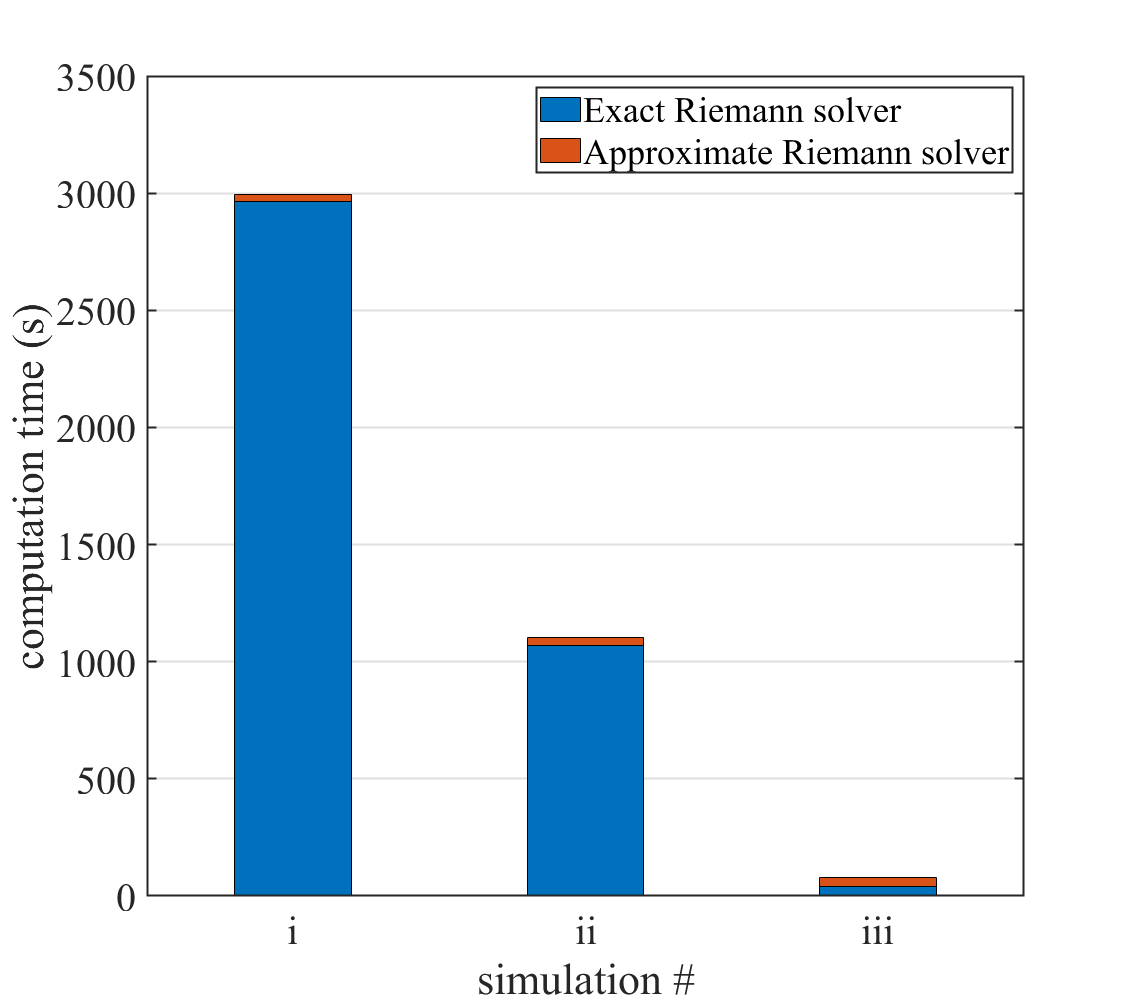}
    \caption{Underwater explosion with the JWL EOS: Comparison of the flux computation time of different simulations.}
    \label{fig:ShyueJWL_accel}
\end{figure}

Figure~\ref{fig:ShyueJWL_accel} shows the performance of the accelerated Riemann problem solver, in the presence of the JWL EOS. Again, the computation time is calculated up to $t = 0.058$. Comparing with Fig.~\ref{fig:ShyueSG_accel}, it is clear that the JWL EOS leads to a higher computational cost. The effect of the acceleration methods is even more significant. When Methods (1), (2), and (3) are combined (i.e.,~Simulation (\romannumeral 3)), the solution of exact Riemann problems is accelerated by $76$ times ($39$ s vs. $2,963$ s). As a result, the total cost of flux computation is reduced by a factor of $38$ ($80$ s vs. $2,994$ s). 

\subsection{Laser-induced cavitation}
\label{sec:laser}
Next, we assess the performance of the acceleration methods using an example problem of cavitation induced by a long-pulse laser, first presented in~\cite{Zhao2023}. A Holmium:YAG laser with a $2.12\times 10^{-3}~\text{mm}$ wavelength, a $10.2~\mu\text{s}$ pulse duration, and a peak power of $2.854~\text{kW}$ is considered. The cylindrical laser fiber has a radius of $0.1825~\text{mm}$, and is placed within a water tank. The energy of the laser pulse is high enough to vaporize a small volume of liquid water in front of the laser fiber, thereby generating a vapor bubble. The two-phase liquid-vapor flow is modeled as a compressible and inviscid flow. A radiative heat source is added to the energy conservation equation. Therefore, the term $\bm{\mathcal{S}}$ in~\eqref{eq:NS_3d_1} has the form
\begin{equation}
\bm{\mathcal{S}} = \left[\begin{array}{c}
0 \\
\bm{0} \\
-\nabla \cdot \bm{q}_r
\end{array}\right],
\end{equation}\
where $\bm{q}_r$ denotes the radiative heat flux, given by
\begin{equation}
    \bm{q}_r = L\bm{s}.
    \label{eq:q_r}
\end{equation}\
Here, $L = L(\bm{x})$ denotes laser radiance at position $\bm{x}$. It characterizes the laser intensity at that position. $\bm{s}$ denotes the propagation direction of the laser light. It is also a function of $\bm{x}$ because the laser beam simulated here has a $7.5^\circ$ diverging angle. $L$ is governed by the following laser radiation equation that enforces conservation of energy.
\begin{equation}
\nabla \cdot(L \bm{s})=\nabla L \cdot \bm{s}+(\nabla \cdot \bm{s}) L=-\mu_\alpha L.
\label{eq:laser_radiation}
\end{equation}\
Here, $\mu_{\alpha}$ is the laser absorption coefficient, which depends on the laser wavelength, the fluid material, and the fluid temperature. Same as in~\cite{Zhao2023}, the method of latent heat reservoir is applied to model the phase transition (i.e.,~vaporization). In addition, the cylindrical symmetry of the fluid flow with respect to the central axis of the laser fiber is leveraged to allow simulations on 2D grids. 

Both the liquid and vapor phases are modeled by the stiffened gas EOS. The parameters are $\gamma = 6.12$, $e_c=0$, $b=0$,  and $p_c = 343~\text{MPa}$ for the liquid phase, and $\gamma = 1.34$, $e_c=0$, $b=0$, and $p_c = 0$ for the vapor. In the latter case, the stiffened gas degenerates to the perfect gas. Additional details about this problem  --- including its background, novelty, the model equations, and the numerical methods  --- can be found in~\cite{Zhao2023}.

The computational domain is 2D, discretized by a non-uniform Cartesian grid with approximately $338,000$ elements. In the most refined region, the element size is around $2.5\times 10^{-3}~\text{mm}$. Simulations (\romannumeral 1), (\romannumeral 2), and (\romannumeral 3) defined in Sec.~\ref{sec:shocktube} are performed for this example problem, on $256$ CPU cores. Again, the error tolerance for integration step adaptation is set to $1.0\times10^{-9}$ in Simulation (\romannumeral 3), which translates to $841$ integration steps for Simulations (\romannumeral 1) and (\romannumeral 2).
    
\begin{figure}[H]
    \centering
    \includegraphics[width=160mm,trim={0cm 0cm 0cm 0cm},clip]{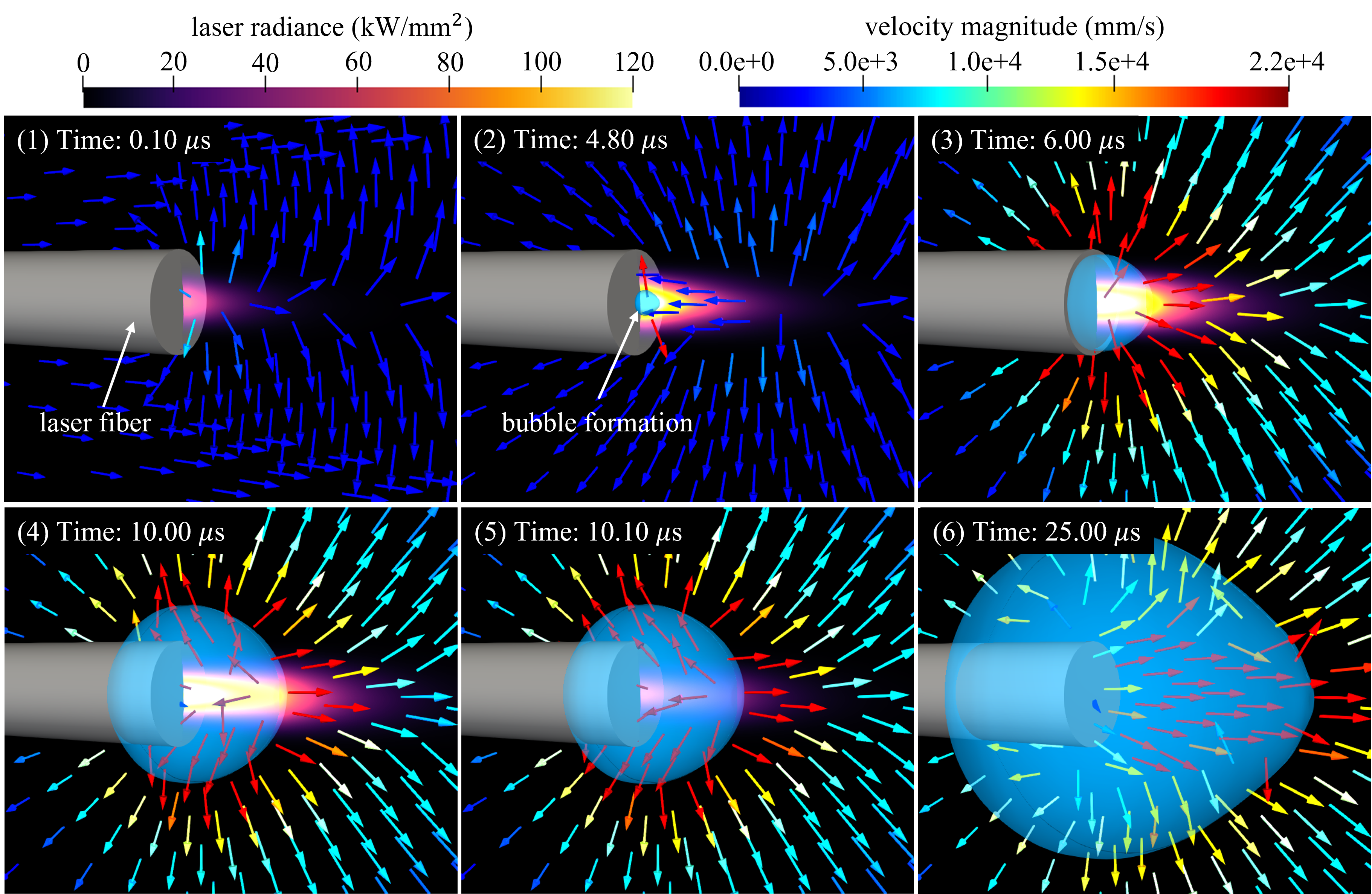}
    \caption{Laser-induced cavitation: Laser radiation, bubble formation, and the fluid velocity field.}
    \label{fig:laserRadiance}
\end{figure}

\begin{figure}[H]
    \centering
    \includegraphics[width=160mm,trim={0cm 0cm 0cm 0cm},clip]{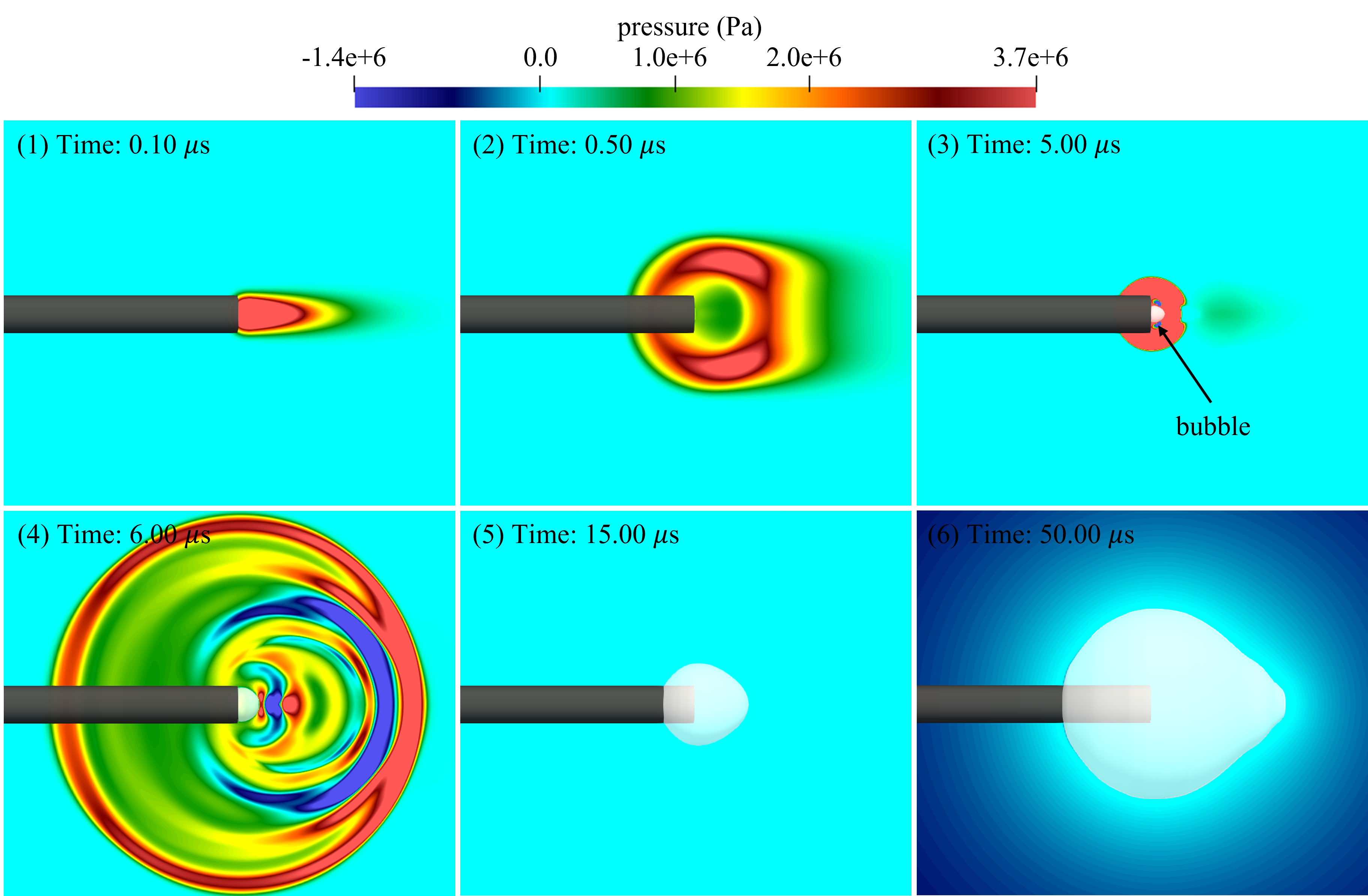}
    \caption{Laser-induced cavitation: The pressure field.}
    \label{fig:laserPressure}
\end{figure}

Figure~\ref{fig:laserRadiance} presents the laser radiance and fluid velocity fields obtained from Simulation (\romannumeral 3). The solutions obtained from the other two simulations are the same. Initially, the entire fluid domain is occupied by the liquid phase. The laser source power rapidly grows from $0$ to its peak value during the first $0.2~\mu\text{s}$. Then, it stays at the peak value until $t = 10.0~\mu\text{s}$. At around $4.8~\mu\text{s}$, the simulation predicts the formation of a vapor bubble right in front of the laser fiber tip. Phase transition lasted around $0.025~\mu\text{s}$. Afterwards, the main flow features are the expansion of the bubble, the propagation of vaporization-induced acoustic waves, and laser heating. Figure~\ref{fig:laserPressure} presents the fluid pressure field at six time instances. Subfigures (1) and (2) show the acoustic waves induced by the sudden increase of the fluid's internal energy due to laser radiation.  Subfigures (3) and (4) demonstrate the formation and expansion of the vapor bubble, as well as the associated acoustic waves. The laser power vanishes at $t = 10.2~\mu\text{s}$. Afterwards, the bubble continues to grow and gradually deforms into a teardrop shape, which is shown in Subfigures (5) and (6). The same shape has been observed in laboratory experiments using the same type of laser~\cite{Ho2021}. This type of non-spherical, beam dependent bubbles is a unique feature of long-pulse laser-induced cavitation.

\begin{figure}[H]
    \centering
    \includegraphics[width=80mm,trim={0cm 0cm 0cm 0cm},clip]{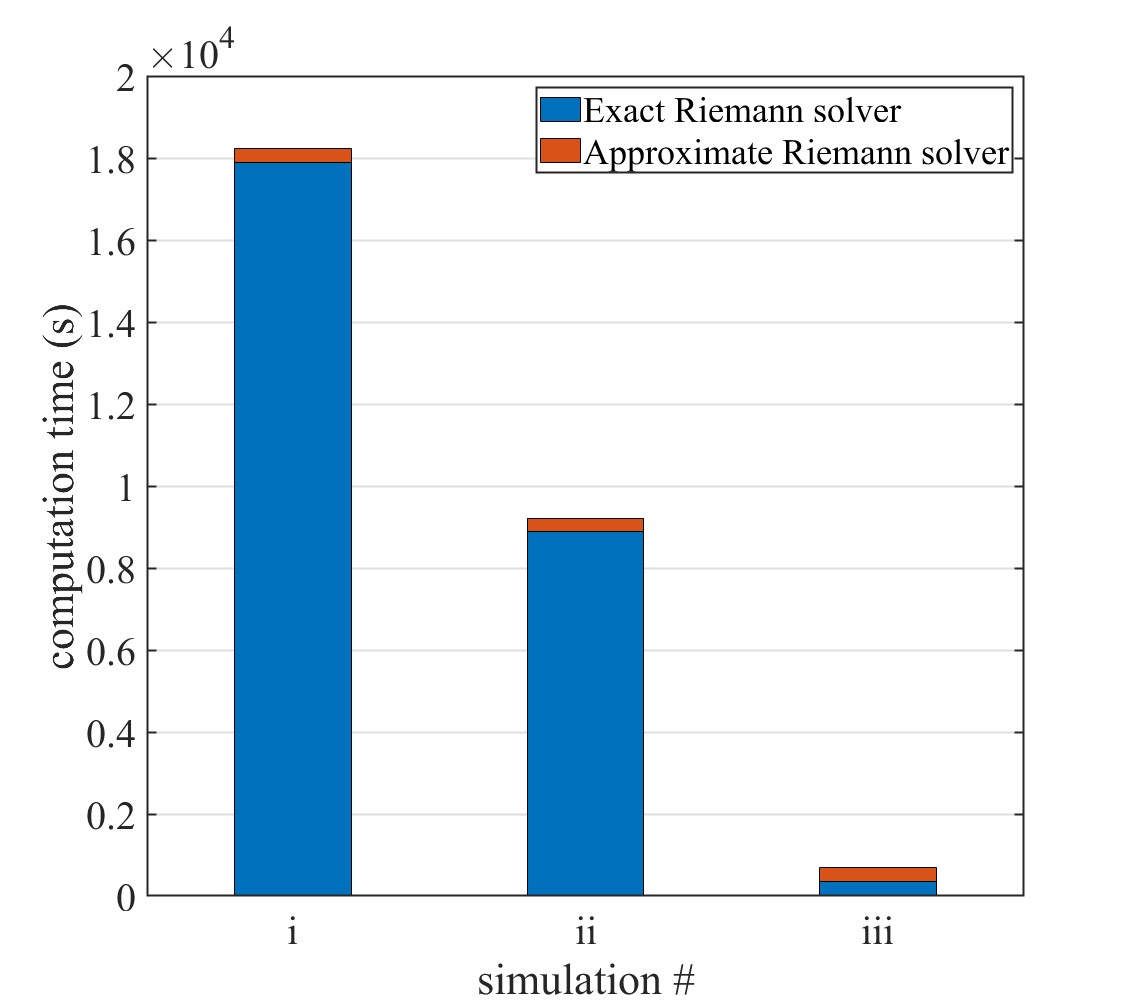}
    \caption{Laser-induced cavitation: Comparison of the flux computation time of different simulations.}
    \label{fig:laserBubble_accel}
\end{figure}

Figure~\ref{fig:laserBubble_accel} compares the computation times of Simulations (\romannumeral 1)(\romannumeral 2)(\romannumeral 3) up to $t=50~\mu\text{s}$. The advective fluxes are evaluated using the LLF numerical flux function. It can be observed that the solution of exact Riemann problems is accelerated by $51$ times ($350$ s vs. $17,899$ s), when Methods (1), (2), and (3) are used. In the baseline Simulation (\romannumeral 1), over $95\%$ of the flux computation time is spent on solving the exact Riemann problems. As a result, after the acceleration, the total flux computation time is reduced by a factor of $26$ (from $18,233$ s to $696$ s).

\subsection{Hypervelocity impact in a fluid environment}
\label{sec:HVI} 
Finally, we consider an example problem of hypervelocity impact, first presented in~\cite{Islam2023}. In this problem, a copper projectile impacts onto a soda lime glass (SLG) target in a neon gas environment. If the projectile's velocity is sufficiently high, the peak pressure inside the projectile and the target can be much higher than the material's Hugoniot elastic limit. In this scenario, the solid materials can be modeled as compressible fluids. Unlike conventional impact simulations that consider only the projectile and the target, we perform a fluid-solid coupled analysis that includes the ambient neon gas as a material subdomain. Therefore, this problem involves three material subdomains, and three material interfaces (neon-copper, copper-SLG, and SLG-neon). Two level set equations are solved to track the boundaries of the copper and SLG subdomains, using the method described in Sec.~\ref{sec:numerical_scheme}. This problem also challenges the robustness of the solver as it exhibits extreme mechanical and thermodynamic states, and the EOS vary significantly across the fluid-solid interfaces. 

Following~\cite{Islam2023}, the effects of viscosity and heat diffusion are neglected. The impact problem is analyzed in two stages, namely the flight of the projectile in the fluid medium, followed by the collision of the projectile onto the target. The first stage produces a bow shock that reaches the target before the projectile. This flow field can be obtained by either a conventional body-fitted CFD analysis (\cite{Islam2023}), or a multi-material simulation in which the projectile surface is tracked using the level set method. To demonstrate the accelerated Riemann problem solver, we first present a simulation that employs the second approach. The copper projectile is a cylinder with a spherical leading edge. Its thermodynamics is modeled using the Mie-Gr\"{u}neisen EOS (i.e.,~Eq.~\eqref{eq:MG_EOS}), with parameters $c_0=3.93~\text{km/s}$, $s=1.5$, $\rho_0=8,960~\text{kg/m}^3$, and $\Gamma_0 = 2.12$~\cite{Islam2023}. The neon gas is modeled as a perfect gas, with $\gamma=1.667$. The initial pressure within the copper projectile and neon gas are both $1.0\times10^5~\text{Pa}$. The copper projectile has an initial density of $8,960~\text{kg/m}^3$ and neon gas $0.82~\text{kg/m}^3$. Therefore, the density ratio across the projectile surface reaches $4$ orders of magnitude. 

\begin{figure}[H]
    \centering
    \begin{subfigure}[b]{50mm}
       \caption{}
       \includegraphics[width=45mm,trim={0cm 0cm 0cm 0cm},clip]{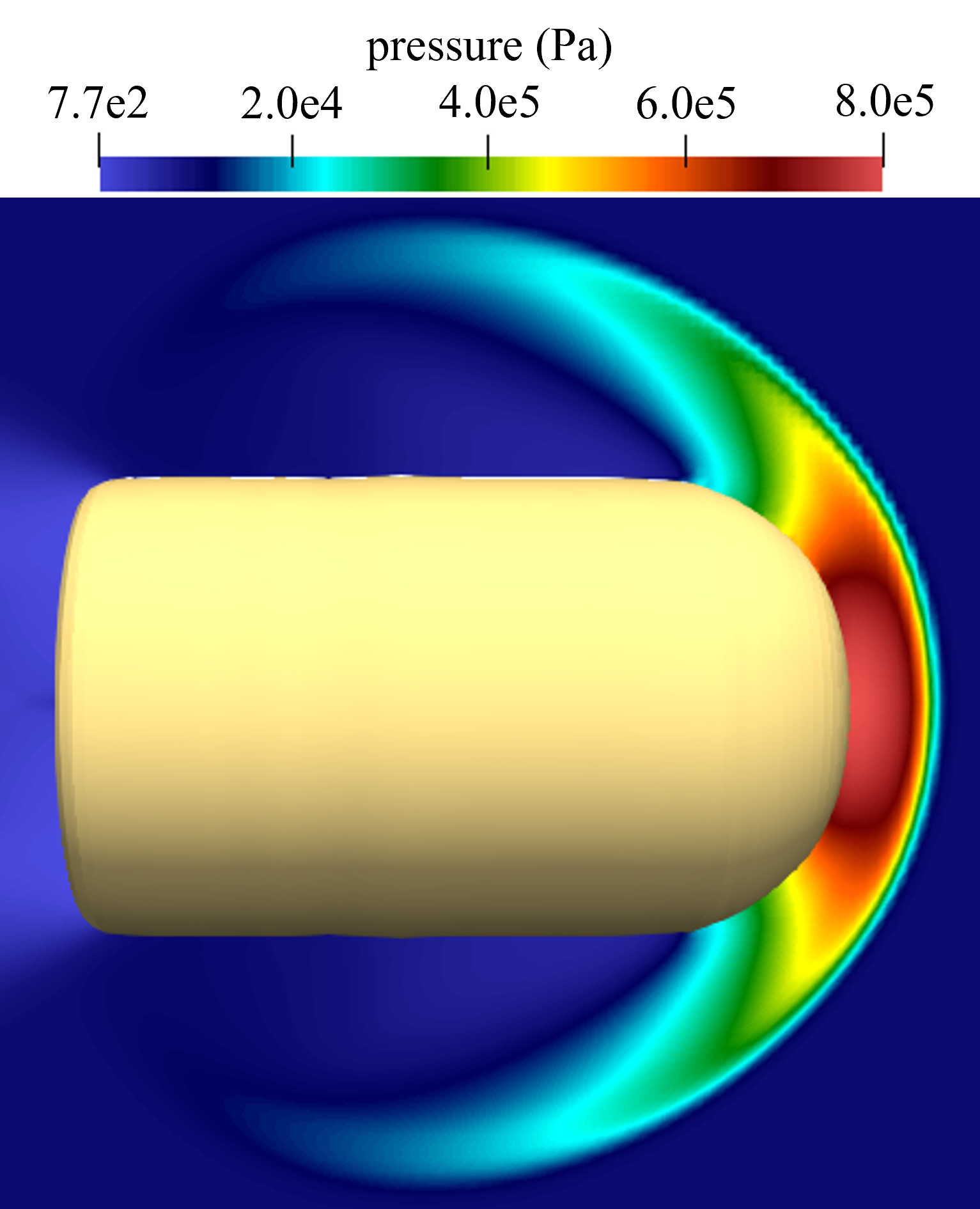}
    \end{subfigure}
    \hspace{0mm}
    \begin{subfigure}[b]{50mm}
       \caption{}
        \includegraphics[width=45mm,trim={0cm 0cm 0cm 0cm},clip]{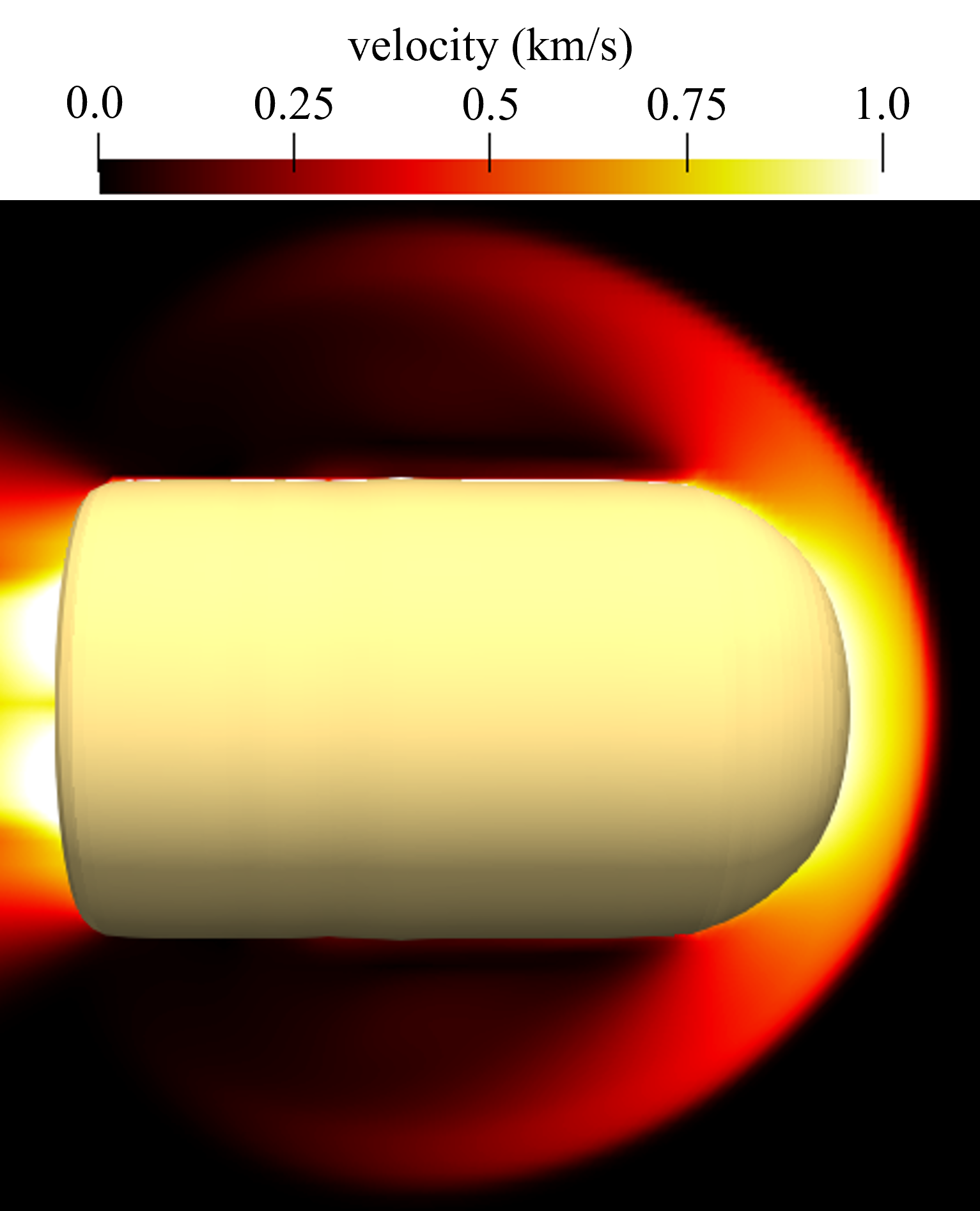}
    \end{subfigure} 
    \caption{Simulation of flow around a copper projectile before collision: Pressure and velocity fields at $t = 8.87~\mu\text{s}$. Velocity of projectile: $1~\text{km}/\text{s}$.}
    \label{fig:HVI_results1}
\end{figure}

Leveraging the cylindrical symmetry of the problem, we use a 2D non-uniform Cartesian mesh. In the most refined region, the element size is around $0.25$ mm. Simulations (i), (ii), (iii), and (iv) defined in Sec.~\ref{sec:shocktube} are performed, each one using $256$ CPU cores. The number of integration steps in Simulations (i) and (ii) is set to $8,915$, which is determined by the step size adaptation of Simulation (iii), to achieve an error tolerance of $1.0 \times 10^{-9}$. This number of integration steps is greater than those in previous test cases, because of the significant discontinuity across the material interface and the extreme physical conditions.

\begin{figure}[H]
    \centering
        \includegraphics[width=80mm,trim={0cm 0cm 0cm 0cm},clip]{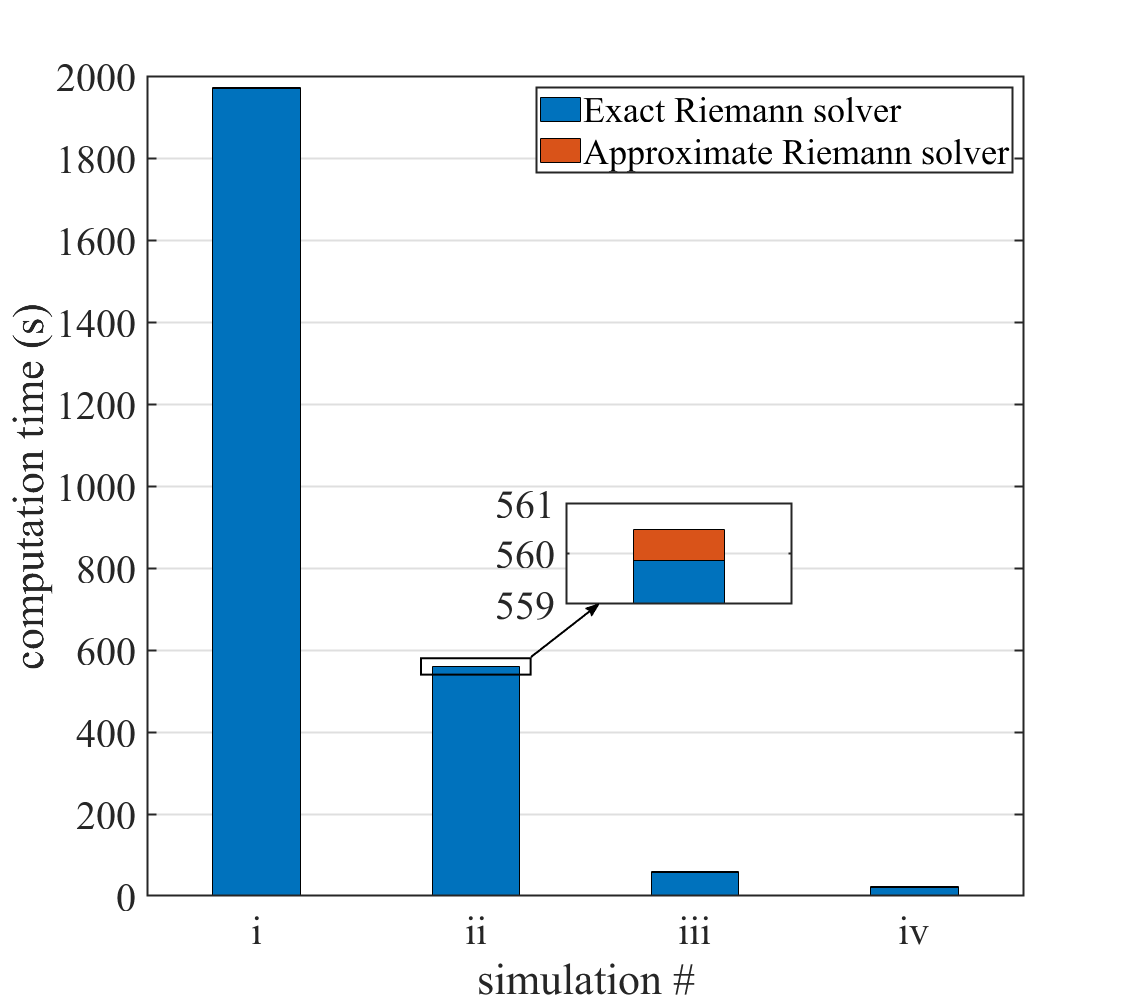}
    \caption{Simulation of flow around a copper projectile before collision: Comparison of the flux computation time.}
    \label{fig:HVI_RtreeAccel}
\end{figure}

Figure~\ref{fig:HVI_results1} presents the pressure and velocity fields obtained from the simulation at $t=8.87~\mu\text{s}$. As expected, a bow shock appears in front of the projectile, and rarefaction waves are generated at the projectile's back surface. The computation times of the four simulations up to $t=40.2~\mu\text{s}$ are presented in Fig.~\ref{fig:HVI_RtreeAccel}. The solution of exact Riemann problems is accelerated by $87$ times (from $1,970.5$ s to $22.7$ s), after applying the four methods. In the baseline simulation, over $99\%$ of the flux computation time is spent on solving the exact Riemann problems. As a result, the total advective flux computation time is reduced by a factor of $81.3$, when the four acceleration methods are applied. In particular, after using the initial guesses searched from R-tree, the flux computation time further drops by $58.7\%$ (from $58.7$ s to $24.2$ s). The computational time spent on updating, normalizing, and accessing the R-tree is only $1.08$ s. Therefore, in this test case with very significant discontinuity across the projectile surface, the benefits of Method $4$ far overtake its cost.

Next, we present a simulation of the second stage of the impact problem. The projectile's velocity is set to $4~\text{km}/\text{s}$. A body-fitted CFD analysis is performed to generate the initial flow field, following~\cite{Islam2023}. In this case, the projectile is a long copper cylinder with a spherical leading edge. Its radius is $5$ mm. The material properties of the copper projectile and the neon gas are identical to the previous test case. The SLG target is modeled using the stiffened gas EOS (i.e.,~Eq.~\eqref{eq:stiffened_EOS}), with $\gamma=3.9$, $e_c=0$, $b=0$, and $p_c = 2.62~\text{GPa}$. Additional details about this test case can be found in~\cite{Islam2023}.

The 2D computational domain is discretized by a non-uniform Cartesian grid with approximately $240,000$ elements. In the most refined region, the grid size is around $0.1$ mm. Simulations (i), (ii), and (iii) defined in Sec.~\ref{sec:shocktube} are performed on $256$ CPU cores. The number of integration steps in Simulations (i) and (ii) is set to $8,596$, which is determined by the step size adaptation of Simulation (iii), to achieve an error tolerance of $1.0 \times 10^{-9}$. The number of integration steps is greater than those in previous sections, because of the extreme physical conditions generated by the impact.

\begin{figure}[H]
    \centering
    \includegraphics[width=160mm,trim={0cm 0cm 0cm 0cm},clip]{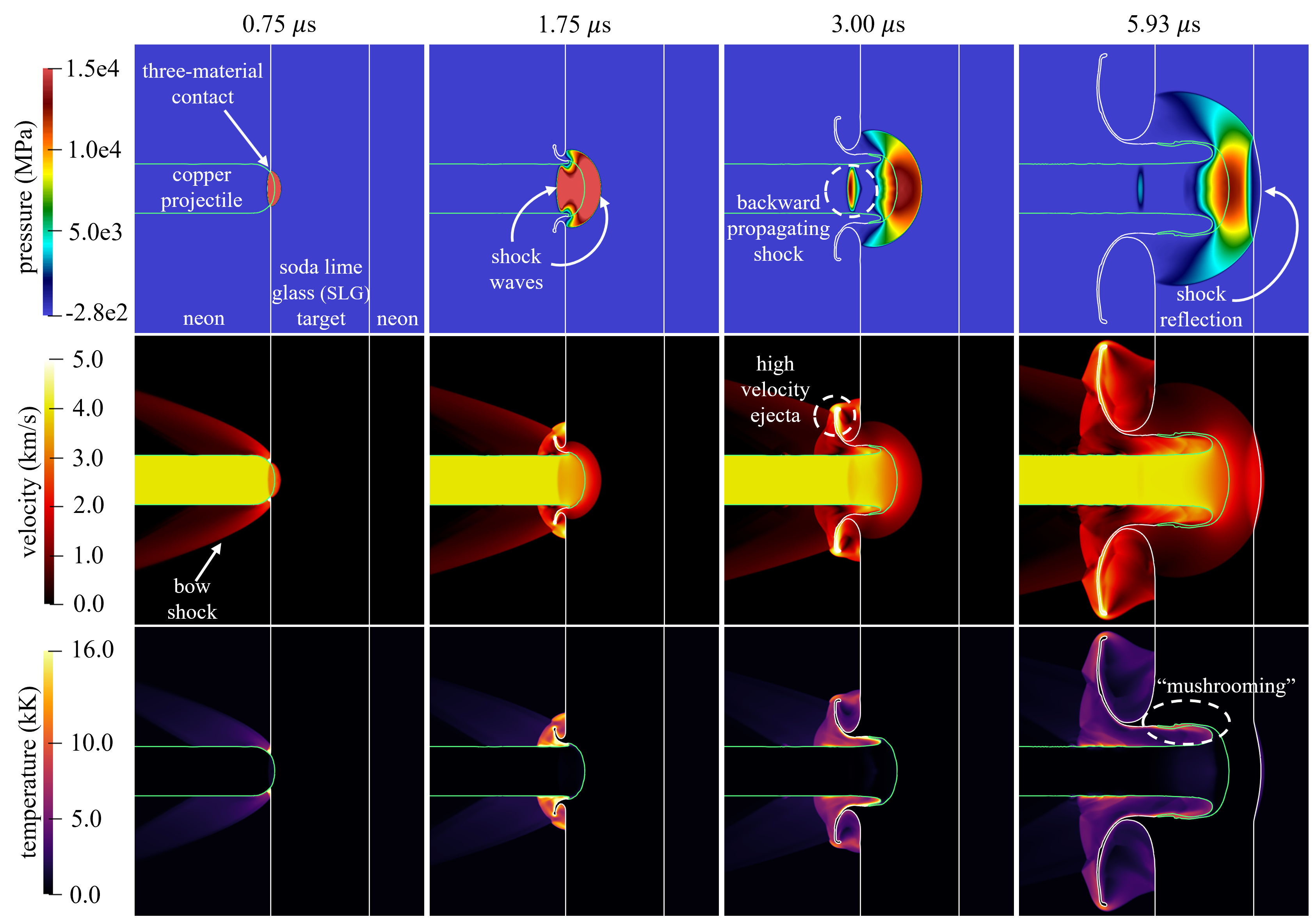}
    \caption{Hypervelocity impact in a fluid environment: pressure, velocity, and temperature fields. Velocity of projectile: $4~\text{km}/\text{s}$.}
    \label{fig:hyperFields}
\end{figure}

Figure~\ref{fig:hyperFields} presents the pressure, velocity, and temperature fields at four different time instances. The surfaces of the copper projectile and the SLG target are obtained by extracting the zero level set of the two level set functions. At $t=0.75~\mu\text{s}$, a time instance shortly after the contact of the two objects, two shock waves propagate in the projectile and the target. The pressure behind the shocks exceeds $20$ GPa. Similar pressure magnitudes were observed in previous experiments~\cite{Kobayashi1998}, also using SLG as the target. In the velocity and temperature fields, the bow shock attached to the projectile is formed due to its hypersonic flight. It is not visible in the pressure field because the scale is selected to highlight the much higher pressure in the solid materials. Because of the impact, the neon gas between the target and the projectile are compressed rapidly, resulting in a region of high pressure and high temperature. As time advances, a crater is formed in the target. The SLG material from the rim of the crater is ejected in the neon gas, at a high velocity that amounts to the impact velocity. At $t=5.93~\mu\text{s}$, the back wall of the target deforms under the impact of the shock wave. In the meantime, the compressive shock wave is reflected as a tensile wave, interfering with the incident shock wave. In the projectile, its leading edge morphs into a ``mushroom head'' shape, due to the rapid deceleration caused by the initially static target. This type of deformation has been observed in previous studies~\cite{sun_shi_liu_wen_2015}.

\begin{figure}[H]
    \centering 
       \includegraphics[width=80mm,trim={0cm 0cm 0cm 0cm},clip]{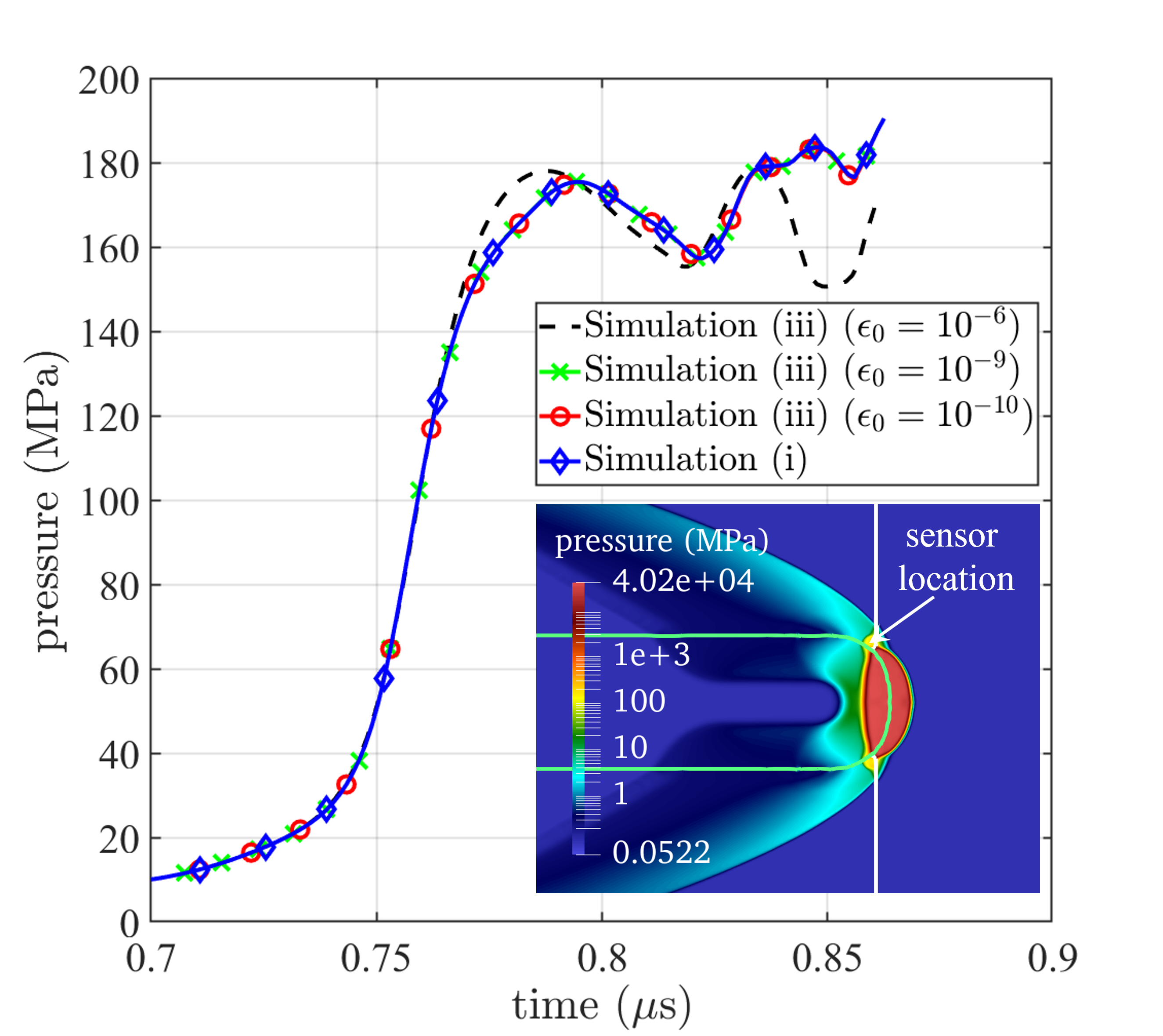}
    \caption{Hypervelocity impact in a fluid environment: the time history of the pressure at the marked sensor location. The pressure field is from $t = 0.825~\mu\text{s}$.}
    \label{fig:HVI_curve}
\end{figure}

In Fig.~\ref{fig:HVI_curve}, the pressure time history at a sensor location is plotted. Specifically, we have performed Simulation (\romannumeral 3) with different error tolerances ($\epsilon_0$) for step adaptation. For comparison, the result from Simulation (\romannumeral 1) with $8,596$ integration steps is also shown. Except for the one with $\epsilon_0 = 10^{-6}$, all the other results match perfectly. This indicates that the flow field predictions can be affected by the error tolerance in the exact Riemann solver; and in certain cases, a very small tolerance  --- such as $10^{-9}$ in this case  --- is needed.

\begin{figure}[H]
    \centering
    \includegraphics[width=80mm,trim={0cm 0cm 0cm 0cm},clip]{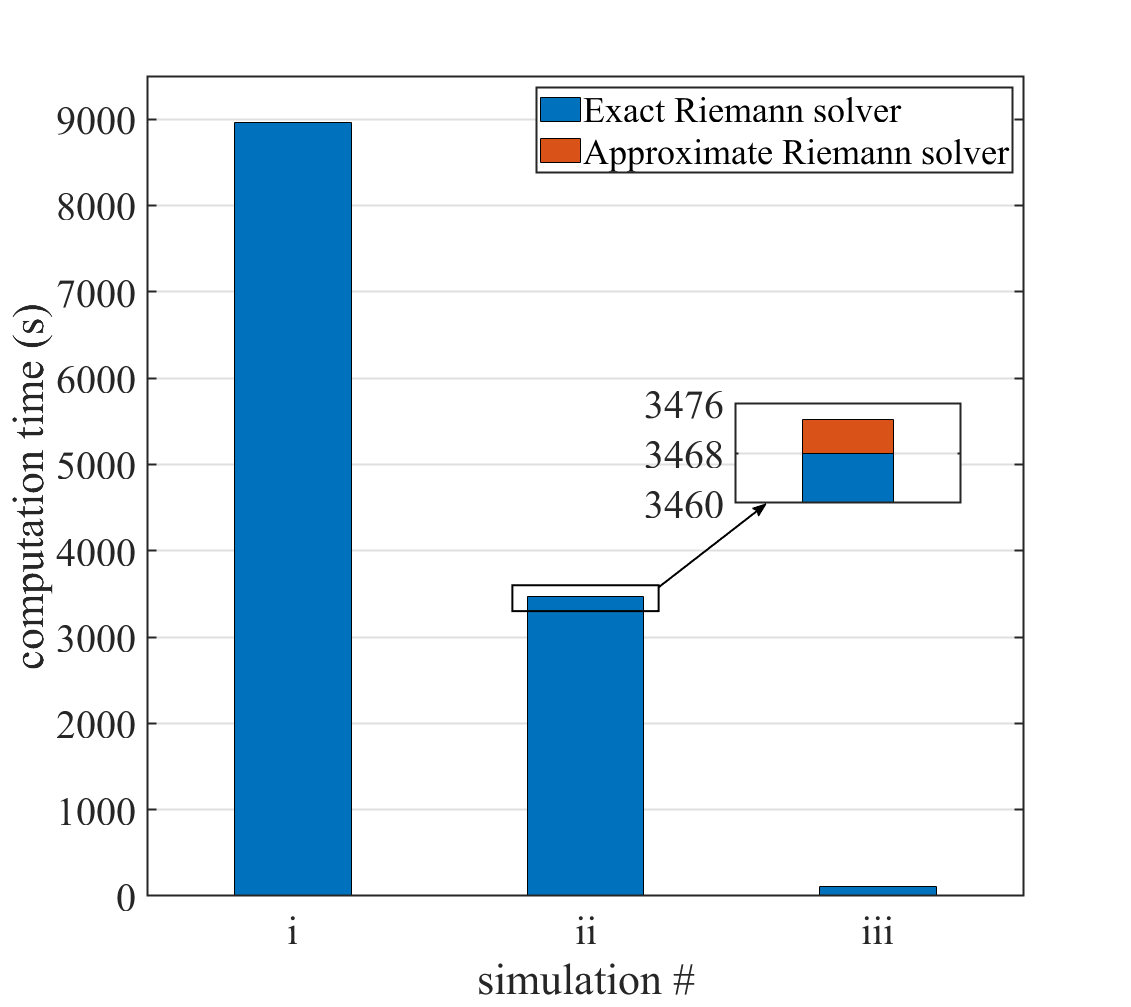}
    \caption{Hypervelocity impact in a fluid environment: Comparison of the flux computation time of different simulations.}
    \label{fig:HVI_accel}
\end{figure}

The computation times of Simulations (i)(ii)(iii) up to $t=0.87~\mu\text{s}$ are presented in Fig.~\ref{fig:HVI_accel}. After applying Methods (1), (2), and (3), the solution of exact Riemann problems is accelerated by $83$ times (from $8,956$ s to $108$ s). The total advective flux computation time is reduced by a factor of $79$.

\section{Summary}
\label{sec:summary}

We consider the numerical solution of compressible multi-material flow problems with sharp interfaces and significant discontinuities. In this context, the advective flux calculator can be viewed as the engine of the solver. Within the engine, the calculation of interfacial fluxes is arguably the most important (yet tricky) part. The idea of locally constructing and solving bimaterial Riemann problems across interfaces has gained popularity in the past two decades. It has a unique advantage in that the discontinuity of both state variables and the thermodynamic EOS are naturally accounted for. This idea entails solving a large number   ---  up to $10^8$ in the test cases presented in Sec.~\ref{sec:numericalExperiments}  --- of bimaterial Riemann problems on the fly. On the other hand, the off-the-shelf method for solving Riemann problems (Sec.~\ref{sec:efficiency}) was developed earlier to solve individual problems on a case-by-case basis, and therefore  not optimized for computational efficiency. As a result, applying it in multi-material flow solvers leads to a drastic  --- sometimes prohibitive  --- increase of computational cost, especially when complicated EOS are involved.

In this paper, we presented four methods to accelerate the solution of bimaterial Riemann problems. The main idea behind these methods is to exploit some special properties of the Riemann problem equations (Sec.~\ref{sec:change_variable},~\ref{sec:trajectory}, and \ref{sec:adaptiveStep}), and to store and reuse previous solutions as much as possible (Sec.~\ref{sec:trajectory} and~\ref{sec:rtree}). The most expensive part of the solution procedure is the numerical integration of state variables through rarefaction fans. Three of the four methods were designed to accelerate this part, exploiting a change of integration variable (Sec.~\ref{sec:change_variable}), the recycle of integration trajectories (Sec.~\ref{sec:trajectory}), and the adoption of adaptive step size (Sec.~\ref{sec:adaptiveStep}). The computational cost also depends heavily on the quality of the initial guess. In this regard, the fourth acceleration method (Sec.~\ref{sec:rtree}) stores previous inputs and solutions in a 5D R-tree, and performs a nearest-neighbor search to generate the initial guess for the next Riemann problem to be solved. Compared to previous efforts on the acceleration of Riemann solvers, the four methods presented in this paper do not introduce new approximations, nor do they require additional pre-processing work. Also, they do not depend on or conflict with each other. These methods have been implemented in a standalone Riemann problem solver~\cite{riemann} and a 3D multi-material flow solver~\cite{m2c}, both of which are open source and under the GPL (Version 3) license.

We presented four example problems to demonstrate and assess these acceleration methods (Sec.~\ref{sec:numericalExperiments}). These problems represent different application areas including underwater explosion (Sec.~\ref{sec:UNDEX}), laser-induced cavitation (Sec.~\ref{sec:laser}), and hypervelocity impact (Sec.~\ref{sec:shocktube} and~\ref{sec:HVI}). They involve some major computational challenges such as large interface motion and deformation, multiple ($>2$) interfaces in contact, drastically different thermodynamic relations, and significant jumps of state variables across interfaces. For two example problems (Sec.~\ref{sec:shocktube} and~\ref{sec:UNDEX}), the proposed numerical methods are verified against reference solutions. For all the problems, the speedup obtained by the proposed acceleration methods was found to be significant. The time spent on solving exact Riemann problems was reduced by a factor of $37$ to $87$. As a result, the total time spent on calculating advective fluxes was also reduced by  a factor of $18$ to $81$. Among the four acceleration methods, the first three were found to be effective in all the test cases. The last one (presented in Sec.~\ref{sec:rtree}) was found to be effective when the local density jump across the material interface is significant. While the simulations presented in this paper were performed using a specific solver that implements the FIVER method, the proposed acceleration methods were not tailored for this solver.  They are generally applicable to multi-material and multiphase flow solvers that utilize the exact solution of Riemann problems; and similar acceleration effects can be expected.

\section*{Declaration of competing interest}

The authors declare that they have no known competing financial
interests or personal relationships that could have appeared
to influence the work reported in this paper.

\section*{Acknowledgement}

The authors gratefully acknowledge the support of the National Science Foundation (NSF) under Award CBET-1751487, the support of the Office of Naval Research (ONR) under Award N00014-19-1-2102, and the support of the National Institutes of Health (NIH) under Award 2R01-DK052985-24A1. W.M. and K.W. also acknowledge the support of U.S. Department of Transportation (DOT) Pipeline and Hazardous Materials Safety Administration under contract number 693JK32250007CAAP.



\end{document}